\documentclass[review]{elsarticle}
\usepackage[utf8]{inputenc}
\usepackage{hyperref,amsmath,amssymb,amsbsy,bbold,color,graphics,graphicx,bigints,relsize,cancel}
\usepackage{lineno,hyperref,appendix}

\modulolinenumbers[5]
\pdfoutput=1
\allowdisplaybreaks

\journal{Journal of \LaTeX\ Templates}

\begin{document}

\begin{frontmatter}

\title{Electrodynamics in geometric algebra}

\author{Sylvain D. Brechet\fnref{myfootnote}}
\address{Institute of Physics, Station 3, Ecole Polytechnique F\'ed\'erale de Lausanne\,-\,EPFL, CH-1015 Lausanne,\,Switzerland}

\ead{sylvain.brechet@epfl.ch}

\begin{abstract}
	
\noindent We consider the electrodynamics of electric charges and currents in vacuum and then generalise our results to the description of a dielectric and magnetic material medium : first in spatial algebra (SA) and then in space-time algebra (STA). Introducing a polarisation multivector $\tilde{P} = \boldsymbol{\tilde{p}} -\,\frac{1}{c}\,\boldsymbol{\tilde{M}}$ and an auxiliary electromagnetic field multivector $G = \varepsilon_0\,F + \tilde{P}$, we express the Maxwell equation in the material medium in SA. Introducing a bound current vector $\tilde{J} = J -\,c\,\nabla\cdot\tilde{P}$ in space-time, the Maxwell equation is then expressed in STA. The wave equation in the material medium is obtained by taking the gradient of the Maxwell equation. For a uniform electromagnetic medium consisting of induced electric and magnetic dipoles, the stress-energy momentum vector is written as $\dot{T}\left(\dot{\nabla}\right) = \frac{1}{c}\,J \cdot F = f$ where $f$ is the electromagnetic force density vector in space-time. Finally, the Maxwell equation in the material medium can be written in STA as a wave equation for the potential vector $A$.
	
\end{abstract}

\end{frontmatter}


\tableofcontents

\section{Introduction}

In his seminal paper of $1865$ entitled ``A dynamical theory of the electromagnetic field''~\cite{Maxwell:1865}, James Clerk Maxwell obtained a fully self consistent field description of electromagnetic phenomena. The electromagnetic fields he used were introduced by Michael Faraday in order to accurately present the results of his experiments. In his earlier attempts to establish a theory of electromagnetic phenomena, Maxwell first used mechanical analogies to explain these phenomena in mechanical terms~\cite{Forbes:2014}. With great insight, he understood that since the mechanical scaffolding of his theory was neither relevant nor needed, he could simply get rid of it. By doing so, Maxwell revealed a beautiful field theory that served as a template for numerous other field theories. It was a paramount paradigm shift in the history of physics that paved the way for the discovery of special relativity presented by Albert Einstein as ``the electrodynamics of moving bodies''~\cite{Einstein:1905}. The prediction of electromagnetic waves played historically a key role for the foundations of quantum mechanics that resulted from the synthesis of the wave mechanics of Erwin Schrödinger~\cite{Schroedinger:1926} and the matrix mechanics of Werner Heisenberg, Max Born and Pascual Jordan~\cite{Born:1925,Born:1926}. The importance of Maxwell's work was clearly stated by Einstein on the centenary of Maxwell's birth : ``We may say that, before Maxwell, physical reality, in so far as it was to represent the process of nature, was thought of as consisting in material particles, whose variations consist only in movements governed by partial differential equations. Since Maxwell’s time, physical reality has been thought of as represented by continuous fields, governed by partial differential equations, and not capable of any mechanical interpretation. This change in the conception of Reality is the most profound and the most fruitful that physics has experienced since the time of Newton.''~\cite{Einstein:1931} Maxwell understood the importance of his theory of electromagnetism and shared his enthusiasm with his cousin Charles Cay : ``I have also a paper afloat, containing an electromagnetic theory of light, which, till I am convinced to the contrary, I hold to be great guns''.~\cite{Longair:2015}

Maxwell's equations are usually presented in textbooks as a set of four vectorial equations~\cite{Jackson:1999,Griffiths:2017}. This was not the direct result of Maxwell's work. It was in fact Oliver Heaviside~\cite{Heaviside:2011} who gathered Maxwell's equations into a set of four vectorial equations in $1884$ using the vector calculus he coformulated with Josiah Willard Gibbs~\cite{Crowe:1967}. In his initial paper published in $1865$, Maxwell wrote $20$ equations in components. Subsequently, in his ``Treatise on Electricity and Magnetism''~\cite{Maxwell:1873}, Maxwell recast his set of $20$ equations in terms of quaternions discovered by William Rowan Hamilton in $1843$. Since, the vector space of Heaviside and Gibbs is ideally suited to describe translations, it was adopted historically as the main geometric framework of physics. However, the quaternion algebra $\mathbb{H}$ is far more suited for the description of rotations than the vector space $\mathbb{R}^3$ where pseudovectors need to be introduced in order to treat rotations. This naturally prompts the question : ``Do we really need to choose or is it possible to take advantage of both framework at the same time ?'' In fact, the good news is that both frameworks belong to a broader mathematical framework called either geometric algebra (GA) $\mathbb{G}^{3}$ or Clifford algebra $C\ell_{3}\left(\mathbb{R}\right)$ that was discovered by William Kingdom Clifford in $1878$ in an article entitled ``Applications of Grassmann's extensive algebra''~\cite{Clifford:1878}. 

In geometric algebra (GA) $\mathbb{G}^{3}$, there are four types of natural geometric entities. First, there are geometric entities of dimension $0$ called scalars, which are oriented points defined by their magnitude and orientation (e.g. positive or negative). Second, there are geometric entities of dimension $1$ called vectors, which are oriented lines defined by their magnitude and orientation. Third, there are geometric entities of dimension $2$ called bivectors, which are oriented surfaces defined by their magnitude and orientation. Fourth, there are geometric entities of dimension $3$ called trivectors or pseudoscalars, which are oriented volume defined by their magnitude and orientation (e.g. positive or negative).  The geometric algebra (GA) $\mathbb{G}^{3}$ is a vector space consisting of multivectors, which are linear combinations of scalars, vectors, bivectors and pseudoscalars. The quaternion algebra $\mathbb{H}$, which consists of linear combination of scalars and bivectors is the even subalgebra of the geometric algebra (GA) $\mathbb{G}^{3}$. The vector space $\mathbb{R}^3$, which consists of linear combination of vectors, is a subspace of the geometric algebra (GA) $\mathbb{G}^{3}$. The geometric algebra (GA) $\mathbb{G}^{3}$ is a vector space endowed with two composition laws : the inner and outer product of two multivectors. The algebraic sum of the inner and outer product of two vectors is called the geometric product, as explained in Appendix~\ref{Spatial algebra}.

The geometric algebra (GA) $\mathbb{G}^{3}$ is also called the spatial algebra (SA). Since special relativity was based on electromagnetism, it is relevant to generalise the spatial algebra (SA) to include an additional temporal dimension. The generalisation of the spatial algebra (SA) $\mathbb{G}^{3}$ to space-time is the geometric algebra (GA) $\mathbb{G}^{1,3}$ or the Clifford algebra $C\ell_{1,3}\left(\mathbb{R}\right)$ called the space-time algebra (STA) developed by Hestenes~\cite{Hestenes:2015}. The spatial algebra (SA) $\mathbb{G}^{3}$ is isomorphic, or equivalent in structure, to the even subalgebra of the space-time algebra (STA) $\mathbb{G}^{1,3}$. Formally, bivectors consisting of the outer product of a space vector and a time vector in $\mathbb{G}^{1,3}$ are isomorphic to spatial vectors in $\mathbb{G}^{3}$, the outer product of these bivectors in $\mathbb{G}^{1,3}$ are isomorphic to spatial bivectors in $\mathbb{G}^{3}$ and the pseudoscalar $I$ in $\mathbb{G}^{3}$ is isomorphic to the pseudoscalar in $\mathbb{G}^{1,3}$.

Geometric algebra in its various forms, namely spatial algebra (SA) $\mathbb{G}^{3}$  or space-time algebra (STA) $\mathbb{G}^{1,3}$, appears to be the most appropriate language to describe physical phenomena since its structure is based on the geometry underlying the natural world. In his memoir entitled ``Reapings and Sowings'', Alexander Grothendieck, one of the greatest mathematician of the $20^{\text{th}}$ century states~\cite{Grothendieck:2015} : ``If there is one thing in mathematics that fascinates me more than anything else (and doubtless always has), it is neither ‘number’ nor ‘size,’ but always form. Among the thousand-and-one faces whereby form chooses to reveal itself to us, the one that fascinates me more than any other and continues to fascinate me, is the structure hidden in mathematical things.''

This article is structured as follows. In the first part, we begin by stating Maxwell's equations in vector space (VS) in Sec.~\ref{Maxwell equations in vector space}. Then, we recast them in spatial algebra (SA) in Sec.~\ref{Maxwell equations in spatial algebra}. In order to do so, we introduce a magnetic induction field bivector $\boldsymbol{B}$ and a auxiliary magnetic field bivector $\boldsymbol{H}$ by duality with the corresponding vectors $\boldsymbol{b}$ and $\boldsymbol{h}$. The duality operation in spatial algebra (SA) is defined in Appendix~\ref{Duality in spatial algebra}. 

In the second part, presented in Sec.~\ref{Maxwell equation in vacuum}-\ref{Poynting theorem in vacuum}, we examine the foundations of electromagnetism in spatial algebra (SA) in the absence of a material medium  which we refer to as ``vacuum''. The term ``vacuum'' does not mean here that there are no electric charges or no electric current but simply that the medium is vacuum. By introducing an electromagnetic multivector $F = \boldsymbol{e} + c\,\boldsymbol{B}$ in spatial algebra (SA) that is algebraically isomorphic to the complex Riemann-Silberstein vector $\boldsymbol{f} = \boldsymbol{e} + i\,c\,\boldsymbol{b}$ in complex vector space $\mathbb{C}^{3}$, we then show in Sec~\ref{Maxwell equation in vacuum}. that the four Maxwell equations are written as a single equation in spatial algebra (SA). In Sec.~\ref{Electromagnetic waves in vacuum}, we show that electromagnetic waves are described quite naturally in spatial algebra (SA). In Sec~\ref{Electromagnetic energy and momentum in vacuum}, we obtain an expression of the energy density and momentum density in terms of the electromagnetic field multivector $F = \boldsymbol{e} + c\,\boldsymbol{B}$, its reverse $F^{\dag} = \boldsymbol{e} -\,c\,\boldsymbol{B}$, the auxiliary electromagnetic field multivector $G = \boldsymbol{d} + \frac{1}{c}\,\boldsymbol{H}$ and its reverse $G^{\dag} = \boldsymbol{d} -\,\frac{1}{c}\,\boldsymbol{H}$, which allows us to establish Poynting's theorem in spatial algebra (SA) in Sec~\eqref{Poynting theorem in vacuum}. In Sec.~\ref{Electric and magnetic potentials}, the electromagnetic field multivector $F$ is expressed in terms of the electric scalar potential $\phi$ and the magnetic vector potential $\boldsymbol{a}$.  

In the third part, shown in Sec.~\ref{Maxwell equation in matter}-\ref{Poynting theorem in matter}, we examine the foundations of electromagnetism in spatial algebra (SA) for a material medium with electric and magnetic dipoles, which we refer to as ``matter'', where we perform a similar analysis as in ``vacuum''.

In the fourth part, presented in Sec.~\ref{Maxwell equation in vacuum STA}-\ref{Stress energy momentum in matter STA}, we recast the results obtained thus far within the space-time algebra (STA). In Sec.~\ref{Maxwell equation in vacuum STA} and~\ref{Maxwell equation in matter STA}, Maxwell's equation in vacuum and matter is written very elegantly in space-time algebra (STA). In Sec.~\ref{Electromagnetic fields STA}, the electromagnetic field multivectors $F$ and $G$ and the electric polarisation multivector $\tilde{P}$ are expressed as bivectors in space-time algebra (STA). In Sec.~\ref{Stress energy momentum in vacuum STA} and~\ref{Stress energy momentum in matter STA}, the stress energy momentum vector is established in vacuum and matter in space-time algebra (STA). In Sec~\ref{Stress energy momentum in matter STA}, we show that the electromagnetic field bivector in space-time $F$ is the curl of the potential vector $A$ in space-time.

Appendix~\ref{Spatial algebra} sets the foundations of spatial algebra (SA) and provides identities for the products of vectors, bivectors and the pseudoscalar. The duality between geometric entities in spatial algebra is presented in Appendix~\ref{Duality in spatial algebra}. The duality of the gradient, the divergence or the curl is establish in Appendix~\ref{Differential duality in spatial algebra}. Dual identities involving geometric entities in spatial algebra is derived in Appendix~\ref{Algebraic identities in spatial algebra}. Dual identities involving the gradient, the divergence or the curl are shown in Appendix~\ref{Differential algebraic identities in spatial algebra}. Finally, Appendix~\ref{Space-time algebra} sets the foundations of space-time algebra (STA) and provides identities for the products of vectors, bivectors and multivectors.

The notation convention adopted in this paper is the following. To distinguish typographically different types of entities in the spatial algebra (SA), we will use in general lower case letters for scalars like the charge density $q$, lower case bold letters for vectors like the electric field $\boldsymbol{e}$, uppercase bold letters for bivectors like the magnetic induction field $\boldsymbol{B}$ and uppercase letters for multivectors like the electromagnetic field $F$. In the space-time algebra (STA), we will in general use uppercase letters for vectors like the electric current density $J$ and for bivectors like the auxiliary electromagnetic field $G$.


\section{Maxwell equations in vector space (VS)}
\label{Maxwell equations in vector space}

\noindent In vector space (VS), electromagnetism is described by a set of four vector fields : the electric field vector $\boldsymbol{e}$, the magnetic field pseudovector $\boldsymbol{b}$, the electric displacement field vector $\boldsymbol{d}$, the auxiliary magnetic field pseudovector $\boldsymbol{h}$. The Maxwell equations relate these fields to the electric charge distribution described by the scalar density field $q$ and the electric current distribution described by the vector density field $\boldsymbol{j}$,~\cite{Jackson:1999,Griffiths:2017}
\begin{align}
\label{Electric Gauss 0}
&\boldsymbol{\nabla}\cdot\boldsymbol{d} = q\\
\label{Maxwell Ampere 0}
&\boldsymbol{\nabla}\times\boldsymbol{h} = \partial_{t}\,\boldsymbol{d} + \boldsymbol{j}\\
\label{Faraday 0}
&\boldsymbol{\nabla}\times\boldsymbol{e} = -\,\partial_{t}\,\boldsymbol{b}\\
\label{Magnetic Gauss 0}
&\boldsymbol{\nabla}\cdot\boldsymbol{b} = 0
\end{align}
The electric Gauss equation~\eqref{Electric Gauss 0} is a scalar equation representing the local electric flux and the magnetic Gauss equation~\eqref{Magnetic Gauss 0} is a pseudoscalar equation representing the local magnetic flux. The Maxwell-Ampère equation~\eqref{Maxwell Ampere 0} is a vectorial equation representing the local electric circulation and the Faraday equation~\eqref{Faraday 0} is pseudovectorial equation representing the local magnetic circulation. 

\noindent The divergence of the Maxwell-Ampère relation~\eqref{Maxwell Ampere 0} is written as,
\begin{equation}\label{div Maxwell Ampere 0}
\boldsymbol{\nabla}\cdot\left(\boldsymbol{\nabla}\times\boldsymbol{h}\right) = \partial_{t}\left(\boldsymbol{\nabla}\cdot\boldsymbol{d}\right) + \boldsymbol{\nabla}\cdot\boldsymbol{j} = 0
\end{equation}
Substituting the electric Gauss equation~\eqref{Electric Gauss 0} into relation~\eqref{div Maxwell Ampere 0}, we obtain the electric continuity equation,
\begin{equation}\label{continuity eq 0}
\partial_{t}\,q + \boldsymbol{\nabla}\cdot\boldsymbol{j} = 0
\end{equation}
The Lorentz force density vector is given by,~\cite{Jackson:1999,Griffiths:2017}
\begin{equation}\label{Lorentz force 0}
\boldsymbol{f} = q\left(\boldsymbol{e} + \boldsymbol{v}\times\boldsymbol{b}\right)
\end{equation}
%


\section{Maxwell equations in spatial algebra (SA)}
\label{Maxwell equations in spatial algebra}

\noindent In spatial algebra (SA), pseudovectors are replaced by bivectors~\cite{Jancewicz:1989}. Pseudovectors are the dual of bivectors. Bivectors respresent oriented plane elements that are orthogonal to the dual pseudovectors. The area or ``magnitude'' of the bivector corresponds to the length or ``norm'' of the dual pseudovector. The orientation of this duality is given by the right hand rule. If the right hand rotates along the oriented plane element defined by the bivector then the dual pseudovector is oriented along the direction of the thumb of the right hand. The converse duality is given by the left hand rule. If the thumb of the left hand is oriented along the pseudovector then the left hand rotates along the oriented plane element defined the dual bivector. Note that this orientation is the opposite of the one that would be obtained with the right hand rule. Thus, the dual of the dual of a bivector yields a bivector with the opposite orientation, which is the opposite of the initial bivector. Similarly, the dual of the dual of a vector yields a vector with the opposite orientation, which is the opposite of the initial vector.~\cite{Macdonald:2011} In view of this duality, we will now recast the Maxwell equations~\eqref{Electric Gauss 0}-\eqref{Magnetic Gauss 0} in terms of vectors and bivectors.

\noindent The auxiliary magnetic field is generated by an electric current circulating in a loop, which is an oriented plane element. The auxiliary magnetic field pseudovector in vector space $\boldsymbol{h}$ is orthogonal to the plane of the loop. It is the dual~\eqref{dual bivector B} of the auxiliary magnetic field bivector $\boldsymbol{H}$ oriented along the circulation of the current density $\boldsymbol{j}$~\cite{Macdonald:2011},
\begin{equation}\label{duality h H}
\boldsymbol{H}^{\ast} = \boldsymbol{h} \qquad\text{where}\qquad \vert\boldsymbol{H}\vert = \vert\boldsymbol{h}\vert
\end{equation}
where the duality is denoted with a $\vphantom{a}\ast$. The dual of this duality~\eqref{dual dual B} is,
\begin{equation}\label{duality H h}
\boldsymbol{h}^{\ast} = \left(\boldsymbol{H}^{\ast}\right)^{\ast} = -\,\boldsymbol{H}
\end{equation}
According to the vectorial duality~\eqref{curl divergence duality ter}, the curl of the auxiliary magnetic field pseudovector $\boldsymbol{h}$ is the opposite of the divergence of the auxiliary magnetic field bivector $\boldsymbol{H}$,
\begin{equation}\label{curl h div H}
\boldsymbol{\nabla}\times\boldsymbol{h} = -\,\boldsymbol{\nabla}\cdot\boldsymbol{H}
\end{equation}
Substituting relation~\eqref{curl h div H} into the Maxwell-Ampère equation~\eqref{Maxwell Ampere 0}, it becomes,
\begin{equation}\label{Maxwell Ampere 00}
\boldsymbol{\nabla}\cdot\boldsymbol{H} = -\,\partial_{t}\,\boldsymbol{d} -\,\boldsymbol{j}
\end{equation}
In vacuum, the magnetic induction field is generated by an electric current circulating in a loop, which is an oriented plane element. The magnetic induction field pseudovector in vector space $\boldsymbol{b}$ is orthogonal to the plane of the loop. It is the dual~\eqref{dual bivector B} of the magnetic induction field bivector $\boldsymbol{B}$ oriented along the circulation of the current density $\boldsymbol{j}$,
\begin{equation}\label{duality b B}
\boldsymbol{B}^{\ast} = \boldsymbol{b} \qquad\text{where}\qquad \vert\boldsymbol{B}\vert = \vert\boldsymbol{b}\vert
\end{equation}
The dual of this duality~\eqref{dual dual B} is,
\begin{equation}\label{duality B b}
\boldsymbol{b}^{\ast} = \left(\boldsymbol{B}^{\ast}\right)^{\ast} = -\,\boldsymbol{B}
\end{equation}
According to the vectorial duality~\eqref{curl divergence duality ter}, the curl pseudovector of the electric field vector $\boldsymbol{e}$ is the dual of the curl bivector of the electric field vector $\boldsymbol{e}$,
\begin{equation}\label{duality curl e}
\boldsymbol{\nabla}\times\boldsymbol{e} = \left(\boldsymbol{\nabla}\wedge\boldsymbol{e}\right)^{\ast}
\end{equation}
In view of relations~\eqref{duality b B} and~\eqref{duality curl e}, the Faraday equation~\eqref{Faraday 0} is recast as,
\begin{equation}\label{Faraday 000}
\left(\boldsymbol{\nabla}\wedge\boldsymbol{e}\right)^{\ast} = -\,\partial_{t}\,\boldsymbol{B}^{\ast}
\end{equation}
The opposite of the dual of the Faraday equation~\eqref{Faraday 000} is given by,
\begin{equation}\label{Faraday 00}
\boldsymbol{\nabla}\wedge\boldsymbol{e} = -\,\partial_{t}\,\boldsymbol{B}
\end{equation}
Using the vectorial duality~\eqref{curl divergence duality more} and~\eqref{curl divergence duality quad}, the dual of the divergence of the magnetic induction field pseudovector $\boldsymbol{b}$ is the opposite of the curl of the magnetic induction field bivector $\boldsymbol{B}$,
\begin{equation}\label{div b curl B}
\left(\boldsymbol{\nabla}\cdot\boldsymbol{b}\right)^{\ast} = \boldsymbol{\nabla}\wedge\boldsymbol{b}^{\ast} = -\,\boldsymbol{\nabla}\wedge\boldsymbol{B}
\end{equation}
In view of relation~\eqref{div b curl B}, the dual the magnetic Gauss equation~\eqref{Magnetic Gauss 0} is given by,
\begin{equation}\label{Magnetic Gauss 00}
\boldsymbol{\nabla}\wedge\boldsymbol{B} = 0
\end{equation}
Thus, in spatial algebra (SA), the Maxwell equations are written as,
\begin{align}
\label{Electric Gauss}
&\boldsymbol{\nabla}\cdot\boldsymbol{d} = q\\
\label{Maxwell Ampere}
&\boldsymbol{\nabla}\cdot\boldsymbol{H} = -\,\partial_{t}\,\boldsymbol{d} -\,\boldsymbol{j}\\
\label{Faraday}
&\boldsymbol{\nabla}\wedge\boldsymbol{e} = -\,\partial_{t}\,\boldsymbol{B}\\
\label{Magnetic Gauss}
&\boldsymbol{\nabla}\wedge\boldsymbol{B} = 0
\end{align}
The electric Gauss equation~\eqref{Electric Gauss} is a scalar equation, the Maxwell-Ampère equation~\eqref{Maxwell Ampere} is a vectorial equation, the Faraday equation~\eqref{Faraday} is a bivectorial equation or pseudovectorial equation, and the magnetic Gauss equation~\eqref{Magnetic Gauss} is a trivectorial equation, or pseudoscalar equation.

\noindent The divergence of the Maxwell-Ampère relation~\eqref{Maxwell Ampere} is written as,
\begin{equation}\label{div Maxwell Ampere}
\boldsymbol{\nabla}\cdot\left(\boldsymbol{\nabla}\cdot\boldsymbol{H}\right) = -\,\partial_{t}\left(\boldsymbol{\nabla}\cdot\boldsymbol{d}\right) -\,\boldsymbol{\nabla}\cdot\boldsymbol{j} = 0
\end{equation}
Substituting the electric Gauss equation~\eqref{Electric Gauss} into relation~\eqref{div Maxwell Ampere}, we obtain the electric continuity equation,
\begin{equation}\label{continuity eq}
\partial_{t}\,q + \boldsymbol{\nabla}\cdot\boldsymbol{j} = 0
\end{equation}
According to the vectorial duality~\eqref{wedge and cross products duality bis},~\eqref{wedge and cross products duality ter} and~\eqref{duality B b}, and to the antisymmetry of the inner product with a bivector~\eqref{inner product v B antisymmetric}, the cross product of the velocity vector $\boldsymbol{v}$ and the magnetic induction field pseudovector $\boldsymbol{b}$ is the inner product of the magnetic induction field bivector $\boldsymbol{B}$ and the velocity vector $\boldsymbol{v}$,
\begin{equation}\label{Lorentz 0}
\boldsymbol{v}\times\boldsymbol{b} = \left(\boldsymbol{v}\wedge\boldsymbol{b}\right)^{\ast} = \boldsymbol{v}\cdot\boldsymbol{b}^{\ast} = -\,\boldsymbol{v}\cdot\boldsymbol{B} = \boldsymbol{B}\cdot\boldsymbol{v}
\end{equation}
In view of the vectorial duality~\eqref{Lorentz 0}, the Lorentz force density vector~\eqref{Lorentz force 0} is recast as,
\begin{equation}\label{Lorentz force}
\boldsymbol{f} = q\left(\boldsymbol{e} + \boldsymbol{B}\cdot\boldsymbol{v}\right)
\end{equation}
%


\section{Maxwell equation in vacuum (SA)}
\label{Maxwell equation in vacuum}

\noindent We now show that in vacuum, the Maxwell equations~\eqref{Electric Gauss}-\eqref{Magnetic Gauss} reduce to a single equation. The term ``vacuum'' does not mean here that there are no electric charges or no electric current but simply that the medium is vacuum. In vacuum, the vectorial linear electric constitutive relation is given by,~\cite{Jackson:1999,Griffiths:2017}
\begin{equation}\label{electric constitutive relation}
\boldsymbol{d} = \varepsilon_0\,\boldsymbol{e}
\end{equation}
and the pseudovectorial linear magnetic constitutive relation is given by,~\cite{Jackson:1999,Griffiths:2017}
\begin{equation}\label{magnetic constitutive relation 0}
\boldsymbol{b} = \mu_0\,\boldsymbol{h}
\end{equation}
The dual of the constitutive relation~\eqref{magnetic constitutive relation 0} is written as,
\begin{equation}\label{magnetic constitutive dual}
\boldsymbol{b}^{\ast} = \mu_0\,\boldsymbol{h}^{\ast}
\end{equation}
Using the duality relations~\eqref{duality H h} and~\eqref{duality B b}, relation~\eqref{magnetic constitutive dual} yields the bivectorial linear magnetic constitutive relation,
\begin{equation}\label{magnetic constitutive relation}
\boldsymbol{B} = \mu_0\,\boldsymbol{H}
\end{equation}
where the vacuum electric permittivity $\varepsilon_0$ and the vacuum magnetic permeability $\mu_0$ are related to the speed of light $c$ by,~\cite{Jackson:1999,Griffiths:2017}
\begin{equation}\label{light}
\varepsilon_0\,\mu_0 = \frac{1}{c^2}
\end{equation}
Substituting the linear constitutive relations~\eqref{electric constitutive relation} and~\eqref{magnetic constitutive relation} into the electric Gauss equation~\eqref{Electric Gauss} and the Maxwell-Ampère equation~\eqref{Maxwell Ampere} and taking into account the definition~\eqref{light} of the speed of light, we obtain,
\begin{align}
\label{Electric Gauss bis}
&\boldsymbol{\nabla}\cdot\boldsymbol{e} = \frac{q}{\varepsilon_0}\\
\label{Maxwell Ampere bis}
&\boldsymbol{\nabla}\cdot\boldsymbol{B} = -\,\frac{1}{c^2}\,\partial_{t}\,\boldsymbol{e} -\,\mu_0\,\boldsymbol{j}
\end{align}
In vector space, the Riemann-Silberstein vector $\boldsymbol{f}$ is defined as,~\cite{Silberstein:1907}
\begin{equation}\label{Riemann Silberstein}
\boldsymbol{f} = \boldsymbol{e} + i\,c\,\boldsymbol{b}
\end{equation}
where $i$ is the unit imaginary number. The Riemann-Silberstein vector $\boldsymbol{f} \in \mathbb{C}^{3}$ is a complex vector field. In spatial algebra (SA), the unit imaginary number $i$ is replaced by the unit pseudoscalar $I$~\eqref{pseudoscalar squared} that has the same essential property, namely $i^2 = I^2 = -\,1$. Thus, the Riemann-Silberstein vector $\boldsymbol{f}$ becomes a multivector called the electromagnetic multivector field,
\begin{equation}\label{Riemann Silberstein multivector 0}
F = \boldsymbol{e} + c\,I\,\boldsymbol{b}
\end{equation}
In three dimensions, the pseudoscalar $I$ commutes with the vector field $\boldsymbol{b}$. Thus, according to the duality~\eqref{dual vector v} and~\eqref{duality B b},
\begin{equation}\label{identity Silberstein}
I\,\boldsymbol{b} = \boldsymbol{b}\,I = -\,\boldsymbol{b}\,I^{-1} = -\,\boldsymbol{b}^{\ast} = \boldsymbol{B}
\end{equation}
This duality operation is defined in terms of the pseudoscalar in Appendix~\eqref{Duality in spatial algebra}. Substituting identity~\eqref{identity Silberstein} into relation~\eqref{Riemann Silberstein multivector 0}, the electromagnetic multivector field~\eqref{Riemann Silberstein multivector 0} becomes,
\begin{equation}\label{Riemann Silberstein multivector}
F = \boldsymbol{e} + c\,\boldsymbol{B}
\end{equation}
The very elegant expression~\eqref{Riemann Silberstein multivector} of the electromagnetic multivector $F$ was used only by few authors like Arthur~\cite{Arthur:2011} and Macdonald~\cite{Macdonald:2011} whereas most authors like Doran and Lasenby~\cite{Lasenby:2003} or Hestenes~\cite{Hestenes:2015} write the electromagnetic field multivector~\eqref{Riemann Silberstein multivector 0} in terms of the magnetic induction field pseudovector $\boldsymbol{b}$. The choice made here seem much more natural in a sense since it reflects the underlying geometry of space. In view of the Faraday equation~\eqref{Faraday}, the Maxwell-Ampère relation~\eqref{Maxwell Ampere} and the speed of light~\eqref{light}, the partial time derivative of the electromagnetic multivector field~\eqref{Riemann Silberstein multivector} yields,
\begin{equation}\label{F t derivative}
\partial_t\,F = \partial_t\,\boldsymbol{e} + c\,\partial_t\,\boldsymbol{B}
\end{equation}
The gradient of the electromagnetic multivector field~\eqref{Riemann Silberstein multivector} is given by,
\begin{equation}\label{F gradient}
\boldsymbol{\nabla}\,F = \boldsymbol{\nabla}\cdot F + \boldsymbol{\nabla}\wedge F = \boldsymbol{\nabla}\cdot\boldsymbol{e} + \boldsymbol{\nabla}\wedge\boldsymbol{e} + c\,\boldsymbol{\nabla}\cdot\boldsymbol{B} + c\,\boldsymbol{\nabla}\wedge\boldsymbol{B}
\end{equation}
In view of the electric Gauss equation~\eqref{Electric Gauss bis}, the Faraday equation~\eqref{Faraday}, the Maxwell-Ampère relation~\eqref{Maxwell Ampere bis}, the magnetic Gauss equation~\eqref{Magnetic Gauss}, the linear electric constitutive relation~\eqref{electric constitutive relation}, the linear magnetic constitutive relation~\eqref{magnetic constitutive relation} and the speed of light~\eqref{light}, the gradient of the electromagnetic multivector field~\eqref{F gradient} becomes, 
\begin{equation}\label{F gradient bis}
\boldsymbol{\nabla}\,F = \frac{q}{\varepsilon_0} -\,\partial_t\,\boldsymbol{B} -\,\frac{1}{c}\,\partial_{t}\,\boldsymbol{e} -\,\frac{1}{\varepsilon_0\,c}\,\boldsymbol{j}
\end{equation}
Taking into account the partial time derivative~\eqref{F t derivative} and the gradient~\eqref{F gradient bis} of the electromagnetic multivector field and the definition~\eqref{light} of the propagation velocity, we obtain,~\cite{Lasenby:2003}
\begin{equation}\label{F eq}
\left(\frac{1}{c}\,\partial_t + \boldsymbol{\nabla}\right)\varepsilon_0\,F = q -\,\frac{1}{c}\,\boldsymbol{j}
\end{equation}
which is the Maxwell equation in spatial algebra. In vacuum, the auxiliary electromagnetic multivector field is defined as,
\begin{equation}\label{G constitutive relation vacuum}
G = \varepsilon_0\,F
\end{equation}
which is the linear electromagnetic constitutive relation. Using the definition~\eqref{G constitutive relation vacuum} of the auxiliary electromagnetic multivector field $G$ in vacuum, the Maxwell equation~\eqref{F eq} is recast as, 
\begin{equation}\label{G eq}
\left(\frac{1}{c}\,\partial_t + \boldsymbol{\nabla}\right)G = q -\,\frac{1}{c}\,\boldsymbol{j}
\end{equation}
Using the constitutive relations~\eqref{electric constitutive relation} and~\eqref{magnetic constitutive relation} and the speed of light~\eqref{light} the auxiliary electromagnetic multivector field $G$ is recast as,~\cite{Lasenby:2003}
\begin{equation}\label{Riemann Silberstein multivector G}
G = \boldsymbol{d} + \frac{1}{c}\,\boldsymbol{H}
\end{equation}
The electric Gauss equation~\eqref{Electric Gauss} and the Maxwell-Ampère equation~\eqref{Maxwell Ampere} describing the electrodynamic phenomena driven by electric charges and currents are expressed as the divergence of the auxiliary electromagnetic multivector field $G$ in vacuum,
\begin{equation}\label{div G}
\boldsymbol{\nabla}\cdot G = \boldsymbol{\nabla}\cdot\boldsymbol{d} + \frac{1}{c}\,\boldsymbol{\nabla}\cdot\boldsymbol{H} = q -\,\frac{1}{c}\left(\boldsymbol{j} + \partial_{t}\,\boldsymbol{d}\right)
\end{equation}
Similarly, the Faraday equation~\eqref{Faraday} and the Magnetic Gauss equation~\eqref{Magnetic Gauss} describing the electrodynamic phenomena that do not involve directly electric charges and currents are expressed as the curl of the electromagnetic multivector field $F$,
\begin{equation}\label{curl F}
\boldsymbol{\nabla}\wedge F = \boldsymbol{\nabla}\wedge\boldsymbol{e} + c\,\boldsymbol{\nabla}\wedge\boldsymbol{B} = -\,\partial_{t}\,\boldsymbol{B}
\end{equation}
%


\section{Electromagnetic waves in vacuum (SA)}
\label{Electromagnetic waves in vacuum}

\noindent In the spatial algebra (SA), electromagnetic waves in vacuum are a direct and straightforward consequence of the Maxwell equation~\eqref{G eq}, as expected, but it is nonetheless quite beautiful. Indeed, multiplying the Maxwell equation~\eqref{G eq} by $c^{-1}\partial_t -\,\boldsymbol{\nabla}$ and using the relations,
\begin{equation}\label{G eq quad left}
\left(\frac{1}{c}\,\partial_t -\,\boldsymbol{\nabla}\right)\left(\frac{1}{c}\,\partial_t + \boldsymbol{\nabla}\right)G = \left(\frac{1}{c^2}\,\partial_t^2 -\,\boldsymbol{\nabla}^2\right)G
\end{equation}
and
\begin{equation}\label{G eq quad right}
\left(\frac{1}{c}\,\partial_t -\,\boldsymbol{\nabla}\right)\left(q -\,\frac{1}{c}\,\boldsymbol{j}\right) = \frac{1}{c}\,\partial_t\,q -\,\boldsymbol{\nabla}\,q -\,\frac{1}{c^2}\,\partial_t\,\boldsymbol{j} + \frac{1}{c}\,\boldsymbol{\nabla}\cdot\boldsymbol{j} + \frac{1}{c}\,\boldsymbol{\nabla}\wedge\boldsymbol{j}
\end{equation}
together with the electric continuity equation~\eqref{continuity eq}, we obtain,
\begin{equation}\label{wave eq G}
\left(\frac{1}{c^2}\,\partial_t^2 -\,\boldsymbol{\nabla}^2\right)G = -\,\boldsymbol{\nabla}\,q -\,\frac{1}{c^2}\,\partial_t\,\boldsymbol{j} + \frac{1}{c}\,\boldsymbol{\nabla}\wedge\boldsymbol{j}
\end{equation}
which is the wave equation for the auxiliary electromagnetic multivector field $G$ in vacuum. Substituting the auxiliary electromagnetic multivector field~\eqref{Riemann Silberstein multivector G} in vacuum into the wave equation~\eqref{wave eq G} yields,
\begin{equation}\label{wave eq G bis}
\left(\frac{1}{c^2}\,\partial_t^2 -\,\boldsymbol{\nabla}^2\right)\left(\boldsymbol{d} + \frac{1}{c}\,\boldsymbol{H}\right) = -\,\boldsymbol{\nabla}\,q -\,\frac{1}{c^2}\,\partial_t\,\boldsymbol{j} + \frac{1}{c}\,\boldsymbol{\nabla}\wedge\boldsymbol{j}
\end{equation}
Identifying the vectorial terms in relation~\eqref{wave eq G bis}, we obtain the wave equation for the electric displacement field $\boldsymbol{d}$,
\begin{equation}\label{wave eq d}
\left(\frac{1}{c^2}\,\partial_t^2 -\,\boldsymbol{\nabla}^2\right)\boldsymbol{d} = -\,\boldsymbol{\nabla}\,q -\,\frac{1}{c^2}\,\partial_t\,\boldsymbol{j}
\end{equation}
Identifying the bivectorial terms in relation~\eqref{wave eq G bis}, we obtain the wave equation for the auxiliary magnetic field $\boldsymbol{H}$,
\begin{equation}\label{wave eq H}
\left(\frac{1}{c^2}\,\partial_t^2 -\,\boldsymbol{\nabla}^2\right)\boldsymbol{H} = \boldsymbol{\nabla}\wedge\boldsymbol{j}
\end{equation}
Dividing the wave equation~\eqref{wave eq d} for the electric displacement field $\boldsymbol{d}$ by $\varepsilon_0$ yields the wave equation for the electric field $\boldsymbol{e}$,
\begin{equation}\label{wave eq e}
\left(\frac{1}{c^2}\,\partial_t^2 -\,\boldsymbol{\nabla}^2\right)\boldsymbol{e} = -\,\frac{1}{\varepsilon_0}\,\boldsymbol{\nabla}\,q -\,\frac{1}{\varepsilon_0\,c^2}\,\partial_t\,\boldsymbol{j}
\end{equation}
Multiplying the wave equation~\eqref{wave eq H} for the auxiliary magnetic field $\boldsymbol{H}$ by $\mu_0$ yields the wave equation for the magnetic field $\boldsymbol{B}$,
\begin{equation}\label{wave eq B}
\left(\frac{1}{c^2}\,\partial_t^2 -\,\boldsymbol{\nabla}^2\right)\boldsymbol{B} = \mu_0\,\boldsymbol{\nabla}\wedge\boldsymbol{j}
\end{equation}
%


\section{Electromagnetic energy and momentum in vacuum (SA)}
\label{Electromagnetic energy and momentum in vacuum}

\noindent The electromagnetic field multivector $F$ is a linear combination of the electric field vector $\boldsymbol{e}$ and the magnetic induction field bivector $\boldsymbol{B}$ according to relation~\eqref{Riemann Silberstein multivector}. Similarly, the auxiliary electromagnetic field multivector $G$ is a linear combination of the electric displacement field vector $\boldsymbol{d}$ and the auxiliary magnetic field bivector $\boldsymbol{H}$ according to relation~\eqref{Riemann Silberstein multivector G}. Since the electromagnetic energy density $e$ and the electromagnetic momentum density $\boldsymbol{p}$ in vacuum are determined by these fields, they can be recast in terms of the multivectors $F$ and $G$. The electromagnetic energy density in vacuum is written in vector space (VS) as,~\cite{Jackson:1999,Griffiths:2017}
\begin{equation}\label{electromagnetic energy density 00}
e = \frac{1}{2}\,\boldsymbol{d}\cdot\boldsymbol{e} + \frac{1}{2}\,\boldsymbol{h}\cdot\boldsymbol{b}
\end{equation}
Using the linear electric constitutive relation~\eqref{electric constitutive relation} and the linear magnetic constitutive relation~\eqref{magnetic constitutive relation 0}, relation~\eqref{electromagnetic energy density 00} is recast as,
\begin{equation}\label{electromagnetic energy density 0 bis}
e = \frac{1}{2}\,\varepsilon_0\,\boldsymbol{e}^2 + \frac{1}{2\,\mu_0}\,\boldsymbol{b}^2
\end{equation}
According to relations~\eqref{duality b B},~\eqref{modulus v},~\eqref{modulus B}, the square of the magnetic induction field bivector $\boldsymbol{B}$ is the opposite of the square of the magnetic induction field pseudovector $\boldsymbol{b}$,
\begin{equation}\label{bivector B squared}
\boldsymbol{b}^2 = \vert\boldsymbol{b}\vert^2 = \vert\boldsymbol{B}\vert^2 = \boldsymbol{B}\,\boldsymbol{B}^{\dag} = -\,\boldsymbol{B}^2
\end{equation}
where $\boldsymbol{B}^{\dag} = -\,\boldsymbol{B}$ is the reverse~\eqref{reverse B} of the magnetic field bivector $\boldsymbol{B}$. In view of relation~\eqref{bivector B squared}, the electromagnetic energy density~\eqref{electromagnetic energy density} is recast in spatial algebra (SA) as,
\begin{equation}\label{electromagnetic energy density}
e = \frac{1}{2}\,\varepsilon_0\,\boldsymbol{e}^2 -\,\frac{1}{2\,\mu_0}\,\boldsymbol{B}^2
\end{equation}
Taking into account the linear electric constitutive relation~\eqref{electric constitutive relation} and the linear magnetic constitutive relation~\eqref{magnetic constitutive relation 0}, the electromagnetic energy density~\eqref{electromagnetic energy density} is recast as,
\begin{equation}\label{electromagnetic energy density deHB}
e = \frac{1}{2}\,\boldsymbol{d}\cdot\boldsymbol{e} -\,\frac{1}{2}\,\boldsymbol{H}\cdot\boldsymbol{B}
\end{equation}
The electromagnetic momentum density in vacuum is written in vector space (VS) as,~\cite{Jackson:1999,Griffiths:2017}
\begin{equation}\label{electromagnetic momentum}
\boldsymbol{p} = \frac{1}{c^2}\,\boldsymbol{e}\times\boldsymbol{h}
\end{equation}
where $\boldsymbol{e}\times\boldsymbol{h}$ is the Poynting vector. By duality using the identities \eqref{wedge and cross products duality}, \eqref{vector duality}, \eqref{duality H h} and~\eqref{inner product v B antisymmetric}, the electromagnetic momentum density~\eqref{electromagnetic momentum} in recast in spatial algebra (SA) as,
\begin{equation}\label{electromagnetic momentum density 00}
\boldsymbol{p} = \frac{1}{c^2}\left(\boldsymbol{e}\wedge\boldsymbol{h}\right)^{\ast} = \frac{1}{c^2}\,\boldsymbol{e}\cdot\boldsymbol{h}^{\ast} = -\,\frac{1}{c^2}\,\boldsymbol{e}\cdot\boldsymbol{H} = \frac{1}{c^2}\,\boldsymbol{H}\cdot\boldsymbol{e}
\end{equation}
Thus, in spatial algebra, in view of the magnetic constitutive relation~\eqref{magnetic constitutive relation}, the electromagnetic momentum density in vacuum~\eqref{electromagnetic momentum density 00} is written as,
\begin{equation}\label{electromagnetic momentum density}
\boldsymbol{p} = \frac{1}{c^2}\,\boldsymbol{H}\cdot\boldsymbol{e} = \frac{1}{c^2}\,\frac{\boldsymbol{B}}{\mu_0}\cdot\boldsymbol{e}
\end{equation}
where $\boldsymbol{H}\cdot\boldsymbol{e}$ is the Poynting vector. According to the definition~\eqref{Riemann Silberstein multivector}, the reverse of the electromagnetic multivector field $F$ is written as,
\begin{equation}\label{Riemann Silberstein multivector F reverse}
F^{\dag} = \boldsymbol{e}^{\dag} + c\,\boldsymbol{B}^{\dag} = \boldsymbol{e} -\,c\,\boldsymbol{B}
\end{equation}
The reversion operation, defined in Appendix~\eqref{Duality in spatial algebra}, reverses the order of the basis vectors in the different geometric entities of a multivector in spatial algebra. In view of the multivectors~\eqref{Riemann Silberstein multivector} and~\eqref{Riemann Silberstein multivector F reverse}, the electric vector field is recast as,
\begin{equation}\label{e as multivector}
\boldsymbol{e} = \frac{1}{2}\left(F + F^{\dag}\right)
\end{equation}
and the magnetic induction bivector field is recast as,
\begin{equation}\label{B as multivector}
\boldsymbol{B} = \frac{1}{2\,c}\left(F -\,F^{\dag}\right)
\end{equation}
According to the definition~\eqref{Riemann Silberstein multivector G}, the reverse of the auxiliary electromagnetic multivector field $G$ is written as,
\begin{equation}\label{Riemann Silberstein multivector G reverse}
G^{\dag} = \boldsymbol{d}^{\dag} + \frac{1}{c}\,\boldsymbol{H}^{\dag} = \boldsymbol{d} -\,\frac{1}{c}\,\boldsymbol{H}
\end{equation}
The reverse of the linear electromagnetic constitutive relation~\eqref{G constitutive relation vacuum} is written as,
\begin{equation}\label{G constitutive relation vacuum reverse}
G^{\dag} = \varepsilon_0\,F^{\dag}
\end{equation}
In view of the multivectors~\eqref{Riemann Silberstein multivector} and~\eqref{Riemann Silberstein multivector F reverse}, the electric displacement vector field is recast as,
\begin{equation}\label{d as multivector}
\boldsymbol{d} = \frac{1}{2}\left(G + G^{\dag}\right)
\end{equation}
and the auxiliary magnetic bivector field is recast as,
\begin{equation}\label{H as multivector}
\boldsymbol{H} = \frac{c}{2}\left(G -\,G^{\dag}\right)
\end{equation}
Using the symmetry of the inner product of two vectors~\eqref{inner product space u v} and two bivectors~\eqref{geometric product bivectors inner}, the electromagnetic energy density~\eqref{electromagnetic energy density deHB} is recast as,
\begin{equation}\label{electromagnetic energy density 0}
e = \frac{1}{2}\left(\boldsymbol{d}\,\boldsymbol{e} + \boldsymbol{e}\,\boldsymbol{d}\right) -\,\frac{1}{2}\left(\boldsymbol{H}\,\boldsymbol{B} + \boldsymbol{B}\,\boldsymbol{H}\right)
\end{equation}
Using relations~\eqref{e as multivector},~\eqref{e as multivector},~\eqref{d as multivector} and~\eqref{H as multivector}, the electromagnetic energy density~\eqref{electromagnetic energy density 0} is expressed in terms of the geometric product of the electromagnetic multivectors as,
\begin{equation}\label{electromagnetic energy density 1}
\begin{split}
&e = \frac{1}{16}\Big(\left(G + G^{\dag}\right)\left(F + F^{\dag}\right) + \left(F + F^{\dag}\right)\left(G + G^{\dag}\right)\Big)\\
&\phantom{e =} -\,\frac{1}{16}\Big(\left(G -\,G^{\dag}\right)\left(F -\,F^{\dag}\right) + \left(F -\,F^{\dag}\right)\left(G -\,G^{\dag}\right)\Big)
\end{split}
\end{equation}
which reduces to,
\begin{equation}\label{electromagnetic energy density G}
e = \frac{1}{8}\left(G\,F^{\dag} + F\,G^{\dag} + G^{\dag}F + F^{\dag}G\right)
\end{equation}
Using the antisymmetry of the inner product of a vector and a bivector~\eqref{inner product v B antisymmetric} the electromagnetic momentum density~\eqref{electromagnetic momentum} is recast as,
\begin{equation}\label{electromagnetic momentum density 0}
\boldsymbol{p} = \frac{1}{2\,c^2}\left(\boldsymbol{H}\,\boldsymbol{e} -\,\boldsymbol{e}\,\boldsymbol{H}\right)
\end{equation}
Using relations~\eqref{e as multivector},~\eqref{e as multivector},~\eqref{d as multivector} and~\eqref{H as multivector}, the electromagnetic momentum density~\eqref{electromagnetic momentum density 0} is expressed in terms of the geometric product of the electromagnetic multivectors as,
\begin{equation}\label{electromagnetic momentum density 1}
\boldsymbol{p} = \frac{1}{8\,c}\Big(\left(G -\,G^{\dag}\right)\left(F + F^{\dag}\right) -\,\left(F + F^{\dag}\right)\left(G -\,G^{\dag}\right)\Big)
\end{equation}
which is expressed as,
\begin{equation}\label{electromagnetic momentum density bis}
\begin{split}
&\boldsymbol{p} = \frac{1}{8\,c}\left(G\,F^{\dag} + F\,G^{\dag} -\,G^{\dag}F -\,F^{\dag}G\right)\\
&\phantom{\boldsymbol{p} =} + \frac{1}{8\,c}\left(G\,F + F^{\dag}G^{\dag} -\,F\,G -\,G^{\dag}F^{\dag}\right)
\end{split}
\end{equation}
In vacuum, in view of the linear electromagnetic constitutive relations~\eqref{G constitutive relation vacuum},~\eqref{G constitutive relation vacuum reverse}, the electromagnetic energy density~\eqref{electromagnetic energy density G} reduces,
\begin{equation}\label{electromagnetic energy density F}
e = \frac{1}{8}\,\left(G\,F^{\dag} + F\,G^{\dag} + G^{\dag}F + F^{\dag}G\right) = \frac{1}{4}\,\varepsilon_0\left(F\,F^{\dag} + F^{\dag}F\right)
\end{equation}
and the electromagnetic momentum density~\eqref{electromagnetic momentum density} reduces to,~\cite{Arthur:2011}
\begin{equation}\label{electromagnetic momentum density F}
\boldsymbol{p} = \frac{1}{8\,c}\left(G\,F^{\dag} + F\,G^{\dag} -\,G^{\dag}F -\,F^{\dag}G\right) = \frac{1}{4}\,\frac{\varepsilon_0}{c}\left(F\,F^{\dag} -\,F^{\dag}F\right)
\end{equation}
Thus, according to relations~\eqref{electromagnetic energy density F} and~\eqref{electromagnetic momentum density F} in vacuum,~\cite{Arthur:2011}
\begin{equation}\label{electromagnetic invariant F}
e + \boldsymbol{p}\,c = \frac{1}{4}\left(G\,F^{\dag} + F\,G^{\dag}\right) = \frac{1}{2}\,\varepsilon_0\,F\,F^{\dag}
\end{equation}
%


\section{Poynting theorem in vacuum (SA)}
\label{Poynting theorem in vacuum}

\noindent The continuity equation for the electromagnetic energy, called the Poynting theorem, is written in vacuum in terms of the time derivative of the energy density and the energy flux which is a multiple of the electromagnetic momentum density. The time derivative of the electromagnetic energy density~\eqref{electromagnetic energy density} in vacuum is given by,
\begin{equation}\label{time derivative energy density vacuum}
\partial_t\,e = \varepsilon_0\,\boldsymbol{e}\cdot\partial_t\,\boldsymbol{e} -\,\frac{1}{\mu_0}\,\boldsymbol{B}\cdot\partial_t\,\boldsymbol{B}
\end{equation}
Using the electric constitutive equation~\eqref{electric constitutive relation} and the magnetic constitutive equation~\eqref{magnetic constitutive relation}, relation~\eqref{time derivative energy density vacuum} is recast as,
\begin{equation}\label{time derivative energy density vacuum bis}
\partial_t\,e = \boldsymbol{e}\cdot\partial_t\,\boldsymbol{d} -\,\boldsymbol{H}\cdot\partial_t\,\boldsymbol{B}
\end{equation}
Using the Mawell-Ampère equation~\eqref{Maxwell Ampere} and the Faraday equation~\eqref{Faraday}, the time derivative of the electromagnetic energy density~\eqref{time derivative energy density vacuum bis} is recast as,
\begin{equation}\label{time derivative energy density vacuum ter}
\partial_t\,e = -\,\boldsymbol{e}\cdot\left(\boldsymbol{\nabla}\cdot\boldsymbol{H} + \boldsymbol{j}\right) + \boldsymbol{H}\cdot\left(\boldsymbol{\nabla}\wedge\boldsymbol{e}\right)
\end{equation}
Thus,
\begin{equation}\label{time derivative energy density vacuum quad}
\partial_t\,e + \left(\boldsymbol{\nabla}\cdot\boldsymbol{H}\right)\cdot\boldsymbol{e} -\,\boldsymbol{H}\cdot\left(\boldsymbol{\nabla}\wedge\boldsymbol{e}\right) = -\,\boldsymbol{j}\cdot\boldsymbol{e}
\end{equation}
Using the algebraic identity~\eqref{div vec bivec hex},
\begin{equation}\label{id poynting}
\boldsymbol{\nabla}\cdot\left(\boldsymbol{H}\cdot\boldsymbol{e}\right) = \left(\boldsymbol{\nabla}\cdot\boldsymbol{H}\right)\cdot\boldsymbol{e} -\,\boldsymbol{H}\cdot\left(\boldsymbol{\nabla}\wedge\boldsymbol{e}\right)
\end{equation}
relation yields~\eqref{time derivative energy density vacuum quad} the Poynting's theorem,
\begin{equation}\label{Poynting's theorem vacuum}
\partial_t\,e + \boldsymbol{\nabla}\cdot\left(\boldsymbol{H}\cdot\boldsymbol{e}\right) = -\,\boldsymbol{j}\cdot\boldsymbol{e}
\end{equation}
where $\boldsymbol{H}\cdot\boldsymbol{e}$ is the Poynting vector, which is the electromagnetic current density. In view of the electromagnetic momentum density~\eqref{electromagnetic momentum density}, Poynting's theorem~\eqref{Poynting's theorem vacuum} in vacuum is recast as,~\cite{Jackson:1999,Griffiths:2017}
\begin{equation}\label{Poynting's theorem vacuum bis}
\frac{1}{c}\,\partial_t\,e + \boldsymbol{\nabla}\cdot\left(\boldsymbol{p}\,c\right) = -\,\frac{1}{c}\,\boldsymbol{j}\cdot\boldsymbol{e}
\end{equation}
%


\section{Electric and magnetic potentials (SA)}
\label{Electric and magnetic potentials}

\noindent In vector space (VS), the electric vector field $\boldsymbol{e}$ and the magnetic induction pseudovector field $\boldsymbol{b}$ can be expressed entirely in terms of the electric scalar potential $\phi$ and magnetic vector potential $\boldsymbol{a}$. This implies that the electromagnetic multivector field $F$ can be entirely expressed in terms of these potentials. The electric vector field $\boldsymbol{e}$ is expressed in terms of the electric scalar potential $\phi$ and the magnetic vector potential $\boldsymbol{a}$ as,~\cite{Jackson:1999,Griffiths:2017}
\begin{equation}\label{e potentials}
\boldsymbol{e} = -\,\boldsymbol{\nabla}\,\phi -\,\partial_t\,\boldsymbol{a}
\end{equation}
and the magnetic induction vector field $\boldsymbol{b}$ is expressed in terms of the vector potential $\boldsymbol{a}$ as,~\cite{Jackson:1999,Griffiths:2017}
\begin{equation}\label{b potentials}
\boldsymbol{b} = \boldsymbol{\nabla}\times\boldsymbol{a}
\end{equation}
According to relation~\eqref{duality B b} and identity~\eqref{curl duality}, the magnetic induction bivector field $\boldsymbol{B}$ is the dual of the magnetic induction vector field $\boldsymbol{b}$,
\begin{equation}\label{B potentials}
\boldsymbol{B} = -\,\boldsymbol{b}^{\ast} = -\,\left(\boldsymbol{\nabla}\times\boldsymbol{a}\right)^{\ast} = \boldsymbol{\nabla}\wedge\boldsymbol{a}
\end{equation}
In view of relations~\eqref{Riemann Silberstein multivector} and~\eqref{B potentials}, the electromagnetic multivector field $F$ is written in terms of the potentials as,
\begin{equation}\label{F potentials}
F = \boldsymbol{e} + c\,\boldsymbol{B} = -\,\boldsymbol{\nabla}\,\phi -\,\partial_t\,\boldsymbol{a} + c\,\boldsymbol{\nabla}\wedge\boldsymbol{a}
\end{equation}
The Lorentz gauge in vector space (VS) is defined as,~\cite{Jackson:1999,Griffiths:2017}
\begin{equation}\label{Lorentz gauge SA}
\frac{1}{c^2}\,\partial_t\,\phi + \boldsymbol{\nabla}\cdot\boldsymbol{a} = 0
\end{equation}
%


\section{Maxwell equation in matter (SA)}
\label{Maxwell equation in matter}

\noindent We now show that in matter, the Maxwell equations~\eqref{Electric Gauss}-\eqref{Magnetic Gauss} reduce to a single equation. The term ``matter'' refers here to a material medium consisting of continuum of electric and magnetic dipoles. In a dielectric and magnetic medium, the vectorial linear electric constitutive relation is given by,~\cite{Jackson:1999,Griffiths:2017}
\begin{equation}\label{electric constitutive relation matter}
\boldsymbol{d} = \varepsilon_0\,\boldsymbol{e} + \boldsymbol{\tilde{p}}
\end{equation}
where $\boldsymbol{\tilde{p}}$ is the matter electric polarisation vector field, and the pseudovectorial linear magnetic constitutive relation is given by,~\cite{Jackson:1999,Griffiths:2017}
\begin{equation}\label{magnetic constitutive relation matter 0}
\boldsymbol{b} = \mu_0\left(\boldsymbol{h}+\boldsymbol{\tilde{m}}\right)
\end{equation}
where $\boldsymbol{\tilde{m}}$ is the matter magnetisation pseudovector field. The dual of the constitutive relation~\eqref{magnetic constitutive relation matter 0} is written as,
\begin{equation}\label{magnetic constitutive matter dual}
\boldsymbol{b}^{\ast} = \mu_0\left(\boldsymbol{h}^{\ast}+\boldsymbol{\tilde{m}}^{\ast}\right)
\end{equation}
The magnetisation field pseudovector $\boldsymbol{\tilde{m}}$ in vector space is the dual of the auxiliary magnetisation field bivector $\boldsymbol{\tilde{M}}$ in spatial algebra~\eqref{dual vector v},
\begin{equation}\label{duality m H}
\boldsymbol{\tilde{M}}^{\ast} = \boldsymbol{\tilde{m}} \qquad\text{where}\qquad \vert\boldsymbol{\tilde{M}}\vert = \vert\boldsymbol{\tilde{m}}\vert
\end{equation}
The dual of this duality is,
\begin{equation}\label{duality M m}
\boldsymbol{\tilde{m}}^{\ast} = \left(\boldsymbol{\tilde{M}}^{\ast}\right)^{\ast} = -\,\boldsymbol{\tilde{M}}
\end{equation}
Using the dualities~\eqref{duality B b},\eqref{duality H h} and~\eqref{duality M m}, relation~\eqref{magnetic constitutive matter dual} becomes the bivectorial linear magnetic constitutive relation,
\begin{equation}\label{magnetic constitutive relation matter}
\boldsymbol{B} = \mu_0\left(\boldsymbol{H}+\boldsymbol{\tilde{M}}\right)
\end{equation}
Substituting the constitutive relations~\eqref{electric constitutive relation matter} and~\eqref{magnetic constitutive relation matter} into the auxiliary electromagnetic multivector field~\eqref{Riemann Silberstein multivector G} in vacuum and taking into account the speed of light~\eqref{light}, we obtain,
\begin{equation}\label{G matter}
G = \left(\varepsilon_0\,\boldsymbol{e} + \boldsymbol{\tilde{p}}\right) + \frac{1}{c}\left(\frac{\boldsymbol{B}}{\mu_0} -\,\boldsymbol{\tilde{M}}\right) = \varepsilon_0\left(\boldsymbol{e} + c\,\boldsymbol{B}\right) + \left(\boldsymbol{\tilde{p}} -\,\frac{1}{c}\,\boldsymbol{\tilde{M}}\right)
\end{equation}
Defining the electromagnetic polarisation multivector as,
\begin{equation}\label{P matter}
\tilde{P} = \boldsymbol{\tilde{p}} -\,\frac{1}{c}\,\boldsymbol{\tilde{M}}
\end{equation}
and substituting the multivectors~\eqref{G constitutive relation vacuum} and~\eqref{P matter} into relation~\eqref{G matter} we obtain the multivectorial linear electromagnetic constitutive relation,
\begin{equation}\label{G matter constitutive relation}
G = \varepsilon_0\,F + \tilde{P}
\end{equation}
Substituting the multivectorial linear electromagnetic constitutive relation~\eqref{G matter constitutive relation} into the Maxwell equation~\eqref{G eq}, we obtain the Maxwell equation for a dielectric and magnetic medium,
\begin{equation}\label{G eq matter}
\left(\frac{1}{c}\,\partial_t + \boldsymbol{\nabla}\right)\left(\varepsilon_0\,F + \tilde{P}\right) = q -\,\frac{1}{c}\,\boldsymbol{j}
\end{equation}
which is in agreement with the result derived by Arthur~\cite{Arthur:2011}. The dynamics of the electromagnetic polarisation of the medium is described by the equation,
\begin{equation}\label{P eq matter 0}
\left(\frac{1}{c}\,\partial_t + \boldsymbol{\nabla}\right)\tilde{P} = \frac{1}{c}\,\partial_t\,\tilde{P} + \boldsymbol{\nabla}\cdot \tilde{P} + \boldsymbol{\nabla}\wedge \tilde{P}
\end{equation}
Using the definition~\eqref{P matter} of the electromagnetic polarisation multivector, we obtain the following relation,
\begin{equation}\label{P eq matter}
\frac{1}{c}\,\partial_t\,\tilde{P} + \boldsymbol{\nabla}\cdot \tilde{P} = \frac{1}{c}\,\partial_t\,\boldsymbol{\tilde{p}} + \boldsymbol{\nabla}\cdot\boldsymbol{\tilde{p}} -\,\frac{1}{c^2}\,\partial_t\,\boldsymbol{\tilde{M}} -\,\frac{1}{c}\,\boldsymbol{\nabla}\cdot\boldsymbol{\tilde{M}}
\end{equation}
Using relations~\eqref{P matter},~\eqref{P eq matter 0} and~\eqref{P eq matter}, the Maxwell equation~\eqref{G eq matter} is recast as,
\begin{equation}\label{G eq matter bis}
\begin{split}
&\left(\frac{1}{c}\,\partial_t + \boldsymbol{\nabla}\right)\varepsilon_0\,F = \left(q -\,\boldsymbol{\nabla}\cdot\boldsymbol{\tilde{p}}\right) -\,\frac{1}{c}\left(\boldsymbol{j} + \partial_t\,\boldsymbol{\tilde{p}} -\,\boldsymbol{\nabla}\cdot\boldsymbol{\tilde{M}}\right)\\
&\phantom{\left(\frac{1}{c}\,\partial_t + \boldsymbol{\nabla}\right)\varepsilon_0\,F =} -\,\boldsymbol{\nabla}\wedge\boldsymbol{\tilde{p}} + \frac{1}{c}\,\boldsymbol{\nabla}\wedge\boldsymbol{\tilde{M}}
\end{split}
\end{equation}
The total electric charge density $\tilde{q}$ in matter is the sum of the free electric charge density $q$ and the bound electric charge density $-\,\boldsymbol{\nabla}\cdot\boldsymbol{\tilde{p}}$,
\begin{equation}\label{bound charge density}
\tilde{q} = q -\,\boldsymbol{\nabla}\cdot\boldsymbol{\tilde{p}}
\end{equation}
and total electric current density $\boldsymbol{\tilde{j}}$ in matter is the sum of the free electric currant density $\boldsymbol{j}$ and the bound electric current density $\partial_t\,\boldsymbol{\tilde{p}} -\,\boldsymbol{\nabla}\cdot\boldsymbol{\tilde{M}} $,
\begin{equation}\label{bound current density}
\boldsymbol{\tilde{j}} = \boldsymbol{j} + \partial_t\,\boldsymbol{\tilde{p}} -\,\boldsymbol{\nabla}\cdot\boldsymbol{\tilde{M}} 
\end{equation}
Using the definitions~\eqref{P matter},~\eqref{bound charge density} and~\eqref{bound current density}, the Maxwell equation~\eqref{G eq matter bis} reduces to,
\begin{equation}\label{G eq matter bound 0}
\left(\frac{1}{c}\,\partial_t + \boldsymbol{\nabla}\right)\varepsilon_0\,F = \tilde{q} -\,\frac{1}{c}\,\boldsymbol{\tilde{j}} -\,\boldsymbol{\nabla}\wedge\boldsymbol{\tilde{p}} + \frac{1}{c}\,\boldsymbol{\nabla}\wedge\boldsymbol{\tilde{M}}
\end{equation}
Taking into account the definition~\eqref{Riemann Silberstein multivector} of the electromagnetic multivector field $F$, the trivectorial part of the Maxwell equation~\eqref{G eq matter bound 0} reduces to,
\begin{equation}\label{trivector B M}
c\,\varepsilon_0\,\boldsymbol{\nabla}\wedge\boldsymbol{B} = \frac{1}{c}\,\boldsymbol{\nabla}\wedge\boldsymbol{\tilde{M}}
\end{equation}
In view of the magnetic Gauss equation~\eqref{Magnetic Gauss}, relation~\eqref{trivector B M} yields the condition,
\begin{equation}\label{trivector B M 0}
\boldsymbol{\nabla}\wedge\boldsymbol{\tilde{M}} = 0
\end{equation}
Taking into account the definition~\eqref{Riemann Silberstein multivector} of the electromagnetic multivector field $F$, the bivectorial part of the Maxwell equation~\eqref{G eq matter bound 0} reduces to,
\begin{equation}\label{bivector B M}
\varepsilon_0\left(\partial_t\,\boldsymbol{B} + \boldsymbol{\nabla}\wedge\boldsymbol{e}\right)  = -\,\boldsymbol{\nabla}\wedge\boldsymbol{\tilde{p}}
\end{equation}
Using the Faraday equation~\eqref{Faraday}, relation~\eqref{bivector B M} yields the condition,
\begin{equation}\label{bivector B M bis}
\boldsymbol{\nabla}\wedge\boldsymbol{\tilde{p}} = 0
\end{equation}
In view of the conditions\eqref{trivector B M 0} and\eqref{bivector B M bis}, the Maxwell equation~\eqref{G eq matter bound 0} reduces to,
\begin{equation}\label{G eq matter bound}
\left(\frac{1}{c}\,\partial_t + \boldsymbol{\nabla}\right)\varepsilon_0\,F = \tilde{q} -\,\frac{1}{c}\,\boldsymbol{\tilde{j}}
\end{equation}
which differs from the result derived by Arthur~\cite{Arthur:2011} since the conditions~\eqref{trivector B M 0} and~\eqref{bivector B M bis} were not properly identified. The Maxwell equation~\eqref{G eq matter bound} in a dielectric and magnetic medium can be obtained by replacing the electric charge density $q$ by the total electric charge density $\tilde{q}$ and the electric current density $\boldsymbol{j}$ by the total electric current density $\boldsymbol{\tilde{j}}$ in the Maxwell equation~\eqref{F eq} in vacuum. The difference between the Maxwell equations~\eqref{G eq matter} and~\eqref{G eq matter bound} yields the relation,
\begin{equation}\label{P eq matter bound 0}
\left(\frac{1}{c}\,\partial_t + \boldsymbol{\nabla}\right)\tilde{P} = \left(q-\,\tilde{q}\right) -\,\frac{1}{c}\left(\boldsymbol{j}-\,\boldsymbol{\tilde{j}}\right)
\end{equation}
In view of relation of the electric charge density in matter~\eqref{bound charge density} and the electric current density in matter~\eqref{bound current density}, relation~\eqref{P eq matter bound 0} is recast as,
\begin{equation}\label{P eq matter bound 00}
\left(\frac{1}{c}\,\partial_t + \boldsymbol{\nabla}\right)\tilde{P} = \boldsymbol{\nabla}\cdot\boldsymbol{\tilde{p}} + \frac{1}{c}\left(\partial_t\,\boldsymbol{\tilde{p}} -\,\boldsymbol{\nabla}\cdot\boldsymbol{\tilde{M}}\right)
\end{equation}
In view of relations~\eqref{trivector B M 0} and~\eqref{bivector B M bis}, the curl of the electromagnetic polarisation multivector~\eqref{P matter} vanishes,
\begin{equation}\label{curl P}
\boldsymbol{\nabla}\wedge\tilde{P} = 0
\end{equation}
According to relation~\eqref{curl P}, the geometric product on the left-hand side of relation~\eqref{P eq matter bound 00} becomes a inner product,
\begin{equation}\label{P eq matter bound}
\left(\frac{1}{c}\,\partial_t + \boldsymbol{\nabla}\right)\cdot\tilde{P} = \boldsymbol{\nabla}\cdot\boldsymbol{\tilde{p}} + \frac{1}{c}\left(\partial_t\,\boldsymbol{\tilde{p}} -\,\boldsymbol{\nabla}\cdot\boldsymbol{\tilde{M}}\right)
\end{equation}
%


\section{Electromagnetic waves in matter (SA)}
\label{Electromagnetic waves in matter}

\noindent In the spatial algebra (SA), electromagnetic waves in vacuum are a direct and straightforward consequence of the Maxwell equation~\eqref{G eq}, as expected, but it is nonetheless quite beautiful. Indeed, multiplying the Maxwell equation~\eqref{G eq matter bound} by $c^{-1}\partial_t -\,\boldsymbol{\nabla}$ yields,
\begin{equation}\label{G eq matter bound product}
\left(\frac{1}{c}\,\partial_t -\,\boldsymbol{\nabla}\right)\left(\frac{1}{c}\,\partial_t + \boldsymbol{\nabla}\right)\varepsilon_0\,F = \left(\frac{1}{c}\,\partial_t -\,\boldsymbol{\nabla}\right)\left(\tilde{q} -\,\frac{1}{c}\,\boldsymbol{\tilde{j}}\right)
\end{equation}
which is written explicitly as,
\begin{equation}\label{G eq matter bound product bis}
\left(\frac{1}{c^2}\,\partial_t^2 -\,\boldsymbol{\nabla}^2\right)\varepsilon_0\,F = \frac{1}{c}\,\partial_t\,\tilde{q} -\,\boldsymbol{\nabla}\,\tilde{q} -\,\frac{1}{c^2}\,\partial_t\,\boldsymbol{\tilde{j}} + \frac{1}{c}\,\boldsymbol{\nabla}\cdot\boldsymbol{\tilde{j}} + \frac{1}{c}\,\boldsymbol{\nabla}\wedge\boldsymbol{\tilde{j}}
\end{equation}
In view of the definitions of the total electric charge density~\eqref{bound charge density} and the total electric current density~\eqref{bound current density}, we obtain, 
\begin{equation}\label{Electric continuity eq total 0}
\partial_t\,\tilde{q} + \boldsymbol{\nabla}\cdot\boldsymbol{\tilde{j}} = \partial_t\,q -\,\partial_t\left(\boldsymbol{\nabla}\cdot\boldsymbol{\tilde{p}}\right) + \boldsymbol{\nabla}\cdot\boldsymbol{j} + \partial_t\left(\boldsymbol{\nabla}\cdot\boldsymbol{\tilde{p}}\right) -\,\boldsymbol{\nabla}\cdot\left(\boldsymbol{\nabla}\cdot\boldsymbol{\tilde{M}}\right)
\end{equation}
Taking into account the electric continuity equation~\eqref{continuity eq} for the free electric charge density $q$, relation~\eqref{Electric continuity eq total 0} yields the electric continuity equation for the total electric charge density $\tilde{q}$,
\begin{equation}\label{Electric continuity eq total}
\partial_t\,\tilde{q} + \boldsymbol{\nabla}\cdot\boldsymbol{\tilde{j}} = 0
\end{equation}
Using the continuity equation~\eqref{Electric continuity eq total}, relation~\eqref{G eq matter bound product bis} yields the electromagnetic wave equation for the electromagnetic multivector field $F$ in a dielectric and magnetic medium,
\begin{equation}\label{wave eq F matter}
\left(\frac{1}{c^2}\,\partial_t^2 -\,\boldsymbol{\nabla}^2\right)\varepsilon_0\,F = -\,\boldsymbol{\nabla}\,\tilde{q} -\,\frac{1}{c^2}\,\partial_t\,\boldsymbol{\tilde{j}} + \frac{1}{c}\,\boldsymbol{\nabla}\wedge\boldsymbol{\tilde{j}}
\end{equation}
Taking into account the definition~\eqref{Riemann Silberstein multivector} of the electromagnetic multivector field $F$, the vectorial part of the wave equation~\eqref{wave eq F matter} yields the wave equation for the electric field $\boldsymbol{e}$,
\begin{equation}\label{wave eq e matter}
\left(\frac{1}{c^2}\,\partial_t^2 -\,\boldsymbol{\nabla}^2\right)\boldsymbol{e} = -\,\frac{1}{\varepsilon_0}\,\boldsymbol{\nabla}\,\tilde{q} -\,\frac{1}{\varepsilon_0\,c^2}\,\partial_t\,\boldsymbol{\tilde{j}}
\end{equation}
Taking into account the definition~\eqref{Riemann Silberstein multivector} of the electromagnetic multivector field $F$ and using the speed of light~\eqref{light}, the bivectorial part of the wave equation~\eqref{wave eq F matter} yields the wave equation for the magnetic field $\boldsymbol{B}$,
\begin{equation}\label{wave eq B matter}
\left(\frac{1}{c^2}\,\partial_t^2 -\,\boldsymbol{\nabla}^2\right)\boldsymbol{B} = \mu_0\,\boldsymbol{\nabla}\wedge\boldsymbol{\tilde{j}}
\end{equation}
%


\section{Electromagnetic energy and momentum in matter (SA)}
\label{Electromagnetic energy and momentum in matter}

\noindent The electromagnetic field multivector $F$ is a linear combination of the electric field vector $\boldsymbol{e}$ and the magnetic induction field bivector $\boldsymbol{B}$ according to relation~\eqref{Riemann Silberstein multivector}. Similarly, the auxiliary electromagnetic field multivector $G$ is a linear combination of the electric displacement field vector $\boldsymbol{d}$ and the auxiliary magnetic field bivector $\boldsymbol{H}$ according to relation~\eqref{Riemann Silberstein multivector G}. Moreover, the electric polarisation multivector $\tilde{P}$ is a linear combination of the electric polarisation vector $\boldsymbol{\tilde{p}}$ and the magnetisation bivector $\boldsymbol{\tilde{M}}$ according to relation~\eqref{P matter}. Since the electromagnetic energy density $e$ and the electromagnetic momentum density $\boldsymbol{p}$ in matter are determined by these fields, they can be recast in terms of the multivectors $F$, $G$ and $\tilde{P}$. In view of the linear electric constitutive relation in matter~\eqref{electric constitutive relation matter} and the linear magnetic constitutive relation~\eqref{magnetic constitutive relation matter}, relation~\eqref{electromagnetic energy density deHB} yields the electromagnetic energy density in a dielectric and magnetic medium,~\cite{Jackson:1999,Griffiths:2017}
\begin{equation}\label{electromagnetic energy density matter}
e = \frac{1}{2}\,\boldsymbol{d}\cdot\boldsymbol{e} -\,\frac{1}{2}\,\boldsymbol{H}\cdot\boldsymbol{B} = \frac{1}{2}\,\varepsilon_0\,\boldsymbol{e}^2 -\,\frac{1}{2\,\mu_0}\,\boldsymbol{B}^2 + \frac{1}{2}\,\boldsymbol{\tilde{p}}\cdot\boldsymbol{e} + \frac{1}{2}\,\boldsymbol{\tilde{M}}\cdot\boldsymbol{B}
\end{equation}
In view of the linear magnetic constitutive relation in matter~\eqref{magnetic constitutive relation matter}, relation~\eqref{electromagnetic momentum density} yields the electromagnetic momentum density in matter,
\begin{equation}\label{electromagnetic momentum density matter}
\boldsymbol{p} = \frac{1}{c^2}\,\boldsymbol{H}\cdot\boldsymbol{e} = \frac{1}{c^2}\left(\frac{\boldsymbol{B}}{\mu_0}-\,\boldsymbol{\tilde{M}}\right)\cdot\boldsymbol{e}
\end{equation}
The reverse of the linear electromagnetic constitutive relation in matter~\eqref{G matter constitutive relation} is written as,
\begin{equation}\label{G constitutive relation reverse matter}
G^{\dag} = \varepsilon_0\,F^{\dag} + \tilde{P}^{\dag}
\end{equation}
According the definition~\eqref{P matter}, the reverse of the electromagnetic polarisation tensor is given by,
\begin{equation}\label{P reverse matter}
\tilde{P}^{\dag} = \boldsymbol{\tilde{p}}^{\dag} -\,\frac{1}{c}\,\boldsymbol{\tilde{M}}^{\dag} = \boldsymbol{\tilde{p}} + \frac{1}{c}\,\boldsymbol{\tilde{M}}
\end{equation}
In view of the multivectors~\eqref{P matter} and~\eqref{P reverse matter}, the electric polarisation vector field is recast as,
\begin{equation}\label{p as multivector}
\boldsymbol{\tilde{p}} = \frac{1}{2}\left(\tilde{P} + \tilde{P}^{\dag}\right)
\end{equation}
and the magnetisation bivector field is recast as,
\begin{equation}\label{M as multivector}
\boldsymbol{\tilde{M}} = -\,\frac{c}{2}\left(\tilde{P} -\,\tilde{P}^{\dag}\right)
\end{equation}
Comparing the expressions~\eqref{electromagnetic energy density deHB} and~\eqref{electromagnetic energy density G} for the electromagnetic energy density in vacuum to the expression~\eqref{electromagnetic energy density matter} of the electromagnetic energy density in matter, we conclude that the electromagnetic energy density in matter~\eqref{electromagnetic energy density matter} is recast as,
\begin{equation}\label{electromagnetic energy density G matter}
e = \frac{1}{8}\left(G\,F^{\dag} + F\,G^{\dag} + G^{\dag}F + F^{\dag}G\right)
\end{equation}
In view of the electromagnetic constitutive equations~\eqref{G matter constitutive relation} and~\eqref{G constitutive relation reverse matter}, the electromagnetic energy density in matter~\eqref{electromagnetic energy density G matter} is recast as,
\begin{equation}\label{electromagnetic energy density G matter bis}
e = \frac{1}{4}\,\varepsilon_0\left(F\,F^{\dag} + F^{\dag}F\right) + \frac{1}{8}\left(P\,F^{\dag} + F\,P^{\dag} + P^{\dag}F + F^{\dag}P\right)
\end{equation}
Comparing the expressions~\eqref{electromagnetic momentum density} and~\eqref{electromagnetic momentum density bis} for the electromagnetic momentum density in vacuum to the expression~\eqref{electromagnetic momentum density matter} of the electromagnetic momentum density in matter, we conclude that the electromagnetic momentum density in matter~\eqref{electromagnetic momentum density matter} is recast as,
\begin{equation}\label{electromagnetic momentum density G matter}
\begin{split}
&\boldsymbol{p} = \frac{1}{8\,c}\left(G\,F^{\dag} + F\,G^{\dag} -\,G^{\dag}F -\,F^{\dag}G\right)\\
&\phantom{\boldsymbol{p} =} + \frac{1}{8\,c}\left(G\,F -\,F\,G + F^{\dag}G^{\dag} -\,G^{\dag}F^{\dag}\right)
\end{split}
\end{equation}
In view of the electromagnetic constitutive equations~\eqref{G matter constitutive relation} and~\eqref{G constitutive relation reverse matter}, the electromagnetic momentum density in matter~\eqref{electromagnetic energy density G matter} is recast as,
\begin{equation}\label{electromagnetic momentum density G matter bis}
\boldsymbol{p} = \frac{1}{4\,c}\,\varepsilon_0\left(F\,F^{\dag} -\,F^{\dag}F\right) + \frac{1}{8\,c}\left(P\,F -\,F\,P + F^{\dag}P^{\dag} -\,P^{\dag}F^{\dag}\right)
\end{equation}
According to relations~\eqref{electromagnetic energy density G matter} and~\eqref{electromagnetic momentum density G matter},
\begin{equation}\label{electromagnetic invariant matter}
e + \boldsymbol{p}\,c = \frac{1}{4}\left(G\,F^{\dag} + F\,G^{\dag}\right) + \frac{1}{8}\left(G\,F -\,F\,G + F^{\dag}G^{\dag}  -\,G^{\dag}F^{\dag}\right)
\end{equation}
In view of the electromagnetic constitutive equations~\eqref{G matter constitutive relation} and~\eqref{G constitutive relation reverse matter}, relation~\eqref{electromagnetic invariant matter} in matter is recast as,
\begin{equation}\label{electromagnetic invariant matter bis bis}
e + \boldsymbol{p}\,c = \frac{1}{2}\,\varepsilon_0\,F\,F^{\dag} + \frac{1}{8}\left(P\,F -\,F\,P + F^{\dag}P^{\dag} -\,P^{\dag}F^{\dag}\right)
\end{equation}
%


\section{Poynting theorem in matter (SA)}
\label{Poynting theorem in matter}

\noindent The continuity equation for the electromagnetic energy, called the Poynting theorem, is written in matter in terms of the time derivative of the energy density and the energy flux which is a multiple of the electromagnetic momentum density. In linear electromagnetism, the matter electric polarisation vector field $\boldsymbol{\tilde{p}}\left(\boldsymbol{e}\right)$ induced by the electric field $\boldsymbol{e}$ is a linear map of the electric vector field $\boldsymbol{e}$,
\begin{equation}\label{electric polarisation matter}
\boldsymbol{\tilde{p}}\left(\boldsymbol{e}\right) = \varepsilon_0\,\boldsymbol{\chi}_{\boldsymbol{e}}\left(\boldsymbol{e}\right)
\end{equation}
where $\boldsymbol{\chi}_{\boldsymbol{e}}\left(\boldsymbol{e}\right)$ is the electric susceptibility vector field, and the matter magnetisation bivector field $\boldsymbol{\tilde{M}}\left(\boldsymbol{B}\right)$ induced by the magnetic induction field $\boldsymbol{B}$ is a linear map of the magnetic induction bivector field $\boldsymbol{B}$,
\begin{equation}\label{magnetisation matter}
\boldsymbol{\tilde{M}}\left(\boldsymbol{B}\right) = \frac{1}{\mu_0}\,\boldsymbol{\chi}_{\boldsymbol{B}}\left(\boldsymbol{B}\right)
\end{equation}
where $\boldsymbol{\chi}_{\boldsymbol{B}}\left(\boldsymbol{B}\right)$ is the electric susceptibility bivector field. In view of the linear maps~\eqref{electric polarisation matter} and~\eqref{magnetisation matter} the electric displacement vector field~\eqref{electric constitutive relation matter} is recast as,
\begin{equation}\label{linear electric displacement}
\boldsymbol{d}\left(\boldsymbol{e}\right) = \varepsilon_0\,\Big(\boldsymbol{e} + \boldsymbol{\chi}_{\boldsymbol{e}}\left(\boldsymbol{e}\right)\Big)
\end{equation}
and the magnetic auxiliary bivector field~\eqref{magnetic constitutive relation matter} is recast as,
\begin{equation}\label{linear auxiliary magnetic}
\boldsymbol{H}\left(\boldsymbol{B}\right) = \frac{1}{\mu_0}\,\Big(\boldsymbol{B} -\,\boldsymbol{\chi}_{\boldsymbol{B}}\left(\boldsymbol{B}\right)\Big)
\end{equation}
In view of linear relations~\eqref{electric polarisation matter} and~\eqref{magnetisation matter}, the electromagnetic energy density~\eqref{electromagnetic energy density matter} is recast as,
\begin{equation}\label{electromagnetic energy density matter linear}
e\left(\boldsymbol{e},\boldsymbol{B}\right) = \frac{1}{2}\,\varepsilon_0\,\Big(\boldsymbol{e} + \boldsymbol{\chi}_{\boldsymbol{e}}\left(\boldsymbol{e}\right)\Big)\cdot\boldsymbol{e}
-\,\frac{1}{2\,\mu_0}\,\Big(\boldsymbol{B} -\,\boldsymbol{\chi}_{\boldsymbol{B}}\left(\boldsymbol{B}\right)\Big)\cdot\boldsymbol{B}
\end{equation}
The time derivative of the linear electromagnetic energy density in matter~\eqref{electromagnetic energy density matter linear} reads,
\begin{align}\label{electromagnetic energy density matter time derivative}
&\partial_t\,e\left(\boldsymbol{e},\boldsymbol{B}\right) = \varepsilon_0\,\boldsymbol{e}\cdot\partial_t\,\boldsymbol{e} + \frac{1}{2}\,\varepsilon_0\,\boldsymbol{\chi}_{\boldsymbol{e}}\left(\boldsymbol{e}\right)\cdot\partial_t\,\boldsymbol{e} + \frac{1}{2}\,\varepsilon_0\,\partial_t\Big(\boldsymbol{\chi}_{\boldsymbol{e}}\left(\boldsymbol{e}\right)\Big)\cdot\boldsymbol{e}\\
&\phantom{\partial_t\,e\left(\boldsymbol{e},\boldsymbol{B}\right) =} -\,\frac{1}{\mu_0}\,\boldsymbol{B}\cdot\partial_t\,\boldsymbol{B} + \frac{1}{2\,\mu_0}\,\boldsymbol{\chi}_{\boldsymbol{B}}\left(\boldsymbol{B}\right)\cdot\partial_t\,\boldsymbol{B} + \frac{1}{2\,\mu_0}\,\partial_t\Big(\boldsymbol{\chi}_{\boldsymbol{B}}\left(\boldsymbol{B}\right)\Big)\cdot\boldsymbol{B}\nonumber
\end{align}
The electric susceptibility vector $\boldsymbol{\chi}_{\boldsymbol{e}}\left(\boldsymbol{e}\right)$ is a linear map of the electric field vector $\boldsymbol{e}$ and the magnetic susceptibility bivector $\boldsymbol{\chi}_{\boldsymbol{B}}\left(\boldsymbol{B}\right)$ is a linear map of the magnetic induction field bivector $\boldsymbol{B}$. Furthermore, we assume that the electric and magnetic properties of the material medium are constant, which implies that,
\begin{equation}\label{time derivatives susceptibilities}
\begin{split}
&\partial_t\Big(\boldsymbol{\chi}_{\boldsymbol{e}}\left(\boldsymbol{e}\right)\Big)\cdot\boldsymbol{e} = \boldsymbol{\chi}_{\boldsymbol{e}}\left(\boldsymbol{e}\right)\cdot\partial_t\,\boldsymbol{e}\\
&\boldsymbol{\chi}_{\boldsymbol{B}}\left(\boldsymbol{B}\right)\cdot\partial_t\,\boldsymbol{B} = \partial_t\Big(\boldsymbol{\chi}_{\boldsymbol{B}}\left(\boldsymbol{B}\right)\Big)\cdot\boldsymbol{B} 
\end{split}
\end{equation}
In view of the conditions~\eqref{time derivatives susceptibilities}, relation~\eqref{electromagnetic energy density matter time derivative} reduces to,
\begin{equation}\label{time derivative energy density matter}
\partial_t\,e\left(\boldsymbol{e},\boldsymbol{B}\right) = \varepsilon_0\,\boldsymbol{e}\cdot\partial_t\,\Big(\boldsymbol{e} + \boldsymbol{\chi}_{\boldsymbol{e}}\left(\boldsymbol{e}\right)\Big) -\,\frac{1}{\mu_0}\,\Big(\boldsymbol{B} -\,\boldsymbol{\chi}_{\boldsymbol{B}}\left(\boldsymbol{B}\right)\Big)\cdot\partial_t\,\boldsymbol{B}
\end{equation}
Using the electric constitutive equation~\eqref{electric constitutive relation matter} and the magnetic constitutive equation~\eqref{magnetic constitutive relation matter}, relation~\eqref{time derivative energy density matter} is recast as,
\begin{equation}\label{time derivative energy density matter bis}
\partial_t\,e = \boldsymbol{e}\cdot\partial_t\,\boldsymbol{d} -\,\boldsymbol{H}\cdot\partial_t\,\boldsymbol{B}
\end{equation}
Using the Mawell-Ampère equation~\eqref{Maxwell Ampere} and the Faraday equation~\eqref{Faraday}, the time derivative of the electromagnetic energy density~\eqref{time derivative energy density matter bis} is recast as,
\begin{equation}\label{time derivative energy density matter ter}
\partial_t\,e = -\,\boldsymbol{e}\cdot\left(\boldsymbol{\nabla}\cdot\boldsymbol{H} + \boldsymbol{j}\right) + \boldsymbol{H}\cdot\left(\boldsymbol{\nabla}\wedge\boldsymbol{e}\right)
\end{equation}
Thus,
\begin{equation}\label{time derivative energy density matter quad}
\partial_t\,e + \left(\boldsymbol{\nabla}\cdot\boldsymbol{H}\right)\cdot\boldsymbol{e} -\,\boldsymbol{H}\cdot\left(\boldsymbol{\nabla}\wedge\boldsymbol{e}\right) = -\,\boldsymbol{j}\cdot\boldsymbol{e}
\end{equation}
Using the algebraic identity~\eqref{div vec bivec hex},
\begin{equation}\label{id poynting bis}
\boldsymbol{\nabla}\cdot\left(\boldsymbol{H}\cdot\boldsymbol{e}\right) = \left(\boldsymbol{\nabla}\cdot\boldsymbol{H}\right)\cdot\boldsymbol{e} -\,\boldsymbol{H}\cdot\left(\boldsymbol{\nabla}\wedge\boldsymbol{e}\right)
\end{equation}
relation yields~\eqref{time derivative energy density matter quad} the Poynting's theorem,
\begin{equation}\label{Poynting's theorem matter}
\partial_t\,e + \boldsymbol{\nabla}\cdot\left(\boldsymbol{H}\cdot\boldsymbol{e}\right) = -\,\boldsymbol{j}\cdot\boldsymbol{e}
\end{equation}
where $\boldsymbol{H}\cdot\boldsymbol{e}$ is the Poynting vector, which is the electromagnetic current density. In view of the electromagnetic momentum density~\eqref{electromagnetic momentum density matter}, Poynting's theorem~\eqref{Poynting's theorem matter} in vacuum is recast as,
\begin{equation}\label{Poynting's theorem matter bis}
\frac{1}{c}\,\partial_t\,e + \boldsymbol{\nabla}\cdot\left(\boldsymbol{p}\,c\right) = -\,\frac{1}{c}\,\boldsymbol{j}\cdot\boldsymbol{e}
\end{equation}
Poynting's theorem~\eqref{Poynting's theorem matter bis} in matter is valid under the assumption that the electric and magnetic properties of the material medium are constant. In linear electromagnetism, according to relations~\eqref{linear electric displacement} and~\eqref{linear auxiliary magnetic}, the auxiliary electromagnetic field multivector~\eqref{Riemann Silberstein multivector G} is written as,
\begin{equation}\label{G susceptibility 0}
G = \boldsymbol{d}\left(\boldsymbol{e}\right) + \frac{1}{c}\,\boldsymbol{H}\left(\boldsymbol{B}\right) = \varepsilon_0\,\Big(\boldsymbol{e} + \boldsymbol{\chi}_{\boldsymbol{e}}\left(\boldsymbol{e}\right)\Big) + \frac{1}{\mu_0\,c}\,\Big(\boldsymbol{B} -\,\boldsymbol{\chi}_{\boldsymbol{B}}\left(\boldsymbol{B}\right)\Big)
\end{equation}
In view of identity~\eqref{light}, the auxiliary electromagnetic field multivector~\eqref{G susceptibility 0} is recast as,
\begin{equation}\label{G susceptibility 1}
G = \varepsilon_0\left(\boldsymbol{e} + c\,\boldsymbol{B}\right) + \varepsilon_0\,\Big(\boldsymbol{\chi}_{\boldsymbol{e}}\left(\boldsymbol{e}\right) -\,c\,\boldsymbol{\chi}_{\boldsymbol{B}}\left(\boldsymbol{B}\right)\Big)
\end{equation}
The susceptibilities are linear maps of the electromagnetic fields $\boldsymbol{e}$ and $\boldsymbol{B}$ that are expressed in relations~\eqref{e as multivector} and~\eqref{B as multivector} as linear combinations of the electromagnetic multivector $F$ and its reverse $F^{\dag}$. Thus, we define an electromagnetic susceptibility multivector in space-time as a linear map of the electromagnetic field bivector $F$,
\begin{equation}\label{F susceptibility}
\chi_{F}\left(F\right) = \boldsymbol{\chi}_{\boldsymbol{e}}\left(\boldsymbol{e}\right) -\,c\,\boldsymbol{\chi}_{\boldsymbol{B}}\left(\boldsymbol{B}\right)
\end{equation}
In view of relations~\eqref{Riemann Silberstein multivector} and~\eqref{F susceptibility}, the auxiliary electromagnetic field multivector~\eqref{G susceptibility 1} is written as a linear map of the electromagnetic field multivector $F$,
\begin{equation}\label{G susceptibility}
G\left(F\right) = \varepsilon_0\,\Big(F + \chi_{F}\left(F\right)\Big)
\end{equation}
In view of expressions~\eqref{G matter constitutive relation} and~\eqref{G susceptibility} for the electromagnetic field multivector $G\left(F\right)$, the electromagnetic polarisation multivector $\tilde{P}\left(F\right)$ is a linear map of the electromagnetic field multivector $F$,
\begin{equation}\label{P susceptibility}
\tilde{P}\left(F\right) = \varepsilon_0\,\chi_{F}\left(F\right)
\end{equation}
%


\section{Maxwell equation in vacuum (STA)}
\label{Maxwell equation in vacuum STA}

Special relativity, that is rooted in electromagnetism, is described in space-time. To describe relativistic theories like electromagnetism in vector space (VS), an additional time dimension is added such that the generalisation of the spatial vector space $\mathbb{R}^{3}$ becomes the space-time vector space $\mathbb{R}^{1,3}$. Similarly, the spatial algebra (SA) $\mathbb{G}^{3}$ is generalised by including an additional time dimension and called the space-time algebra (STA) $\mathbb{G}^{1,3}$.~\cite{Hestenes:2015} As we will show, the fundamental equations of electromagnetism reach the highest degree of simplicity and beauty because it turns out that the natural language of relativistic physical phenomena is precisely space-time algebra (STA). We consider an orthonormal vector frame $\{e_0,e_1,e_2,e_3\}$ in space-time in order to recast Maxwell's equation in vacuum. The gradient operator $\nabla$ is a covariant vector in space-time is written in coordinates as,
\begin{equation}\label{gradient operator}
\nabla = e^{\mu}\,\partial_{\mu} = e^{0}\,\partial_{0} + e^{i}\,\partial_{i} = e^{0}\,\frac{1}{c}\,\partial_{t} + e^{i}\,\partial_{i}
\end{equation}
In view of relation~\eqref{inner product covariant vector}, the inner product of the gradient $\nabla$ with the time vector $e_{0}$ yields,
\begin{equation}\label{inner product gradient}
\nabla \cdot e_{0} = e_{0} \cdot \nabla = \partial_{0} = \frac{1}{c}\,\partial_{t}
\end{equation}
According to relation~\eqref{outer product covariant vector}, the outer product of the gradient $\nabla$ with the time vector $e_{0}$ yields,
\begin{equation}\label{outer product gradient}
\nabla \wedge e_{0} = -\,e_{0} \wedge \nabla = -\,\boldsymbol{e}_{i}\,\partial^{i} = -\,\boldsymbol{\nabla}
\end{equation}
In view of the inner product~\eqref{inner product covariant vector} and the outer product~\eqref{outer product covariant vector}, the geometric product of the gradient $\nabla$ and the time vector $e_{0}$ yields,
\begin{equation}\label{geometric product gradient}
\nabla\,e_{0} = \nabla \cdot e_{0} + \nabla \wedge e_{0} = \frac{1}{c}\,\partial_{t} -\,\boldsymbol{\nabla}
\end{equation}
Taking into account the properties~\eqref{inner product space-time} and~\eqref{outer product space-time}, the geometric product in reverse order is given by,
\begin{equation}\label{geometric product gradient bis}
e_{0}\,\nabla = e_{0} \cdot \nabla + e_{0} \wedge \nabla = \nabla \cdot e_{0} -\,\nabla \wedge e_{0} = \frac{1}{c}\,\partial_{t} + \boldsymbol{\nabla}
\end{equation}
%


\noindent The electric current density $J$ is a contravariant vector in space-time defined as,
\begin{equation}\label{electric current density}
J = J^{\mu}\,e_{\mu} = J^{0}\,e_{0} + J^{i}\,e_{i} = qc\,e_{0} + j^{i}\,e_{i}
\end{equation}
In view of relation~\eqref{inner product contravariant vector}, the inner product of the electric current density $J$ with the time vector $e_{0}$ yields,
\begin{equation}\label{inner product electric current density}
J \cdot e_{0} = e_{0} \cdot J = J^{0} = qc
\end{equation}
According to relation~\eqref{outer product contravariant vector}, the outer product of the electric current density $J$ with the time vector $e_{0}$ yields,
\begin{equation}\label{outer product electric current density}
J \wedge e_{0} = -\,e_{0} \wedge J = J^{i}\left(e_{i} \wedge e_{0}\right) = j^{i}\,\boldsymbol{e}_{i} = \boldsymbol{j}
\end{equation}
In view of the inner product~\eqref{inner product contravariant vector} and the outer product~\eqref{outer product contravariant vector}, the geometric product of the electric current density $J$ and the time vector $e_{0}$ yields,
\begin{equation}\label{geometric product electric current density}
J\,e_{0} = J \cdot e_{0} + J \wedge e_{0} = qc + \boldsymbol{j}
\end{equation}
Taking into account the properties~\eqref{inner product space-time} and~\eqref{outer product space-time}, the geometric product in reverse order is given by,
\begin{equation}\label{geometric product electric current density bis}
e_{0}\,J = e_{0} \cdot J + e_{0} \wedge J = J \cdot e_{0} -\,J \wedge e_{0} = qc -\,\boldsymbol{j}
\end{equation}
The Maxwell equation~\eqref{G eq} in vacuum is written in spatial algebra as,
\begin{equation}\label{G eq bis}
\left(\frac{1}{c}\,\partial_t + \boldsymbol{\nabla}\right)G = \frac{1}{c}\left(qc -\,\boldsymbol{j}\right)
\end{equation}
In view of the geometric product~\eqref{geometric product gradient bis} between the time vector $e_0$ and the gradient operator $\nabla$ and the geometric product~\eqref{geometric product electric current density bis} of the time vector $e_0$ and the electric current density $J$, the Maxwell equation~\eqref{G eq bis} in vacuum is recast in space-time algebra as,
\begin{equation}\label{G eq ter}
e_{0}\,\nabla\,G = \frac{1}{c}\,e_{0}\,J
\end{equation}
The time vector $e_0$ defines a relative spatial frame $\{\boldsymbol{e}_1,\boldsymbol{e}_2,\boldsymbol{e}_3\}$ that is orthogonal to it. Taking the geometric product of the time vector $e_0$ and the Maxwell equation~\eqref{G eq ter} in spatial algebra, the latter is projected onto an orthonormal frame $\{e_0,e_1,e_2,e_3\}$ with a specific time vector $e_0$. Thus, in space-time algebra the frame independent Maxwell equation in vacuum reads,
\begin{equation}\label{Maxwell equation STA vacuum}
\nabla\,G = \frac{1}{c}\,J
\end{equation}
The gradient of the auxiliary electromagnetic field on the left-hand side of the Maxwell equation~\eqref{Maxwell equation STA vacuum} in vacuum can be written as the sum of the divergence and the curl,
\begin{equation}\label{Maxwell equation STA vacuum bis}
\nabla \cdot G + \nabla \wedge G = \frac{1}{c}\,J
\end{equation}
Since the electric current density $J$ is a vector in space-time and the the auxiliary electromagnetic multivector field $G$ is the sum of a linear combination~\eqref{Riemann Silberstein multivector G} of the vector field $\boldsymbol{d}$ and the auxiliary magnetic bivector field $\boldsymbol{H}$, the curl of the auxiliary electromagnetic multivector field, which is a bivector, has to vanish,
\begin{equation}\label{Maxwell equation STA homogeneous vacuum}
\nabla \wedge G = 0
\end{equation}
which is the homogeneous Maxwell equation in vacuum that is independent of the electric current density $J$. In view of the homogeneous Maxwell equation~\eqref{Maxwell equation STA homogeneous vacuum} in vacuum, the Maxwell equation~\eqref{Maxwell equation STA vacuum bis} in vacuum yields the inhomogeneous Maxwell equation in vacuum,
\begin{equation}\label{Maxwell equation STA inhomogeneous vacuum}
\nabla \cdot G = \frac{1}{c}\,J
\end{equation}
In view of the electromagnetic constitutive relation in vacuum~\eqref{G constitutive relation vacuum}, the homogeneous Maxwell equation~\eqref{Maxwell equation STA homogeneous vacuum} in vacuum is recast as,
\begin{equation}\label{Maxwell equation STA homogeneous vacuum F}
\nabla \wedge F = 0
\end{equation}
and the inhomogeneous Maxwell equation~\eqref{Maxwell equation STA inhomogeneous vacuum} in vacuum is recast as,
\begin{equation}\label{Maxwell equation STA inhomogeneous vacuum F}
\nabla \cdot F = \frac{1}{\varepsilon_0\,c}\,J
\end{equation}
The Maxwell equation in vacuum is the sum of the homogeneous and inhomogeneous Maxwell equations~\eqref{Maxwell equation STA homogeneous vacuum} and~\eqref{Maxwell equation STA inhomogeneous vacuum} in vacuum,
\begin{equation}\label{Maxwell equation STA vacuum F}
\nabla\,F = \frac{1}{\varepsilon_0\,c}\,J
\end{equation}
%


\section{Maxwell equation in matter (STA)}
\label{Maxwell equation in matter STA}

\noindent In order to recast Maxwell's equation in matter, we follow a similar approach as in vacuum except that the electromagnetic constitutive equation~\eqref{G matter constitutive relation} is different due to the electric polarisation multivector $\tilde{P}$. In view of the gradient~\eqref{geometric product gradient bis} and the electric current density~\eqref{geometric product electric current density bis}, the Maxwell equation~\eqref{G eq matter} is recast as,
\begin{equation}\label{Maxwell equation G matter 0}
e_0\,\nabla\left(\varepsilon_0\,F + \tilde{P}\right) = \frac{1}{c}\,e_0\,J
\end{equation}
Since $e_0^2 = 1$, multiplying relation~\eqref{Maxwell equation G matter 0} by $e_0$ yields,
\begin{equation}\label{Maxwell equation G matter 1}
\nabla\left(\varepsilon_0\,F + \tilde{P}\right) = \frac{1}{c}\,J
\end{equation}
In view of the electromagnetic constitutive equation~\eqref{G matter constitutive relation}, the Maxwell equation in matter~\eqref{Maxwell equation G matter 1} reduces to,
\begin{equation}\label{Maxwell equation G matter}
\nabla\,G = \frac{1}{c}\,J
\end{equation}
We now recast the electromagnetic constitutive equation in terms of the electric current density in matter $\tilde{J}$. The electric current density in a dielectric and magnetic medium $\tilde{J}$ is a contravariant vector in space-time defined as,
\begin{equation}\label{electric current density matter}
\tilde{J} = \tilde{J}^{\mu}\,e_{\mu} = \tilde{J}^{0}\,e_{0} + \tilde{J}^{i}\,e_{i} = \tilde{q}c\,e_{0} + \tilde{j}^{i}\,e_{i}
\end{equation}
In view of relation~\eqref{inner product contravariant vector}, the inner product of the electric current density in matter $\tilde{J}$ with the time vector $e_{0}$ yields,
\begin{equation}\label{inner product electric current density matter}
\tilde{J} \cdot e_{0} = e_{0} \cdot \tilde{J} = \tilde{J}^{0} = \tilde{q}c
\end{equation}
According to relation~\eqref{outer product contravariant vector}, the outer product of the electric current density in matter $\tilde{J}$ with the time vector $e_{0}$ yields,
\begin{equation}\label{outer product electric current density matter}
\tilde{J} \wedge e_{0} = -\,e_{0} \wedge \tilde{J} = \tilde{J}^{i}\left(e_{i} \wedge e_{0}\right) = \tilde{j}^{i}\,\boldsymbol{e}_{i} = \boldsymbol{\tilde{j}}
\end{equation}
In view of the inner product~\eqref{inner product contravariant vector} and the outer product~\eqref{outer product contravariant vector}, the geometric product of the electric current density in matter $\tilde{J}$ and the time vector $e_{0}$ yields,
\begin{equation}\label{geometric product electric current density matter}
\tilde{J}\,e_{0} = \tilde{J} \cdot e_{0} + \tilde{J} \wedge e_{0} = \tilde{q}c + \boldsymbol{\tilde{j}}
\end{equation}
Taking into account the properties~\eqref{inner product space-time} and~\eqref{outer product space-time}, the geometric product in reverse order is given by,
\begin{equation}\label{geometric product electric current density bis matter}
e_{0}\,\tilde{J} = e_{0} \cdot \tilde{J} + e_{0} \wedge \tilde{J} = \tilde{J} \cdot e_{0} -\,\tilde{J} \wedge e_{0} = \tilde{q}c -\,\boldsymbol{\tilde{j}}
\end{equation}
In view of relations~\eqref{bound charge density} and~\eqref{bound current density}, relation~\eqref{geometric product electric current density matter} is recast as,
\begin{equation}\label{geometric product electric current density matter bis}
e_{0}\,\tilde{J} = qc -\,\boldsymbol{j} -\,c\,\boldsymbol{\nabla}\cdot\boldsymbol{\tilde{p}} -\,\left(\partial_t\,\boldsymbol{\tilde{p}} -\,\boldsymbol{\nabla}\cdot\boldsymbol{\tilde{M}}\right)
\end{equation}
Using the gradient~\eqref{geometric product gradient bis}, relation~\eqref{P eq matter bound} is recast as,
\begin{equation}\label{electric polarisation gradient}
e_{0}\,\nabla\cdot\tilde{P} = \boldsymbol{\nabla}\cdot\boldsymbol{\tilde{p}} + \frac{1}{c}\left(\partial_t\,\boldsymbol{\tilde{p}} -\,\boldsymbol{\nabla}\cdot\boldsymbol{\tilde{M}}\right)
\end{equation}
In view of the electric current density~\eqref{geometric product electric current density bis} and the divergence of the electromagnetic polarisation multivector~\eqref{electric polarisation gradient}, the electric current density in matter~\eqref{geometric product electric current density matter bis} reduces to,
\begin{equation}\label{geometric product electric current density matter ter}
e_{0}\,\tilde{J} = e_{0}\,J -\,e_{0}\,c\ \nabla\cdot\tilde{P}
\end{equation}
Since $e_0^2 = 1$, taking the geometric of the vector $e_0$ and equation~\eqref{geometric product electric current density matter ter} yields,
\begin{equation}\label{electric current density matter STA}
\tilde{J} = J -\,c\ \nabla\cdot\tilde{P}
\end{equation}
which is quite a beautiful result. The Maxwell equation~\eqref{G eq matter bound bis} in a dielectric and magnetic medium is written in spatial algebra $\mathbb{G}^{3}$ as,
\begin{equation}\label{G eq matter bound bis}
\left(\frac{1}{c}\,\partial_t + \boldsymbol{\nabla}\right)\varepsilon_0\,F = \tilde{q} -\,\frac{1}{c}\,\boldsymbol{\tilde{j}}
\end{equation}
In view of the geometric product~\eqref{geometric product gradient bis} between the time vector $e_0$ and the gradient operator $\nabla$ and the geometric product~\eqref{geometric product electric current density bis matter} of the time vector $e_0$ and the electric current density in matter $\tilde{J}$, the Maxwell equation~\eqref{G eq bis} in matter is recast in space-time algebra $\mathbb{G}^{3}$ as,
\begin{equation}\label{G eq matter bound ter}
e_{0}\,\nabla\,F = \frac{1}{\varepsilon_0\,c}\,e_{0}\,\tilde{J}
\end{equation}
Taking the geometric product of the time vector $e_0$ and the Maxwell equation~\eqref{G eq matter bound ter} in spatial algebra $\mathbb{G}^{3}$, the latter is projected onto an orthonormal frame $\{e_0,e_1,e_2,e_3\}$ with a specific time vector $e_0$. Thus, in space-time algebra $\mathbb{G}^{3}$, the frame independent Maxwell equation in matter reads,
\begin{equation}\label{Maxwell equation STA matter}
\nabla\,F = \frac{1}{\varepsilon_0\,c}\,\tilde{J}
\end{equation}
The gradient of the auxiliary electromagnetic field on the left-hand side of the Maxwell equation~\eqref{Maxwell equation STA matter} in matter can be written as the sum of the divergence and the curl,
\begin{equation}\label{Maxwell equation STA matter bis}
\nabla \cdot F + \nabla \wedge F = \frac{1}{\varepsilon_0\,c}\,\tilde{J}
\end{equation}
Since the electric current density $J$ is a vector in space-time and the the electromagnetic multivector field $F$ is the sum of a linear combination~\eqref{Riemann Silberstein multivector} of the vector field $\boldsymbol{e}$ and the magnetic bivector field $\boldsymbol{B}$, the curl of the electromagnetic multivector field, which is a bivector, has to vanish,
\begin{equation}\label{Maxwell equation STA homogeneous matter}
\nabla \wedge F = 0
\end{equation}
which is the homogeneous Maxwell equation in matter that is independent of the electric current density in matter $\tilde{J}$. In view of the homogeneous Maxwell equation~\eqref{Maxwell equation STA homogeneous matter} in matter, the Maxwell equation~\eqref{Maxwell equation STA matter bis} in matter yields the inhomogeneous Maxwell equation in matter,
\begin{equation}\label{Maxwell equation STA inhomogeneous matter}
\nabla \cdot F = \frac{1}{\varepsilon_0\,c}\,\tilde{J}
\end{equation}
%


\section{Electromagnetic waves in vacuum (STA)}
\label{Electromagnetic waves in vacuum STA}

\noindent In the space-time algebra (STA), electromagnetic waves in vacuum are a direct and straightforward consequence of the Maxwell equation~\eqref{Maxwell equation STA vacuum} as in space-algebra but it is a way that is even simpler. The gradient of the Maxwell equation~\eqref{Maxwell equation STA vacuum} in space-time is in fact the electromagnetic wave equation. The gradient of the Maxwell equation~\eqref{Maxwell equation STA vacuum} in vacuum reads,
\begin{equation}\label{Maxwell equation STA vacuum gradient}
\nabla^2\,G = \frac{1}{c}\,\nabla\,J
\end{equation}
The gradient of the electric current density on the right-hand side of relation~\eqref{Maxwell equation STA vacuum gradient} can be written as the sum of the divergence and the curl,
\begin{equation}\label{Maxwell equation STA vacuum gradient bis}
\nabla^2\,G = \frac{1}{c}\,\nabla \cdot J + \frac{1}{c}\,\nabla \wedge J
\end{equation}
Since the Laplacian $\nabla^2$ is a scalar operator and the auxiliary electromagnetic multivector field $G$ is the sum of a linear combination~\eqref{Riemann Silberstein multivector G} of the electric displacement vector field $\boldsymbol{d}$ and the auxiliary magnetic bivector field $\boldsymbol{H}$, the divergence of the current density $\nabla \cdot J$ on the right-hand side of relation~\eqref{Maxwell equation STA vacuum gradient bis}, which is a scalar, has to vanish,
\begin{equation}\label{conservation electric charge}
\nabla \cdot J = 0
\end{equation}
The electric charge conservation law~\eqref{conservation electric charge} is recast in spatial algebra as the electric continuity equation~\eqref{continuity eq 0} using to the identity~\eqref{inner product coavariant contravariant bis} for the scalar product of two vectors in space-time and the definition of the current density vector~\eqref{G eq ter},
\begin{equation}\label{conservation electric charge 2}
\nabla \cdot J = \partial_{t}\,q + \boldsymbol{\nabla}\cdot\boldsymbol{j} = 0
\end{equation}
In view of the electric charge conservation law~\eqref{conservation electric charge}, relation~\eqref{Maxwell equation STA vacuum gradient bis} reduces to,
\begin{equation}\label{Maxwell equation STA vacuum gradient ter}
\nabla^2\,G = \frac{1}{c}\,\nabla \wedge J
\end{equation}
which is the electromagnetic wave equation in vacuum in space-time algebra. To show this, we recast now this equation in spatial algebra. In view of the geometric products~\eqref{geometric product gradient},~\eqref{geometric product gradient bis} and of the identity $e_0\,e_0 =1$, the Laplacian operator called the d'Alembertian operator is written in a specific space-time frame defined by a time vector $e_0$ as,
\begin{equation}\label{d Alembertian}
\nabla^2 = \nabla\,\nabla = \left(\nabla\,e_0\right)\left(e_0\,\nabla\right) = \left(\frac{1}{c}\,\partial_{t} -\,\boldsymbol{\nabla}\right)\left(\frac{1}{c}\,\partial_{t} + \boldsymbol{\nabla}\right) = \frac{1}{c^2}\,\partial^2_{t} -\,\boldsymbol{\nabla}^2
\end{equation}
In view of identity~\eqref{d Alembertian}, the left-hand side of relation~\eqref{Maxwell equation STA vacuum gradient ter} is recast in spatial algebra as,
\begin{equation}\label{d Alembertian G}
\nabla^2\,G = \left(\frac{1}{c^2}\,\partial^2_{t} -\,\boldsymbol{\nabla}^2\right)G
\end{equation}
In view of identity~\eqref{outer product coavariant contravariant bis}, the right-hand side of relation~\eqref{Maxwell equation STA vacuum gradient ter} is recast in spatial algebra as,
\begin{equation}\label{curl electric current density}
\frac{1}{c}\,\nabla \wedge J = -\,\frac{1}{c^2}\,\partial_t\boldsymbol{j} -\,\boldsymbol{\nabla}\,q + \frac{1}{c}\,\boldsymbol{\nabla} \wedge \boldsymbol{j}
\end{equation}
According to relations~\eqref{d Alembertian G} and~\eqref{curl electric current density}, the gradient of the Maxwell equation~\eqref{Maxwell equation STA vacuum gradient ter} in vacuum is recast in spatial algebra as the electromagnetic wave equation~\eqref{wave eq G},
\begin{equation}\label{electromagnetic wave equation vacuum SA}
\left(\frac{1}{c^2}\,\partial^2_{t} -\,\boldsymbol{\nabla}^2\right)G = -\,\boldsymbol{\nabla}\,q -\,\frac{1}{c^2}\,\partial_t\boldsymbol{j}  + \frac{1}{c}\,\boldsymbol{\nabla} \wedge \boldsymbol{j}
\end{equation}
%


\section{Electromagnetic waves in matter (STA)}
\label{Electromagnetic waves in matter STA}

\noindent In the space-time algebra (STA), electromagnetic waves in matter are a direct and straightforward consequence of the Maxwell equation~\eqref{Maxwell equation G matter} or~\eqref{Maxwell equation STA matter bis} as in space-algebra but it is a way that is even simpler. The gradient of the Maxwell equation~\eqref{Maxwell equation G matter} or~\eqref{Maxwell equation STA matter bis} in space-time is in fact the electromagnetic wave equation. The gradient of the Maxwell equation~\eqref{Maxwell equation G matter} in matter reads,
\begin{equation}\label{Maxwell equation STA matter gradient}
\nabla^2\,G = \frac{1}{c}\,\nabla\,J
\end{equation}
In view of equation~\eqref{conservation electric charge}, the electromagnetic wave equation in matter~\eqref{Maxwell equation STA matter gradient} becomes,
\begin{equation}\label{Maxwell equation STA matter gradient extra}
\nabla^2\,G = \frac{1}{c}\,\nabla \wedge J
\end{equation}
The gradient of the inhomogeneous Maxwell equation~\eqref{Maxwell equation STA matter bis} in matter is given by,
\begin{equation}\label{Maxwell equation STA matter gradient bis}
\nabla^2\,F = \frac{1}{\varepsilon_0\,c}\,\nabla\,\tilde{J}
\end{equation}
Since the Laplacian $\nabla^2$ is a scalar operator and the electromagnetic multivector field $F$ is the sum of a linear combination~\eqref{Riemann Silberstein multivector} of the vector field $\boldsymbol{e}$ and the magnetic bivector field $\boldsymbol{B}$, the divergence of the current density in matter $\nabla \cdot \tilde{J}$ on the right-hand side of relation~\eqref{Maxwell equation STA matter gradient bis}, which is a scalar, has to vanish,
\begin{equation}\label{conservation electric charge matter}
\nabla \cdot \tilde{J} = 0
\end{equation}
In view of the electric charge conservation law~\eqref{conservation electric charge matter}, relation~\eqref{Maxwell equation STA matter gradient bis} reduces to,
\begin{equation}\label{Maxwell equation STA matter gradient ter}
\nabla^2\,F = \frac{1}{\varepsilon_0\,c}\,\nabla \wedge \tilde{J}
\end{equation}
which is the electromagnetic wave equation in matter in space-time algebra. To show this, we recast now this equation in spatial algebra. In view of identity~\eqref{d Alembertian}, the left-hand side of relation~\eqref{Maxwell equation STA matter gradient ter} is recast in spatial algebra as,
\begin{equation}\label{d Alembertian F}
\nabla^2\,F = \left(\frac{1}{c^2}\,\partial^2_{t} -\,\boldsymbol{\nabla}^2\right)F
\end{equation}
In view of identity~\eqref{outer product coavariant contravariant bis}, the right-hand side of relation~\eqref{Maxwell equation STA matter gradient ter} is recast in spatial algebra as,
\begin{equation}\label{curl electric current density matter}
\frac{1}{\varepsilon_0\,c}\,\nabla \wedge \tilde{J} = -\,\frac{1}{\varepsilon_0\,c^2}\,\partial_t\boldsymbol{\tilde{j}} -\,\frac{1}{\varepsilon_0}\,\boldsymbol{\nabla}\,\tilde{q} + \frac{1}{\varepsilon_0\,c}\,\boldsymbol{\nabla} \wedge \boldsymbol{\tilde{j}}
\end{equation}
According to relations~\eqref{d Alembertian F} and~\eqref{curl electric current density matter}, the gradient of the Maxwell equation~\eqref{Maxwell equation STA matter gradient ter} in matter is recast in spatial algebra as the electromagnetic wave equation~\eqref{wave eq G},
\begin{equation}\label{electromagnetic wave equation matter SA}
\left(\frac{1}{c^2}\,\partial^2_{t} -\,\boldsymbol{\nabla}^2\right)F = -\,\frac{1}{\varepsilon_0}\,\boldsymbol{\nabla}\,\tilde{q} -\,\frac{1}{\varepsilon_0\,c^2}\,\partial_t\boldsymbol{\tilde{j}}  + \frac{1}{\varepsilon_0\,c}\,\boldsymbol{\nabla} \wedge \boldsymbol{\tilde{j}}
\end{equation}
%


\section{Electromagnetic fields (STA)}
\label{Electromagnetic fields STA}

\noindent The electromagnetic multivectors $F$ and $G$ and the electromagnetic polarisation multivector $\tilde{P}$ in spatial algebra (SA) are in fact bivectors in space-time algebra (STA), as we will show in this section. We will also determine the components of these bivectors with respect to an orthonormal space-time frame. The electromagnetic multivector~\eqref{Riemann Silberstein multivector} is written in terms of the relative spatial vectors as,
\begin{equation}\label{F spatial vectors}
F = \boldsymbol{e} + c\,\boldsymbol{B} = e^i\,\boldsymbol{e}_i + \frac{1}{2}\,c\,B^{\,ij}\,\boldsymbol{e}_i\wedge\boldsymbol{e}_j
\end{equation}
Using the definition~\eqref{space-time bivectors} of the relative spatial vectors, the electromagnetic multivector~\eqref{F spatial vectors} is recast as,
\begin{equation}\label{F spatial vectors bis}
F = e^i\left(e_i\,e_0\right) + \frac{1}{2}\,c\,B^{\,ij}\left(e_i\,e_0\right)\wedge\left(e_j\,e_0\right)
\end{equation}
The electric field vector $\boldsymbol{e}$ and magnetic induction field bivector $\boldsymbol{B}$ are frame dependent since they are defined in a specific spatial algebra $\mathbb{G}^{3}$. In contrast, the electromagnetic field multivector $F$ is frame independent since it is defined in the space-time algebra $\mathbb{G}^{1,3}$. Using identities~\eqref{square space-time} and~\eqref{outer product space-time}, we obtain,
\begin{equation}\label{identities F}
\begin{split}
&e_i\,e_0 = e_i \wedge e_0\\
&\left(e_i\,e_0\right)\wedge\left(e_j\,e_0\right) = e_i\,e_0\,e_j\,e_0 = -\,e_i\,e_j = -\,e_i \wedge e_j
\end{split}
\end{equation}
In view of identities~\eqref{identities F}, the electromagnetic field multivector~\eqref{F spatial vectors bis} is recast as,
\begin{equation}\label{F spatial vectors ter}
F = e^i\,e_i \wedge e_0 -\,\frac{1}{2}\,c\,B^{\,ij}\,e_i \wedge e_j
\end{equation}
In the space-time algebra, the electromagnetic field multivector~\eqref{F spatial vectors ter} is in fact a bivector written as,
\begin{equation}\label{F space time bivector}
F = \frac{1}{2}\,F^{\mu\nu}\,e_\mu \wedge e_\nu 
\end{equation}
According to relations~\eqref{F space time bivector},~\eqref{inner product basis bivectors} and~\eqref{inner product basis bivectors bis}, the components of the electromagnetic field bivector $F$ are given by,
\begin{equation}\label{F electromagnetic components colon}
F^{\mu\nu} = \left(e^{\nu} \wedge e^{\mu}\right)\cdot F = F\cdot\left(e^{\nu} \wedge e^{\mu}\right)
\end{equation}
According to relations~\eqref{inner product different vectors} and~\eqref{triple product basis vectors STA}, the components of the electromagnetic field bivector $F$ are written as,
\begin{equation}\label{F electromagnetic components dots}
F^{\mu\nu} = e^{\nu}\cdot\left(e^{\mu}\cdot F\right) = \left(F\cdot e^{\nu}\right)\cdot e^{\mu}
\end{equation}
According to identity~\eqref{outer product space-time}, the components of the electromagnetic field bivector $F$ are antisymmetric,
\begin{equation}\label{F electromagnetic components anti}
F^{\mu\nu} = -\,F^{\nu\mu}
\end{equation}
In view of identities~\eqref{outer product space-time} and~\eqref{F electromagnetic components anti}, the electromagnetic field bivector~\eqref{F space time bivector} is split as,
\begin{equation}\label{F space time bivector bis}
F = \frac{1}{2}\left(F^{i0}\,e_i \wedge e_0 + F^{0i}\,e_0 \wedge e_i + F^{ij}\,e_i \wedge e_j\right) = F^{i0}\,e_i \wedge e_0 + \frac{1}{2}\,F^{ij}\,e_i \wedge e_j
\end{equation}
In view of relations~\eqref{F spatial vectors ter},~\eqref{F space time bivector bis} and~\eqref{F electromagnetic components anti}, the components of the electromagnetic field bivector $F$ are given by,
\begin{align}\label{components F}
&F^{\mu\mu} = -\,F^{\mu\mu} = 0\nonumber\\
&F^{i0} = -\,F^{0i} = e^{i}\\
&F^{ij} = -\,F^{ji} = -\,c\,B^{ij}\nonumber
\end{align}
Thus, the components of the electromagnetic bivector field $F$ are written as an antisymmetric matrix,
\begin{equation}\label{F electromagnetic matrix}
F^{\mu\nu} = \begin{pmatrix}
0 & -\,e^{1} & -\,e^{2} & -\,e^{3}\\
e^{1} & 0 & -\,c\,B^{12} & -\,c\,B^{13}\\
e^{2} & c\,B^{12} & 0 & -\,c\,B^{23}\\
e^{3} & c\,B^{13} & c\,B^{23} & 0\\
\end{pmatrix}
\end{equation}
The components of the electromagnetic field bivector $F^{\mu\nu}$ can be recast in terms of the components of the dual magnetic induction field vector $\boldsymbol{b}$ where $\boldsymbol{b} = \boldsymbol{B}^{\ast}$,
\begin{equation}\label{duality components b}
b^1\,\boldsymbol{e}_1 + b^2\,\boldsymbol{e}_2 + b^3\,\boldsymbol{e}_3 = B^{12}\left(\boldsymbol{e}_1\wedge\boldsymbol{e}_2\right)^{\ast} + B^{23}\left(\boldsymbol{e}_2\wedge\boldsymbol{e}_3\right)^{\ast} + B^{31}\left(\boldsymbol{e}_3\wedge\boldsymbol{e}_1\right)^{\ast}
\end{equation}
According to the duality~\eqref{vector duality},
\begin{align}\label{duality basis vectors}
&\left(\boldsymbol{e}_1\wedge\boldsymbol{e}_2\right)^{\ast} = \boldsymbol{e}_1\times\boldsymbol{e}_2 = \boldsymbol{e}_3\nonumber\\
&\left(\boldsymbol{e}_2\wedge\boldsymbol{e}_3\right)^{\ast} = \boldsymbol{e}_2\times\boldsymbol{e}_3 = \boldsymbol{e}_1\\
&\left(\boldsymbol{e}_3\wedge\boldsymbol{e}_1\right)^{\ast} = \boldsymbol{e}_3\times\boldsymbol{e}_1 = \boldsymbol{e}_2\nonumber
\end{align}
which implies that the identity~\eqref{duality components b} is recast as,
\begin{equation}\label{duality components bis}
b^1\,\boldsymbol{e}_1 + b^2\,\boldsymbol{e}_2 + b^3\,\boldsymbol{e}_3 = B^{23}\,\boldsymbol{e}_1 + B^{31}\,\boldsymbol{e}_2 + B^{12}\,\boldsymbol{e}_3
\end{equation}
Thus,
\begin{equation}\label{duality components ter}
B^{12} = b_3 \qquad\text{and}\qquad
B^{13} = -\,B^{31} = -\,b_2 \qquad\text{and}\qquad
B^{23} = b_1
\end{equation}
Using relations~\eqref{duality components ter}, the components of the electromagnetic field bivector~\eqref{F electromagnetic matrix} are recast in vector space as,
\begin{equation}\label{electromagnetic matrix bis}
F^{\mu\nu} = \begin{pmatrix}
0 & -\,e^{1} & -\,e^{2} & -\,e^{3}\\
e^{1} & 0 & -\,c\,b^{3} & c\,b^{2}\\
e^{2} & c\,b^{3} & 0 & -\,c\,b^{1}\\
e^{3} & -\,c\,b^{2} & c\,b^{1} & 0\\
\end{pmatrix}
\end{equation}
The frame dependence of the electric field vector $\boldsymbol{e}$ and magnetic induction field bivector $\boldsymbol{B}$ is made explicit using a geometric product with the time vector $e_0$. In view of relation~\eqref{F spatial vectors ter}, the double geometric product of the electromagnetic multivector $F$ and the time vector $e_0$ yields,
\begin{equation}\label{double geometric product F}
\begin{split}
&e_0\,F\,e_0 = e^i\,e_0\,e_i\,e_0 + \frac{1}{2}\,c\,B^{\,ij}\,e_0\left(e_i\wedge e_j\right)e_0\\
&\phantom{e_0\,F\,e_0} = -\,e^i\,e_i + \frac{1}{2}\,c\,B^{\,ij}\left(e_i\wedge e_j\right)
\end{split}
\end{equation}
In view of relations~\eqref{F spatial vectors},~\eqref{F spatial vectors bis} and~\eqref{double geometric product F}, we obtain,
\begin{equation}\label{time split F}
e_0\,F\,e_0 = -\,\boldsymbol{e} + c\,\boldsymbol{B} = -\,F^{\dag}
\end{equation}
Using relations~\eqref{time split F},~\eqref{e as multivector} and~\eqref{B as multivector}, the electric field vector $\boldsymbol{e}$ is written entirely in terms of the electromagnetic field multivector $F$ and the time vector $e_0$ as,
\begin{equation}\label{e field STA}
\boldsymbol{e} = \frac{1}{2}\left(F -\,e_0\,F\,e_0\right)
\end{equation}
and the magnetic field bivector $\boldsymbol{B}$ is written as,
\begin{equation}\label{B field STA}
\boldsymbol{B} = \frac{1}{2\,c}\left(F + e_0\,F\,e_0\right)
\end{equation}
The auxiliary electromagnetic multivector~\eqref{Riemann Silberstein multivector G} is written in terms of the relative spatial vectors as,
\begin{equation}\label{G spatial vectors}
G = \boldsymbol{d} + \frac{1}{c}\,\boldsymbol{H} = d^{\,i}\,\boldsymbol{e}_i + \frac{1}{2\,c}\,H^{\,ij}\,\boldsymbol{e}_i\wedge\boldsymbol{e}_j
\end{equation}
Using the definition~\eqref{space-time bivectors} of the relative spatial vectors, the auxiliary electromagnetic multivector~\eqref{G spatial vectors} is recast as,
\begin{equation}\label{G spatial vectors bis}
G = d^{\,i}\left(e_i\,e_0\right) + \frac{1}{2\,c}\,H^{\,ij}\left(e_i\,e_0\right)\wedge\left(e_j\,e_0\right)
\end{equation}
The electric displacement field vector $\boldsymbol{d}$ and auxiliary magnetic field bivector $\boldsymbol{H}$ are frame dependent since they are defined in a specific spatial algebra $\mathbb{G}^{3}$. In contrast, the electromagnetic field multivector $G$ is frame independent since it is defined in the space-time algebra $\mathbb{G}^{1,3}$. In view of identities~\eqref{identities F}, the auxiliary electromagnetic field multivector~\eqref{G spatial vectors bis} is recast as,
\begin{equation}\label{G spatial vectors ter}
G = d^{\,i}\,e_i \wedge e_0 -\,\frac{1}{2\,c}\,H^{\,ij}\,e_i \wedge e_j
\end{equation}
In the space-time algebra, the electromagnetic field multivector~\eqref{G spatial vectors ter} is in fact a bivector written as,
\begin{equation}\label{G space time bivector}
G = \frac{1}{2}\,G^{\mu\nu}\,e_\mu \wedge e_\nu 
\end{equation}
According to relations~\eqref{G space time bivector},~\eqref{inner product basis bivectors} and~\eqref{inner product basis bivectors bis}, the components of the electromagnetic field bivector $G$ are given by,
\begin{equation}\label{G electromagnetic components colon}
G^{\mu\nu} = \left(e^{\nu} \wedge e^{\mu}\right)\cdot G = G\cdot\left(e^{\nu} \wedge e^{\mu}\right)
\end{equation}
According to relations~\eqref{inner product different vectors} and~\eqref{triple product basis vectors STA}, the components of the electromagnetic field bivector $G$ are written as,
\begin{equation}\label{G electromagnetic components dots}
G^{\mu\nu} = e^{\nu}\cdot\left(e^{\mu}\cdot G\right) = \left(G\cdot e^{\nu}\right)\cdot e^{\mu}
\end{equation}
According to identity~\eqref{outer product space-time}, the components of the electromagnetic field bivector $G$ are antisymmetric,
\begin{equation}\label{G electromagnetic components anti}
G^{\mu\nu} = -\,G^{\nu\mu}
\end{equation}
In view of identities~\eqref{outer product space-time} and~\eqref{outer product space-time}, the auxiliary electromagnetic field bivector~\eqref{G space time bivector} is split as,
\begin{equation}\label{G space time bivector bis}
G = \frac{1}{2}\left(G^{i0}\,e_i \wedge e_0 + G^{0i}\,e_0 \wedge e_i + G^{ij}\,e_i \wedge e_j\right) = G^{i0}\,e_i \wedge e_0 + \frac{1}{2}\,G^{ij}\,e_i \wedge e_j
\end{equation}
In view of relations~\eqref{G spatial vectors ter},~\eqref{G space time bivector} and~\eqref{G electromagnetic components anti}, the components of the electromagnetic field bivector $G$ are given by,
\begin{align}\label{components G}
&G^{\mu\mu} = -\,G^{\mu\mu} = 0\nonumber\\
&G^{i0} = -\,G^{0i} = d^{i}\\
&G^{ij} = -\,G^{ji} = -\,c^{-1}\,H^{ij}\nonumber
\end{align}
Thus, the components of the electromagnetic bivector field $G$ are written as an antisymmetric matrix,
\begin{equation}\label{G electromagnetic matrix}
G^{\mu\nu} = \begin{pmatrix}
0 & -\,d^{1} & -\,d^{2} & -\,d^{3}\\
d^{1} & 0 & -\,c^{-1}\,H^{12} & -\,c^{-1}\,H^{13}\\
d^{2} & c^{-1}\,H^{12} & 0 & -\,c^{-1}\,H^{23}\\
d^{3} & c^{-1}\,H^{13} & c^{-1}\,H^{23} & 0\\
\end{pmatrix}
\end{equation}
The components of the electromagnetic field bivector $G^{\mu\nu}$ can be recast in terms of the components of the dual auxiliary magnetic field vector $\boldsymbol{h}$ where $\boldsymbol{h} = \boldsymbol{H}^{\ast}$,
\begin{equation}\label{duality components h}
h^1\,\boldsymbol{e}_1 + h^2\,\boldsymbol{e}_2 + h^3\,\boldsymbol{e}_3 = H^{12}\left(\boldsymbol{e}_1\wedge\boldsymbol{e}_2\right)^{\ast} + H^{23}\left(\boldsymbol{e}_2\wedge\boldsymbol{e}_3\right)^{\ast} + H^{31}\left(\boldsymbol{e}_3\wedge\boldsymbol{e}_1\right)^{\ast}
\end{equation}
According to the duality~\eqref{duality basis vectors}, the identity~\eqref{duality components h} is recast as,
\begin{equation}\label{duality components quad}
h^1\,\boldsymbol{e}_1 + h^2\,\boldsymbol{e}_2 + h^3\,\boldsymbol{e}_3 = H^{23}\,\boldsymbol{e}_1 + H^{31}\,\boldsymbol{e}_2 + H^{12}\,\boldsymbol{e}_3
\end{equation}
Thus,
\begin{equation}\label{duality components pent}
H^{12} = h_3 \qquad\text{and}\qquad
H^{13} = -\,H^{31} = -\,h_2 \qquad\text{and}\qquad
H^{23} = h_1
\end{equation}
Using relations~\eqref{duality components pent}, the components of the electromagnetic field bivector~\eqref{G electromagnetic matrix} are recast in vector space as,
\begin{equation}\label{G electromagnetic matrix bis}
G^{\mu\nu} = \begin{pmatrix}
0 & -\,d^{1} & -\,d^{2} & -\,d^{3}\\
d^{1} & 0 & -\,c^{-1}\,h^{3} & c^{-1}\,h^{2}\\
d^{2} & c^{-1}\,h^{3} & 0 & -\,c^{-1}\,h^{1}\\
d^{3} & -\,c^{-1}\,h^{2} & c^{-1}\,h^{1} & 0\\
\end{pmatrix}
\end{equation}
The frame dependence of the electric displacement field vector $\boldsymbol{d}$ and auxiliary magnetic field bivector $\boldsymbol{H}$ is made explicit using a geometric product with the time vector $e_0$. In view of relation~\eqref{G spatial vectors ter}, the double geometric product of the auxiliary electromagnetic multivector $G$ and the time vector $e_0$ yields,
\begin{equation}\label{double geometric product G}
\begin{split}
&e_0\,G\,e_0 = d^i\,e_0\,e_i\,e_0 + \frac{1}{2\,c}\,H^{\,ij}\,e_0\left(e_i\wedge e_j\right)e_0\\
&\phantom{e_0\,G\,e_0} = -\,e^i\,e_i + \frac{1}{2\,c}\,H^{\,ij}\left(e_i\wedge e_j\right)
\end{split}
\end{equation}
In view of relations~\eqref{G spatial vectors},~\eqref{G spatial vectors bis} and~\eqref{double geometric product G}, we obtain,
\begin{equation}\label{time split G}
e_0\,G\,e_0 = -\,\boldsymbol{e} + \frac{1}{c}\,\boldsymbol{H} = -\,G^{\dag}
\end{equation}
Using relations~\eqref{time split G},~\eqref{d as multivector} and~\eqref{H as multivector}, the electric displacement field vector $\boldsymbol{d}$ is written entirely in terms of the auxiliary electromagnetic field multivector $G$ and the time vector $e_0$ as,
\begin{equation}\label{d field STA}
\boldsymbol{d} = \frac{1}{2}\left(G-\,e_0\,G\,e_0\right)
\end{equation}
and the auxiliary magnetic field bivector $\boldsymbol{H}$ is written as,
\begin{equation}\label{H field STA}
\boldsymbol{H} = \frac{c}{2}\left(G + e_0\,G\,e_0\right)
\end{equation}
The electromagnetic polarisation multivector~\eqref{P matter} is written in terms of the relative spatial vectors as,
\begin{equation}\label{P spatial vectors}
\tilde{P} = \boldsymbol{\tilde{p}} -\,\frac{1}{c}\,\boldsymbol{\tilde{M}} = p^{\,i}\,\boldsymbol{e}_i -\,\frac{1}{2\,c}\,M^{\,ij}\,\boldsymbol{e}_i\wedge\boldsymbol{e}_j
\end{equation}
Using the definition~\eqref{space-time bivectors} of the relative spatial vectors, the electromagnetic polarisation multivector~\eqref{P spatial vectors} is recast as,
\begin{equation}\label{P spatial vectors bis}
\tilde{P} = p^{\,i}\left(e_i\,e_0\right) -\,\frac{1}{2\,c}\,M^{\,ij}\left(e_i\,e_0\right)\wedge\left(e_j\,e_0\right)
\end{equation}
The electric polarisation vector $\boldsymbol{\tilde{p}}$ and magnetisation bivector $\boldsymbol{\tilde{M}}$ are frame dependent since they are defined in a specific spatial algebra $\mathbb{G}^{3}$. In contrast, the electromagnetic polarisation multivector $\tilde{P}$ is frame independent since it is defined in the space-time algebra $\mathbb{G}^{1,3}$. In view of identities~\eqref{identities F}, the electric polarisation multivector~\eqref{G spatial vectors bis} is recast as,
\begin{equation}\label{P spatial vectors ter}
\tilde{P} = \tilde{p}^{\,i}\,e_i \wedge e_0 + \frac{1}{2\,c}\,\tilde{M}^{\,ij}\,e_i \wedge e_j
\end{equation}
In the space-time algebra, the electromagnetic polarisation multivector~\eqref{P spatial vectors ter} is in fact a bivector written as,
\begin{equation}\label{P space time bivector}
\tilde{P} = \frac{1}{2}\,\tilde{P}^{\mu\nu}\,e_\mu \wedge e_\nu 
\end{equation}
According to relations~\eqref{P space time bivector},~\eqref{inner product basis bivectors} and~\eqref{inner product basis bivectors bis}, the components of the electromagnetic polarisation bivector $\tilde{P}$ are given by,
\begin{equation}\label{P electromagnetic components colon}
\tilde{P}^{\mu\nu} = \left(e^{\nu} \wedge e^{\mu}\right)\cdot \tilde{P} = \tilde{P}\cdot\left(e^{\nu} \wedge e^{\mu}\right)
\end{equation}
According to relations~\eqref{inner product different vectors} and~\eqref{triple product basis vectors STA}, the components of the electromagnetic polarisation bivector $\tilde{P}$ are written as,
\begin{equation}\label{P electromagnetic components dots}
\tilde{P}^{\mu\nu} = e^{\nu}\cdot\left(e^{\mu}\cdot \tilde{P}\right) = \left(\tilde{P}\cdot e^{\nu}\right)\cdot e^{\mu}
\end{equation}
According to identity~\eqref{outer product space-time}, the components of the electromagnetic polarisation bivector $\tilde{P}$ are antisymmetric,
\begin{equation}\label{P electromagnetic components anti}
\tilde{P}^{\mu\nu} = -\,\tilde{P}^{\nu\mu}
\end{equation}
In view of identities~\eqref{outer product space-time} and~\eqref{outer product space-time}, the electromagnetic polarisation bivector~\eqref{P space time bivector} is split as,
\begin{equation}\label{P space time bivector bis}
\tilde{P} = \frac{1}{2}\left(\tilde{P}^{i0}\,e_i \wedge e_0 + \tilde{P}^{0i}\,e_0 \wedge e_i + \tilde{P}^{ij}\,e_i \wedge e_j\right) = \tilde{P}^{i0}\,e_i \wedge e_0 + \frac{1}{2}\,\tilde{P}^{ij}\,e_i \wedge e_j
\end{equation}
In view of relations~\eqref{P spatial vectors ter},~\eqref{P space time bivector} and~\eqref{P electromagnetic components anti}, the components of the electromagnetic polarisation bivector $\tilde{P}$ are given by,
\begin{align}\label{components P}
&\tilde{P}^{\mu\mu} = -\,\tilde{P}^{\mu\mu} = 0\nonumber\\
&\tilde{P}^{i0} = -\,\tilde{P}^{0i} = \tilde{p}^{i}\\
&\tilde{P}^{ij} = -\,\tilde{P}^{ji} = c^{-1}\,\tilde{M}^{ij}\nonumber
\end{align}
Thus, the components of the electromagnetic polarisation bivector $\tilde{P}$ are written as an antisymmetric matrix,
\begin{equation}\label{P electromagnetic matrix}
\tilde{P}^{\mu\nu} = \begin{pmatrix}
0 & -\,\tilde{p}^{1} & -\,\tilde{p}^{2} & -\,\tilde{p}^{3}\\
\tilde{p}^{1} & 0 & c^{-1}\,\tilde{M}^{12} & c^{-1}\,\tilde{M}^{13}\\
\tilde{p}^{2} & -\,c^{-1}\,\tilde{M}^{12} & 0 & c^{-1}\,\tilde{M}^{23}\\
\tilde{p}^{3} & -\,c^{-1}\,\tilde{M}^{13} & -\,c^{-1}\,\tilde{M}^{23} & 0\\
\end{pmatrix}
\end{equation}
The components of the electromagnetic polarisation bivector $\tilde{P}^{\mu\nu}$ can be recast in terms of the components of the dual auxiliary magnetisation vector $\boldsymbol{\tilde{m}}$ where $\boldsymbol{\tilde{m}} = \boldsymbol{\tilde{M}}^{\ast}$,
\begin{equation}\label{duality components m}
\tilde{m}^1\,\boldsymbol{e}_1 + \tilde{m}^2\,\boldsymbol{e}_2 + \tilde{m}^3\,\boldsymbol{e}_3 = \tilde{M}^{12}\left(\boldsymbol{e}_1\wedge\boldsymbol{e}_2\right)^{\ast} + \tilde{M}^{23}\left(\boldsymbol{e}_2\wedge\boldsymbol{e}_3\right)^{\ast} + \tilde{M}^{31}\left(\boldsymbol{e}_3\wedge\boldsymbol{e}_1\right)^{\ast}
\end{equation}
According to the duality~\eqref{duality basis vectors}, the identity~\eqref{duality components m} is recast as,
\begin{equation}\label{duality components hex}
\tilde{m}^1\,\boldsymbol{e}_1 + \tilde{m}^2\,\boldsymbol{e}_2 + \tilde{m}^3\,\boldsymbol{e}_3 = \tilde{M}^{23}\,\boldsymbol{e}_1 + \tilde{M}^{31}\,\boldsymbol{e}_2 + \tilde{M}^{12}\,\boldsymbol{e}_3
\end{equation}
Thus,
\begin{equation}\label{duality components hep}
\tilde{M}^{12} = \tilde{m}_3 \qquad\text{and}\qquad
\tilde{M}^{13} = -\,\tilde{M}^{31} = -\,\tilde{m}_2 \qquad\text{and}\qquad
\tilde{M}^{23} = \tilde{m}_1
\end{equation}
Using relations~\eqref{duality components hep}, the components of the electromagnetic polarisation bivector~\eqref{P electromagnetic matrix} are recast in vector space as,
\begin{equation}\label{P electromagnetic matrix bis}
\tilde{P}^{\mu\nu} = \begin{pmatrix}
0 & -\,\tilde{p}^{1} & -\,\tilde{p}^{2} & -\,\tilde{p}^{3}\\
\tilde{p}^{1} & 0 & c^{-1}\,\tilde{m}^{3} & -\,c^{-1}\,\tilde{m}^{2}\\
\tilde{p}^{2} & -\,c^{-1}\,\tilde{m}^{3} & 0 & c^{-1}\,\tilde{m}^{1}\\
\tilde{p}^{3} & c^{-1}\,\tilde{m}^{2} & -\,c^{-1}\,\tilde{m}^{1} & 0\\
\end{pmatrix}
\end{equation}
The frame dependence of the electric polarisation vector $\boldsymbol{\tilde{m}}$ and magnetisation bivector $\boldsymbol{\tilde{M}}$ is made explicit using a geometric product with the time vector $e_0$. In view of relation~\eqref{G spatial vectors ter}, the double geometric product of the electromagnetic polarisation multivector $P$ and the time vector $e_0$ yields,
\begin{equation}\label{double geometric product P}
\begin{split}
&e_0\,\tilde{P}\,e_0 = \tilde{p}^i\,e_0\,e_i\,e_0 -\,\frac{1}{2\,c}\,\tilde{M}^{\,ij}\,e_0\left(e_i\wedge e_j\right)e_0\\
&\phantom{e_0\,P\,e_0} = -\,\tilde{p}^i\,e_i -\,\frac{1}{2\,c}\,\tilde{M}^{\,ij}\left(e_i\wedge e_j\right)
\end{split}
\end{equation}
In view of relations~\eqref{P spatial vectors},~\eqref{P spatial vectors bis} and~\eqref{double geometric product G}, we obtain,
\begin{equation}\label{time split P}
e_0\,\tilde{P}\,e_0 = -\,\boldsymbol{\tilde{p}} -\,\frac{1}{c}\,\boldsymbol{\tilde{M}} = -\,\tilde{P}^{\dag}
\end{equation}
Using relations~\eqref{time split P},~\eqref{p as multivector} and~\eqref{M as multivector}, the electric polarisation field vector $\boldsymbol{\tilde{p}}$ is written entirely in terms of the electromagnetic polarisation multivector $P$ and the time vector $e_0$ as,
\begin{equation}\label{p field STA}
\boldsymbol{\tilde{p}} = \frac{1}{2}\left(\tilde{P}-\,e_0\,\tilde{P}\,e_0\right)
\end{equation}
and the magnetisation field bivector $\boldsymbol{\tilde{M}}$ is written as,
\begin{equation}\label{M field STA}
\boldsymbol{\tilde{M}} = -\,\frac{c}{2}\left(\tilde{P} + e_0\,\tilde{P}\,e_0\right)
\end{equation}
%


\section{Stress energy momentum in vacuum (STA)}
\label{Stress energy momentum in vacuum STA}

\noindent In the space-time algebra (STA) as opposed to to tensor calculus, the stress-energy momentum is not a tensor but a vector. In fact it is a self-adjoint linear application of a space-time vector $V$ in vacuum. If this vector $V$ is the space-time gradient $\nabla$, the stress-energy momentum vector $\dot{T}\left(\dot{\nabla}\right)$ reduces to the electromagnetic force density vector $f$ in space-time as we show below. In order to show this, we begin by writing the electromagnetic momentum density $P$ in space-time. The electromagnetic momentum density $P$ is a contravariant vector in space-time defined as,
\begin{equation}\label{P momentum}
P = P^{\mu}\,e_{\mu} = P^{0}\,e_{0} + P^{i}\,e_{i} = \frac{e}{c}\,e_{0} + \tilde{p}^{i}\,e_{i}
\end{equation}
According to the gradient~\eqref{gradient operator} and the identity~\eqref{inner product coavariant contravariant bis} for the scalar product of two vectors in space-time, the divergence of the electromagnetic momentum density~\eqref{P momentum} is given by,
\begin{equation}\label{divergence P momentum}
\nabla \cdot P = \frac{1}{c^2}\,\partial_{t}\,e + \boldsymbol{\nabla}\cdot\boldsymbol{p}
\end{equation}
In view of relations~\eqref{Riemann Silberstein multivector},~\eqref{geometric product electric current density}, and $e_0^2 =1$, the inner product of the electric current density vector $J$ and the electromagnetic multivector $F$ is written as,
\begin{equation}\label{J dot F 0}
J \cdot F = \left(J\,e_0\right)e_0 \cdot F = \left(qc + \boldsymbol{j}\right)e_0\cdot\left(\boldsymbol{e} + c\,\boldsymbol{B}\right)
\end{equation}
Since the time vector $e_0$ anticommutes with the vector $\boldsymbol{e}$ and commutes with the bivector $\boldsymbol{B}$, using the identity~\eqref{inner product v B antisymmetric}, relation~\eqref{J dot F 0} is recast as,
\begin{equation}\label{J dot F 00}
J \cdot F = -\,\left(\boldsymbol{j}\cdot\boldsymbol{e}\right)e_0 -\,c\left(q\,\boldsymbol{e} + \boldsymbol{B}\cdot\boldsymbol{j}\right)e_0
\end{equation}
where $\boldsymbol{j}\cdot\boldsymbol{e}$ is the electromagnetic power density and $q\,\boldsymbol{e} + \,\boldsymbol{B}\cdot\boldsymbol{j}$ is the Lorentz force density in spatial algebra $\mathbb{G}^{3}$. This means that the vector $\frac{1}{c}\,\,J \cdot F$ in the space-time algebra $\mathbb{G}^{1,3}$ represents the electromagnetic force density,
\begin{equation}\label{force density 4D}
f = \frac{1}{c}\,J \cdot F
\end{equation}
In view of relations~\eqref{electric current density} and~\eqref{F space time bivector} and identities~\eqref{F electromagnetic components anti} and~\eqref{triple product basis vectors STA bis}, the electromagnetic force density~\eqref{force density 4D} in space-time is given by,
\begin{equation}\label{force density 4D bis}
f_{\mu}\,e^{\mu} = \frac{1}{2\,c}\,J^{\rho}\,F_{\nu\mu}\,e_{\rho}\cdot\left(e^{\nu} \wedge e^{\mu}\right) = \frac{1}{2\,c}\left(J^{\nu}\,F_{\nu\mu} -\,J^{\nu}\,F_{\mu\nu}\right)e^{\mu} = \frac{1}{c}\,J^{\nu}\,F_{\nu\mu}\,e^{\mu}
\end{equation}
Using relations~\eqref{F spatial vectors} and~\eqref{F spatial vectors ter}, the electromagnetic force density~\eqref{force density 4D} in space-time is written in the orthonormal basis $\{\,e_0,\,e_1,\,e_2,\,e_3\,\}$ as,
\begin{equation}\label{J dot F 4D vector}
f = \frac{1}{c}\,J \cdot F = \frac{1}{c}\,\left(J^{i}\,F_{i0}\right)e_0 + \frac{1}{c}\,\left(J^{i}\,F_{ij}\right)e_j
\end{equation}
In view of relations~\eqref{components F},~\eqref{J dot F 00} and~\eqref{J dot F 4D vector} and identity~\eqref{square space-time},
\begin{equation}\label{J dot F 4D vector bis}
f\cdot e_0 = \frac{1}{c}\,\left(J \cdot F\right)\cdot e_0 = \frac{1}{c}\,\left(J^{i}\,F_{i0}\right) = \frac{1}{c}\,\left(j^{i}\,e_{i}\right) = -\,\frac{1}{c}\,\boldsymbol{j}\cdot\boldsymbol{e}
\end{equation}
In view of relations~\eqref{divergence P momentum} and~\eqref{J dot F 4D vector bis}, Poynting's theorem in vacuum~\eqref{Poynting's theorem vacuum bis} is recast as,
\begin{equation}\label{Poynting's theorem STA 0}
\nabla \cdot P = \frac{1}{c^2}\,\left(J \cdot F\right)\cdot e_0 = \frac{1}{c}\,f\cdot e_0
\end{equation}
The geometric product of the electromagnetic momentum density~\eqref{P momentum} and the time vector $e_0$ yields,
\begin{equation}\label{P_e0}
P\,e_0 = \frac{e}{c}\left(e_{0}\,e_{0}\right) + \tilde{p}^{i}\left(e_{i}\,e_{0}\right) = \frac{1}{c}\left(e + \boldsymbol{p}\,c\right)
\end{equation}
Since $e_0^2 = 1$, multiplying relation~\eqref{P_e0} by $e_0$ we obtain,
\begin{equation}\label{P_e0 bis}
P = \frac{1}{c}\left(e + \boldsymbol{p}\,c\right)e_{0}
\end{equation}
In view of relations~\eqref{electromagnetic invariant F} and~\eqref{time split F},
\begin{equation}\label{electromagnetic invariant vacuum}
e + \boldsymbol{p}\,c = -\,\frac{1}{4}\left(G\,e_{0}\,F\,e_{0} + F\,e_{0}\,G\,e_{0}\right) = -\,\frac{1}{2}\,\varepsilon_{0}\,F\,e_{0}\,F\,e_{0}
\end{equation}
and using the identity $e_0^2 = 1$, the momentum vector~\eqref{P_e0 bis} is recast as,
\begin{equation}\label{P_e0 ter}
P = -\,\frac{1}{4\,c}\left(G\,e_{0}\,F + F\,e_{0}\,G\right) = -\,\frac{1}{2\,c}\,\varepsilon_{0}\,F\,e_{0}\,F
\end{equation}
The stress energy momentum vector $T\left(e_0\right)$ is defined as a linear mapping of the time vector $e_0$,
\begin{equation}\label{T_e0}
T\left(e_0\right) = P\,c = -\,\frac{1}{4}\left(G\,e_{0}\,F + F\,e_{0}\,G\right) = -\,\frac{1}{2}\,\varepsilon_{0}\,F\,e_{0}\,F
\end{equation}
Since the stress energy momentum vector $T\left(e_0\right)$ depends on the time vector $e_0$, it is frame dependent. In view of relation~\eqref{T_e0}, Poynting's theorem~\eqref{Poynting's theorem STA 0} is recast as,
\begin{equation}\label{Poynting's theorem STA}
\nabla\cdot T\left(e_0\right) = \frac{1}{c}\,\left(J \cdot F\right)\cdot e_0 = f\cdot e_0
\end{equation}
To generalise the specific stress energy momentum vector~\eqref{T_e0}, a stress energy momentum vector $T\left(V\right)$ can be defined as a linear mapping of a vector $V$,
\begin{equation}\label{T_V}
T\left(V\right) = -\,\frac{1}{4}\left(G\,V\,F + F\,V\,G\right) = -\,\frac{1}{2}\,\varepsilon_{0}\,F\,V\,F
\end{equation}
The inner product of a vector $U$ and the stress energy momentum vector $T\left(V\right)$ yields a scalar,
\begin{equation}\label{U_T_V}
U \cdot T\left(V\right) = -\,\frac{1}{4}\,U\cdot\left(G\,V\,F + F\,V\,G\right) = -\,\frac{1}{4}\,\Big(\langle\,U\,G\,V\,F\,\rangle + \langle\,U\,F\,V\,G\,\rangle\Big)
\end{equation}
where the angle brackets denote the scalar part of the multivector. Similarly, the inner product of the vector $V$ and the stress energy momentum vector $T\left(U\right)$ also yields a scalar,
\begin{equation}\label{V_T_U}
V \cdot T\left(U\right) = -\,\frac{1}{4}\,V\cdot\left(G\,U\,F + F\,U\,G\right) = -\,\frac{1}{4}\,\Big(\langle\,V\,G\,U\,F\,\rangle + \langle\,V\,F\,U\,G\,\rangle\Big)
\end{equation}
Using the invariance under cyclic permutation of the scalar part of the geometric product of four multivectors~\eqref{multivectors cyclic permutation bis}, we obtain,
\begin{equation}\label{U_F_V_F cyclic}
\langle\,U\,G\,V\,F\,\rangle = \langle\,V\,F\,U\,G\,\rangle \qquad\text{and}\qquad \langle\,U\,F\,V\,G\,\rangle = \langle\,V\,G\,U\,F\,\rangle
\end{equation}
Thus, according to relations~\eqref{U_T_V},~\eqref{V_T_U} and~\eqref{U_F_V_F cyclic} in vacuum, the stress energy momentum vector satisfies the symmetry,~\cite{Lasenby:2003}
\begin{equation}\label{T sym}
U \cdot T\left(V\right) = V \cdot T\left(U\right)
\end{equation}
which means that $T\left(V\right)$ is a self-adjoint linear application of $V$. In particular, for $U = \nabla$ and $V = e_0$, relation~\eqref{T sym} becomes,
\begin{equation}\label{T nabla e_0}
\nabla \cdot T\left(e_0\right) = e_0 \cdot \dot{T}\,(\dot{\nabla}) = \dot{T}\,(\dot{\nabla}) \cdot e_0
\end{equation}
In view of relation~\eqref{T nabla e_0}, Poynting's theorem~\eqref{Poynting's theorem STA} is recast as,
\begin{equation}\label{Poynting's theorem STA bis}
\dot{T}\,(\dot{\nabla}) \cdot e_0 = \frac{1}{c}\,\left(J \cdot F\right) \cdot e_0 = f \cdot e_0
\end{equation}
which means that Poynting's theorem is an explicit expression of the stress energy momentum vector,
\begin{equation}\label{Poynting's theorem STA ter}
\dot{T}\,(\dot{\nabla}) = \frac{1}{c}\,J \cdot F = f
\end{equation}
According to relation~\eqref{T_V}, the stress energy momentum of the gradient is given by,
\begin{equation}\label{T_nabla}
\dot{T}\,(\dot{\nabla}) = -\,\frac{1}{4}\,\dot{G}\,\dot{\nabla}\,F -\,\frac{1}{4}\,G\,\dot{\nabla}\,\dot{F} -\,\frac{1}{4}\,\dot{F}\,\dot{\nabla}\,G -\,\frac{1}{4}\,F\,\dot{\nabla}\,\dot{G}
\end{equation}
where the overdot denotes on which bivector the gradient operates to the left or the right. The stress energy momentum vector~\eqref{T_nabla} is the electromagnetic force density vector in space-time. It is explicitly written as,
\begin{equation}\label{T_nabla_explicit}
\dot{T}\,(\dot{\nabla}) = \frac{1}{4}\left(\nabla\,G\right)F -\,\frac{1}{4}\,G\left(\nabla\,F\right) + \frac{1}{4}\left(\nabla\,F\right)G -\,\frac{1}{4}\,F\left(\nabla\,G\right)
\end{equation}
where the positive signs are due to the anticommutation between the vector $\nabla$ and the bivectors $G$ and $F$. Equivalently, in vacuum, using the electromagnetic constitutive equation~\eqref{G constitutive relation vacuum}, relations~\eqref{T_nabla} and~\eqref{T_nabla_explicit} for the stress energy momentum of the gradient are recast as,~\cite{Lasenby:2003}
\begin{equation}\label{T_nabla sec}
\dot{T}\,(\dot{\nabla}) = -\,\frac{1}{2}\,\varepsilon_0\,\dot{F}\,\dot{\nabla}\,F -\,\frac{1}{2}\,\varepsilon_0\,F\,\dot{\nabla}\,\dot{F}  = \frac{1}{2}\,\varepsilon_0\left(\nabla\,F\right)F -\,\frac{1}{2}\,\varepsilon_0\,F\left(\nabla\,F\right)
\end{equation}
To show the consistency of our approach, we use the Maxwell equation~\eqref{Maxwell equation STA vacuum} and~\eqref{Maxwell equation STA inhomogeneous vacuum F}, the electromagnetic constitutive equation~\eqref{G constitutive relation vacuum} and the identity~\eqref{inner product antisym STA quad} in order to recast the stress energy momentum vector~\eqref{T_nabla_explicit} as,
\begin{align}\label{T_nabla bis}
&\dot{T}\,(\dot{\nabla}) = \frac{1}{4}\left(\frac{1}{c}\,J\right)F -\,\frac{1}{4}\,\varepsilon_0\,F\left(\frac{1}{c\,\varepsilon_0}\,J\right) + \frac{1}{4}\,\left(\frac{1}{c\,\varepsilon_0}\,J\right)\varepsilon_0\,F -\,\frac{1}{4}\,F\left(\frac{1}{c}\,J\right)\nonumber\\
&\phantom{\dot{T}\,(\dot{\nabla})} = \frac{1}{2\,c}\left(J\,F -\,F\,J\right) = \frac{1}{c}\,J \cdot F = f
\end{align}
We now show that we recover the expression of the components of the stress energy momentum tensor in vacuum usually obtained in the framework of tensor calculus. In view of identity~\eqref{T sym}, the symmetric components of the stress energy momentum tensor are written as,
\begin{equation}\label{T components}
T^{\mu\nu} = e^{\mu}\cdot T\left(e^{\nu}\right) = \langle\,e^{\mu}\,T\left(e^{\nu}\right)\,\rangle = \langle\,e^{\nu}\,T\left(e^{\mu}\right)\,\rangle = e^{\nu}\cdot T\left(e^{\mu}\right) = T^{\nu\mu}
\end{equation}
In view of relation~\eqref{T_V}, relation~\eqref{T components} becomes,
\begin{equation}\label{T components bis}
\begin{split}
&T^{\mu\nu} = -\,\frac{1}{4}\,\Big(\,\langle\,e^{\mu}\,G\,e^{\nu}\,F\,\rangle + \langle\,e^{\mu}\,F\,e^{\nu}\,G\,\rangle\,\Big)\\
&\phantom{T^{\mu\nu}} = -\,\frac{1}{4}\,\Big(\,\langle\,e^{\nu}\,G\,e^{\mu}\,F\,\rangle + \langle\,e^{\nu}\,F\,e^{\mu}\,G\,\rangle\,\Big) = T^{\nu\mu}
\end{split}
\end{equation}
Using the invariance under cyclic permutation of the scalar part of the geometric product of four multivectors~\eqref{multivectors cyclic permutation bis}, we obtain,
\begin{equation}\label{e_F_e_F cyclic}
\langle\,e^{\mu}\,G\,e^{\nu}\,F\,\rangle = \langle\,e^{\nu}\,F\,e^{\mu}\,G\,\rangle \qquad\text{and}\qquad \langle\,e^{\mu}\,F\,e^{\nu}\,G\,\rangle = \langle\,e^{\nu}\,G\,e^{\mu}\,F\,\rangle
\end{equation}
In view of identity~\eqref{e_F_e_F cyclic}, relation~\eqref{T components bis} is recast as,
\begin{equation}\label{T components bis bis}
T^{\mu\nu} = -\,\frac{1}{4}\,\Big(\,\langle\,e^{\mu}\,G\,e^{\nu}\,F\,\rangle + \langle\,e^{\nu}\,G\,e^{\mu}\,F\,\rangle\,\Big)
\end{equation}
According to the symmetry~\eqref{e_F_e_F cyclic}, the scalar components~\eqref{T components bis} of the symmetric stress energy momentum tensor are the result of the inner product of vectors and the double inner product of bivectors,
\begin{equation}\label{T components ter}
T^{\mu\nu} = -\,\frac{1}{2}\,\left(e^{\mu} \cdot G\right) \cdot \left(e^{\nu} \cdot F\right) -\,\frac{1}{2}\,\left(e^{\nu} \cdot G\right) \cdot \left(e^{\mu} \cdot F\right) + \frac{1}{2}\,\left(e^{\mu} \cdot e^{\nu}\right)\,G \colon\! F
\end{equation}
where the positive sign in front of the last term is due to the anticommutation of the vectors $e^{\mu}$ and $e^{\nu}$ with the bivector $G$. In view of relations~\eqref{F electromagnetic components anti},~\eqref{F electromagnetic components dots},~\eqref{G electromagnetic components anti} and~\eqref{G electromagnetic components dots},
\begin{equation}\label{F G comp anti sym}
\begin{split}
&e^{\nu}\cdot\left(e^{\mu} \cdot F\right) = F^{\mu\nu} = -\,F^{\nu\mu} = -\,\left(F \cdot e^{\mu}\right)\cdot e^{\nu}\\
&e^{\nu}\cdot\left(e^{\mu} \cdot G\right) = G^{\mu\nu} = -\,G^{\nu\mu} = -\,\left(G \cdot e^{\mu}\right)\cdot e^{\nu}\\
\end{split}
\end{equation}
which implies that,
\begin{equation}\label{F G comp anti sym bis}
e^{\mu} \cdot F = -\,F \cdot e^{\mu} \qquad\text{and}\qquad e^{\mu} \cdot G = -\,G \cdot e^{\mu}
\end{equation}
In view of the symmetry~\eqref{F G comp anti sym bis} and identity~\eqref{inner product space-time}, relation~\eqref{T components ter} reduces to,
\begin{equation}\label{T components quad}
T^{\mu\nu} = \frac{1}{2}\,\left(e^{\mu} \cdot G\right) \cdot \left(F \cdot e^{\nu}\right) + \frac{1}{2}\,\left(e^{\nu} \cdot G\right) \cdot \left(F \cdot e^{\mu}\right) + \frac{1}{2}\,\left(e^{\mu} \cdot e^{\nu}\right)\,G \colon\! F
\end{equation}
According to relations~\eqref{F space time bivector},~\eqref{F electromagnetic components anti},~\eqref{G space time bivector},~\eqref{G electromagnetic components anti},~\eqref{inner product different vectors},~\eqref{triple product basis vectors STA} and~\eqref{triple product basis vectors STA ter},
\begin{equation}\label{F G comp bis}
\begin{split}
&e^{\mu} \cdot G = \frac{1}{2}\,G^{\rho\sigma}\,e^{\mu}\cdot\left(e_{\rho} \wedge e_{\sigma}\right) = \frac{1}{2}\,G^{\rho\sigma}\,\Big(\left(e^{\mu} \cdot e_{\rho}\right)e_{\sigma} -\,\left(e^{\mu} \cdot e_{\sigma}\right)e_{\rho}\Big)\\
&\phantom{e^{\mu} \cdot G} = \frac{1}{2}\,G^{\mu\sigma}\,e_{\sigma} -\,\frac{1}{2}\,G^{\rho\mu}\,e_{\rho} = G^{\mu\sigma}\,e_{\sigma}\\
&F \cdot e^{\nu} = \frac{1}{2}\,F^{\rho\sigma}\,\left(e_{\rho} \wedge e_{\sigma}\right)\cdot e^{\nu} = \frac{1}{2}\,F^{\rho\sigma}\,\Big(\left(e^{\nu} \cdot e_{\sigma}\right)e_{\rho} -\,\left(e^{\nu} \cdot e_{\rho}\right)e_{\sigma}\Big)\\
&\phantom{e^{\nu} \cdot F} = \frac{1}{2}\,F^{\rho\nu}\,e_{\rho} -\,\frac{1}{2}\,F^{\nu\sigma}\,e_{\sigma} = F^{\rho\nu}\,e_{\rho}
\end{split}
\end{equation}
In view of relations~\eqref{F electromagnetic components anti},~\eqref{F G comp anti sym bis},~\eqref{F G comp bis} and~\eqref{inner product space-time},
\begin{equation}\label{F G comp ter}
\begin{split}
&\left(e^{\mu} \cdot F\right) \cdot \left(G \cdot e^{\nu}\right) = \left(G^{\mu\sigma}\,e_{\sigma}\right)\cdot\left(F^{\rho\nu}\,e_{\rho}\right)
= G^{\mu\sigma}\,F^{\rho\nu}\left(e_{\sigma}\cdot e_{\rho}\right)\\
&\phantom{\left(e^{\mu} \cdot F\right) \cdot \left(G \cdot e^{\nu}\right)} = G^{\mu\sigma}\,F^{\rho\nu}\,\eta_{\sigma\rho} = G^{\mu\rho}\,F_{\rho}^{\phantom{\rho}\nu}
\end{split}
\end{equation}
According to relations~\eqref{F space time bivector},~\eqref{F electromagnetic components anti},~\eqref{G space time bivector},~\eqref{inner product different vectors} and~\eqref{inner product basis bivectors bis},
\begin{equation}\label{F G comp quad}
\begin{split}
&G \colon\! F = \frac{1}{4}\,G_{\rho\sigma}\,F^{\mu\nu}\,\left(e_{\rho} \wedge e_{\sigma}\right)\cdot\left(e^{\mu} \wedge e^{\nu}\right)\\
&\phantom{G \cdot F =} = \frac{1}{4}\,G_{\rho\sigma}\,F^{\mu\nu}\,\Big(\left(e_{\sigma} \cdot e^{\mu}\right)\left(e_{\rho} \cdot e^{\nu}\right) -\,\left(e_{\sigma} \cdot e^{\nu}\right)\left(e_{\rho} \cdot e^{\mu}\right)\Big)\\
&\phantom{G \cdot F =} = \frac{1}{4}\,\Big(G_{\rho\sigma}\,F^{\sigma\rho} -\,G_{\rho\sigma}\,F^{\rho\sigma}\,\Big) = -\,\frac{1}{2}\,G_{\rho\sigma}\,F^{\rho\sigma}
\end{split}
\end{equation}
In view of relations~\eqref{F G comp ter} and~\eqref{F G comp quad}, the components of the stress energy momentum tensor~\eqref{T components quad} are explicitly given by,
\begin{equation}\label{T components pent}
T^{\mu\nu} = G^{\mu\rho}\,F_{\rho}^{\phantom{\rho}\nu} -\,\frac{1}{4}\,\eta^{\mu\nu}\,G_{\rho\sigma}\,F^{\rho\sigma}
\end{equation}
or equivalently by lowering the second index of the components $T^{\mu\lambda}$ with the Minkowski metric $\eta_{\lambda\nu}$,
\begin{equation}\label{T components hex}
T^{\mu}_{\phantom{\mu}\nu} = G^{\mu\rho}\,F_{\rho\nu} -\,\frac{1}{4}\,\delta^{\mu}_{\nu}\,G^{\rho\sigma}\,F_{\rho\sigma}
\end{equation}
In vacuum, the components of the electromagnetic constitutive relation~\eqref{G constitutive relation vacuum} are given by,
\begin{equation}\label{EM const rel components}
e_{\nu}\cdot\left(e^{\mu}\cdot G\right)  = \varepsilon_0\,e_{\nu}\cdot\left(e^{\mu}\cdot F\right)
\end{equation}
In view of identities~\eqref{F electromagnetic components dots} and~\eqref{G electromagnetic components dots}, it is recast as,
\begin{equation}\label{EM const rel components bis}
G^{\mu}_{\phantom{\mu}\nu} = \varepsilon_0\,F^{\mu}_{\phantom{\mu}\nu}
\end{equation}
In vacuum, using the constitutive relation~\eqref{EM const rel components bis}, the components of the stress energy momentum tensor~\eqref{T components hex} are recast as,
\begin{equation}\label{T components vacuum}
T^{\mu}_{\phantom{\mu}\nu} = \varepsilon_0\left(F^{\mu\rho}\,F_{\rho\nu} -\,\frac{1}{4}\,\delta^{\mu}_{\nu}\,F^{\rho\sigma}\,F_{\rho\sigma}\right)
\end{equation}
%


%
\section{Stress energy momentum in matter (STA)}
\label{Stress energy momentum in matter STA}

\noindent In matter the stress-energy momentum is also a vector. As we show in this section, it is a self-adjoint linear application of a space-time vector $V$ in matter as in vacuum. If this vector $V$ is the space-time gradient $\nabla$, the stress-energy momentum vector $\dot{T}\left(\dot{\nabla}\right)$ reduces to the electromagnetic force density vector $f$ in space-time as we show below. In spatial algebra (SA), the Poynting theorem in matter~\eqref{Poynting's theorem matter bis} has the same algebraic structure as the Poynting theorem in vacuum~\eqref{Poynting's theorem vacuum bis}. Thus, in space-time algebra, the Poynting theorem in matter has the same structure the Poynting theorem in vacuum~\eqref{Poynting's theorem STA},
\begin{equation}\label{Poynting's theorem STA matter}
\nabla\cdot T\left(e_0\right) = \frac{1}{c}\,\left(J \cdot F\right) \cdot e_0 = f \cdot e_0
\end{equation}
where according to equations~\eqref{P_e0 bis} and~\eqref{T_e0}, the stress energy momentum tensor in matter $T\left(e_0\right)$ is given by,
\begin{equation}\label{T_e0 matter}
T\left(e_0\right) = P\,c = \left(e + \boldsymbol{p}\,c\right)e_0
\end{equation}
In view of the identities~\eqref{electromagnetic invariant matter} and~\eqref{time split F}, we obtain,
\begin{equation}\label{electromagnetic invariant matter bis}
e + \boldsymbol{p}\,c = -\,\frac{1}{4}\left(G\,e_0\,F\,e_0 + F\,e_0\,G\,e_0\right) -\,\frac{1}{8}\left(F\,G -\,G\,F + e_0\,F\,G\,e_0  -\,e_0\,G\,F\,e_0\right)
\end{equation}
Using relation~\eqref{electromagnetic invariant matter bis}, and the identity $e_0^2 = 1$, the stress energy momentum tensor in matter~\eqref{T_e0 matter} is recast as,
\begin{equation}\label{T_e0 matter bis}
T\left(e_0\right) = -\,\frac{1}{4}\left(G\,e_0\,F + F\,e_0\,G\right) -\,\frac{1}{8}\Big(\left(F\,G-\,G\,F\right)e_0 + e_0\left(F\,G -\,G\,F\right)\Big)
\end{equation}
In view of the commutator~\eqref{commutator STA bivectors} of the bivectors $F$ and $G$, the stress energy momentum vector~\eqref{T_e0 matter bis} is recast as,
\begin{equation}\label{T_e0 matter ter}
T\left(e_0\right) = -\,\frac{1}{4}\left(G\,e_0\,F + F\,e_0\,G\right) -\,\frac{1}{4}\Big(\left(F \times G\right)\cdot e_0 + e_0\cdot\left(F \times G\right)\Big)
\end{equation}
Since $F \times G$ is a bivector and $e_0$ is a vector, according to identity~\eqref{inner product antisym STA ter},
\begin{equation}\label{F G antisym}
\left(F \times G\right) \cdot e_0 = -\,e_0 \cdot \left(F \times G\right)
\end{equation}
which implies that the stress energy momentum vector~\eqref{T_e0 matter ter} reduces to,
\begin{equation}\label{T_e0 matter quad}
T\left(e_0\right) = -\,\frac{1}{4}\left(G\,e_0\,F + F\,e_0\,G\right)
\end{equation}
To generalise the specific stress energy momentum vector~\eqref{T_e0 matter bis}, a stress energy momentum vector $T\left(V\right)$ can be defined as a linear mapping of a vector $V$,
\begin{equation}\label{T_V matter}
T\left(V\right) = -\,\frac{1}{4}\left(G\,V\,F + F\,V\,G\right)
\end{equation}
According to relations~\eqref{U_T_V},~\eqref{V_T_U} and~\eqref{U_F_V_F cyclic} in matter, the stress energy momentum vector satisfies the symmetry,
\begin{equation}\label{T sym matter}
U \cdot T\left(V\right) = V \cdot T\left(U\right)
\end{equation}
which means that $T\left(V\right)$ is a self-adjoint linear application of $V$. In particular, for $U = \nabla$ and $V = e_0$, relation~\eqref{T sym matter} becomes,
\begin{equation}\label{T nabla e_0 matter}
\nabla \cdot T\left(e_0\right) = e_0 \cdot \dot{T}\,(\dot{\nabla}) = \dot{T}\,(\dot{\nabla})\cdot e_0
\end{equation}
In view of relation~\eqref{T nabla e_0 matter}, Poynting's theorem~\eqref{Poynting's theorem STA matter} is recast as,
\begin{equation}\label{Poynting's theorem STA matter bis}
\dot{T}\,(\dot{\nabla})\cdot e_0 = \frac{1}{c}\,\left(J \cdot F\right)\cdot e_0 = f \cdot e_0
\end{equation}
which means that Poynting's theorem is an explicit expression of the stress energy momentum vector,
\begin{equation}\label{Poynting's theorem STA matter ter}
\dot{T}\,(\dot{\nabla}) = \frac{1}{c}\,J \cdot F = f
\end{equation}
According to relation~\eqref{T_V}, the stress energy momentum of the gradient is given by,
\begin{equation}\label{T_nabla matter}
\dot{T}\,(\dot{\nabla}) = -\,\frac{1}{4}\,\dot{G}\,\dot{\nabla}\,F -\,\frac{1}{4}\,G\,\dot{\nabla}\,\dot{F} -\,\frac{1}{4}\,\dot{F}\,\dot{\nabla}\,G -\,\frac{1}{4}\,F\,\dot{\nabla}\,\dot{G}
\end{equation}
where the overdot denotes on which bivector the gradient operates to the left or the right. The stress energy momentum vector~\eqref{T_nabla matter} is the electromagnetic force density vector in space-time. It is explicitly written as,
\begin{equation}\label{T_nabla_explicit matter}
\dot{T}\,(\dot{\nabla}) = \frac{1}{4}\left(\nabla\,G\right)F -\,\frac{1}{4}\,G\left(\nabla\,F\right) + \frac{1}{4}\left(\nabla\,F\right)G -\,\frac{1}{4}\,F\left(\nabla\,G\right)
\end{equation}
where the positive signs are due to the anticommutation between the vector $\nabla$ and the bivectors $G$ and $F$. The electromagnetic susceptibility bivector in space-time $\chi_{F}\left(F\right)$ is a linear map of the electromagnetic field bivector $F$. Furthermore, we assume that the electromagnetic properties of the material medium are constant and uniform, which implies that,
\begin{equation}\label{gradient susceptibilities}
F\,\left(\nabla\chi_{F}\left(F\right)\right) = \chi_{F}\left(F\right)\left(\nabla\,F\right)
\end{equation}
According to relations~\eqref{G susceptibility} and~\eqref{gradient susceptibilities},
\begin{equation}\label{gradient susceptibilities F and G}
\begin{split}
&F\,\Big(\nabla\,G\left(F\right)\Big) = F\,\Big(\nabla\left(F + \chi_{F}\left(F\right)\right)\Big) = F\left(\nabla\,F\right) + F\,\left(\nabla\chi_{F}\left(F\right)\right)\\
&G\left(F\right)\left(\nabla\,F\right) = \Big(F + \chi_{F}\left(F\right)\Big)\,\left(\nabla\,F\right) = F\left(\nabla\,F\right) + \chi_{F}\left(F\right)\left(\nabla\,F\right)
\end{split}
\end{equation}
In view of relations~\eqref{gradient susceptibilities} and~\eqref{gradient susceptibilities F and G}, we deduce the following identity,
\begin{equation}\label{gradient susceptibilities FG}
G\left(F\right)\left(\nabla\,F\right) = F\,\Big(\nabla\,G\left(F\right)\Big)
\end{equation}
Thus, the stress energy momentum vector~\eqref{T_nabla matter} in matter reduces to,
\begin{equation}\label{T_nabla_explicit matter bis}
\dot{T}\,(\dot{\nabla}) = \frac{1}{2}\left(\nabla\,G\right)F -\,\frac{1}{2}\,F\left(\nabla\,G\right)
\end{equation}
To show the consistency of our approach, we use the Maxwell equation~\eqref{Maxwell equation G matter}, relation~\eqref{Poynting's theorem STA matter ter} and identity~\eqref{inner product antisym STA quad} to recast the stress energy momentum vector~\eqref{T_nabla_explicit matter bis} as,
\begin{align}\label{T_nabla bis matter}
&\dot{T}\,(\dot{\nabla}) = \frac{1}{2}\left(\frac{1}{c}\,J\right)F -\,\frac{1}{2}\,F\left(\frac{1}{c}\,J\right) = \frac{1}{2\,c}\left(J\,F -\,F\,J\right) = \frac{1}{c}\left(J \cdot F\right) = f\nonumber
\end{align}
For an non-uniform material material medium, an additional ponderomotive force density vector in space-time has to be added to the electromagnetic force density vector $f$ in space-time. Since the stress energy momentum vector in matter~\eqref{T_nabla_explicit matter} has the same structure as the stress energy momentum vector in vacuum~\eqref{T_nabla_explicit}, the components of the stress energy momentum tensor in matter have the same structure as the components of the stress energy momentum vector in vacuum,
\begin{equation}\label{T components matter}
T^{\mu}_{\phantom{\mu}\nu} = G^{\mu\rho}\,F_{\rho\nu} -\,\frac{1}{4}\,\delta^{\mu}_{\nu}\,G^{\rho\sigma}\,F_{\rho\sigma}
\end{equation}
In matter, the components of the electromagnetic constitutive relation~\eqref{G matter constitutive relation} are given by,
\begin{equation}\label{const rel components}
e_{\nu}\cdot\left(e^{\mu}\cdot G\right) = \varepsilon_0\,e_{\nu}\cdot\left(e^{\mu}\cdot F\right) + \varepsilon_0\,e_{\nu}\cdot\left(e^{\mu}\cdot \tilde{P}\right)
\end{equation}
In view of identities~\eqref{F electromagnetic components dots},~\eqref{G electromagnetic components dots} and~\eqref{P electromagnetic components dots}, it is recast as,
\begin{equation}\label{const rel components bis}
G^{\mu}_{\phantom{\mu}\nu} = \varepsilon_0\,F^{\mu}_{\phantom{\mu}\nu} + \tilde{P}^{\mu}_{\phantom{\mu}\nu}
\end{equation}
In matter, using the constitutive relation~\eqref{const rel components bis}, the components of the stress energy momentum tensor~\eqref{T components matter} are recast as,
\begin{equation}\label{T components matter bis}
\begin{split}
&T^{\mu}_{\phantom{\mu}\nu} = \varepsilon_0\left(F^{\mu\rho}\,F_{\rho\nu} -\,\frac{1}{4}\,\delta^{\mu}_{\nu}\,F^{\rho\sigma}\,F_{\rho\sigma}\right)\\
&\phantom{T^{\mu}_{\phantom{\mu}\nu} =} + \tilde{P}^{\mu\rho}\,F_{\rho\nu} -\,\frac{1}{4}\,\delta^{\mu}_{\nu}\,\tilde{P}^{\rho\sigma}\,F_{\rho\sigma}
\end{split}
\end{equation}
%


\section{Electromagnetic potential (STA)}
\label{Electromagnetic potential STA}

\noindent In vector space (VS), the electric scalar potential $\phi$ and the magnetic vector potential $\boldsymbol{a}$ are the temporal and spatial parts of a quadrivector $A$. Similarly, in the space-time algebra (STA), we define an electromagnetic vector potential $A$ in space-time and express the electromagnetic field bivector $F$ in terms of $A$.  The electromagnetic vector potential $A$ is a contravariant vector in space-time defined as,
\begin{equation}\label{electromagnetic vector potential}
A = A^{\mu}\,e_{\mu} = A^{0}\,e_{0} + A^{i}\,e_{i} = \frac{\phi}{c}\,e_{0} + a^{i}\,e_{i}
\end{equation}
In view of relation~\eqref{inner product contravariant vector}, the inner product of the electromagnetic vector potential $A$ with the time vector $e_{0}$ yields,
\begin{equation}\label{inner product electromagnetic vector potential}
A \cdot e_{0} = e_{0} \cdot A = A^{0} = \frac{\phi}{c}
\end{equation}
According to relation~\eqref{outer product contravariant vector}, the outer product of the electromagnetic vector potential $A$ with the time vector $e_{0}$ yields,
\begin{equation}\label{outer product electromagnetic vector potential}
A \wedge e_{0} = -\,e_{0} \wedge A = A^{i}\left(e_{i} \wedge e_{0}\right) = a^{i}\,\boldsymbol{e}_{i} = \boldsymbol{a}
\end{equation}
In view of the inner product~\eqref{inner product contravariant vector} and the outer product~\eqref{outer product contravariant vector}, the geometric product of the electromagnetic vector potential $A$ and the time vector $e_{0}$ yields,
\begin{equation}\label{geometric product electromagnetic vector potential}
A\,e_{0} = A \cdot e_{0} + A \wedge e_{0} = \frac{\phi}{c} + \boldsymbol{a}
\end{equation}
Taking into account the properties~\eqref{inner product space-time} and~\eqref{outer product space-time}, the geometric product in reverse order is given by,
\begin{equation}\label{geometric product electromagnetic vector potential bis}
e_{0}\,A = e_{0} \cdot A + e_{0} \wedge A = A \cdot e_{0} -\,A \wedge e_{0} = \frac{\phi}{c} -\,\boldsymbol{a}
\end{equation}
In view of identity~\eqref{outer product coavariant contravariant bis}, the curl of the electromagnetic vector potential $A$ is written as,
\begin{equation}\label{geometric product electromagnetic vector potential bis sec}
\nabla \wedge A = -\,\frac{1}{c}\,\partial_t\,\boldsymbol{a} -\,\boldsymbol{\nabla}\left(\frac{\phi}{c}\right) + \boldsymbol{\nabla}\cdot\boldsymbol{a}
\end{equation}
According to relation~\eqref{geometric product electromagnetic vector potential bis sec}, the electromagnetic field bivector~\eqref{F potentials} is the curl of the electromagnetic potential up to a factor $c$,
\begin{equation}\label{F potential A}
F = c\,\nabla \wedge A
\end{equation}
In view of identity~\eqref{inner product coavariant contravariant bis}, the divergence of the vector potential $A$ is written as,
\begin{equation}\label{geometric product electromagnetic vector potential ter}
\nabla \cdot A = \frac{1}{c}\,\partial_t\left(\frac{\phi}{c}\right) + \boldsymbol{\nabla}\cdot\boldsymbol{a}
\end{equation}
In view of relations~\eqref{geometric product electromagnetic vector potential ter} and~\eqref{Lorentz gauge SA}, the Lorentz gauge in space-time algebra is written as,
\begin{equation}\label{Lorentz gauge STA}
\nabla \cdot A = 0
\end{equation}
In the Lorentz gauge~\eqref{Lorentz gauge STA}, the electromagnetic field bivector~\eqref{F potential A} is recast as,
\begin{equation}\label{F potential A bis}
F = c\,\nabla \wedge A = F = c\,\nabla\,A -\,c\,\nabla \cdot A = c\,\nabla\,A
\end{equation}
Thus, in the Lorentz gauge~\eqref{Lorentz gauge STA}, the gradient of the electromagnetic field bivector~\eqref{F potential A bis} is given by,
\begin{equation}\label{gradient F STA ter}
\nabla\,F = c\,\nabla^2 A
\end{equation}
In view of relations~\eqref{gradient F STA ter} and~\eqref{light}, the Maxwell equation in vacuum~\eqref{Maxwell equation STA vacuum F} is recast in terms of the electromagnetic potential vector $A$ as,
\begin{equation}\label{Maxwell eq A vacuum}
\nabla^2\,A = \mu_0\,J
\end{equation}
In view of relations~\eqref{gradient F STA ter} and~\eqref{light}, the Maxwell equation in matter~\eqref{Maxwell equation STA matter} is recast in terms of the electromagnetic potential vector $A$ as,
\begin{equation}\label{Maxwell eq A matter}
\nabla^2\,A = \mu_0\,\tilde{J}
\end{equation}
%


\section{Conclusion}
\label{Conclusion}

Geometric algebra (GA) has been the subject of intense study for the past half century~\cite{Hestenes:1999,Hestenes:2015,Lasenby:2003,Macdonald:2011,Macdonald:2012}. Electrodynamics has been examined in spatial algebra (SA) and space-time algebra (STA) by numerous authors~\cite{Hestenes:2015,Lasenby:2003,Jancewicz:1989,Baylis:1999,Arthur:2011,Joot:2019}. This prompts the following question : ``Is there any need to discuss the matter further ?'' The answer is ``yes'' for two main reasons. First, the magnetic induction field $\boldsymbol{b}$ and the auxiliary magnetic field $\boldsymbol{h}$ are treated in general as pseudovectorial fields by the vast majority of authors with the notable exception of MacDonald~\cite{Macdonald:2011} who introduced magnetic induction field bivector $\boldsymbol{B}$ and the auxiliary magnetic field bivector $\boldsymbol{H}$ but he did not investigate the foundations of electromagnetism in spatial algebra (SA) or space-time (STA) further. Since the magnetic fields clearly geometrically as bivectors, it is necessary to treat them accordingly and this has been done in this publication. Second, the electric polarisation and magnetisation in matter was not studied in depth so far in spatial algebra (SA) and space-time (STA). Although, there was an attempt by Arthur~\cite{Arthur:2011} to describe electromagnetism in matter in spatial algebra (SA) but these results remain inconclusive.

In this publication, we considered an electric polarisation vector $\boldsymbol{\tilde{p}}$ induced linearly by the electric field vector $\boldsymbol{e}$ and a magnetisation bivector $\boldsymbol{\tilde{M}}$ induced linearly by the magnetic induction field bivector $\boldsymbol{B}$.

First, in the spatial algebra (SA), we introduced a polarisation multivector $\tilde{P} = \boldsymbol{\tilde{p}} -\,\frac{1}{c}\,\boldsymbol{\tilde{M}}$ and an auxiliary electromagnetic field multivector $G = \varepsilon_0\,F + \tilde{P}$ we were able to express the Maxwell equation in matter~\eqref{Maxwell equation G matter} very elegantly. Using the bound electric charge density $\tilde{q}$ and the bound electric current density $\boldsymbol{\tilde{j}}$, the Maxwell equation~\eqref{Maxwell equation STA matter} could then be written entirely in terms of the electromagnetic field multivector $F$. This allowed us to derive the electromagnetic wave equation in matter~\eqref{wave eq F matter} straightforwardly in the spatial algebra (SA). Writing the electromagnetic energy density $e$ and momentum density $\boldsymbol{p}$ in terms of the electromagnetic multivectors $F$, $G$ and $\tilde{P}$, we show that the electromagnetic polarisation multivector $\tilde{P}$ is a linear map of the electromagnetic multivector $F$ written as $\tilde{P}\left(F\right) = \varepsilon_0\,\chi\left(F\right)$ where $\chi\left(F\right)$ is the electromagnetic susceptibility multivector.

Second, in the space-time algebra (STA), the electromagnetic multivectors $F$, $G$ and $\tilde{P}$ defined in the spatial algebra (SA) could be interpreted as bivectors in the space-time algebra (STA). Writing the Maxwell equation~\eqref{Maxwell equation G matter} in terms of the auxiliary electromagnetic field multivector $G$, the stunningly elegant equation $\nabla\,G = \frac{1}{c}\,J$ had the same analytical expression as in vacuum~\eqref{Maxwell equation STA vacuum}. By introducing a bound current vector $\tilde{J} = J -\,c\,\nabla\cdot\tilde{P}$ in space-time, the Maxwell equation in matter~\eqref{Maxwell equation STA matter} could be written also in terms of the electromagnetic field bivector $F$ as $\nabla\,F = \frac{1}{\varepsilon_0\,c}\,\tilde{J}$. In the space-time algebra (STA), the wave equation in matter~\eqref{Maxwell equation STA matter gradient extra} is obtained by simply taking the gradient of the Maxwell equation in matter~\eqref{Maxwell equation G matter}, namely $\nabla^2\,G = \frac{1}{c}\,\nabla\wedge J$ since $\nabla\cdot J = 0$, or equivalently $\nabla^2\,F = \frac{1}{\varepsilon_0\,c}\,\nabla\wedge\tilde{J}$ since $\nabla\cdot\tilde{J} = 0$. The stress energy momentum vector $T\left(V\right)$ in space-time (STA) describes the energy flux across the hypersurface orthogonal to $V$. For a uniform electromagnetic medium consisting of induced electric and magnetic dipoles, the stress-energy momentum vector is a linear mapping~\eqref{T_V matter} of the vector $V$ given by $T\left(V\right) = -\,\frac{1}{4}\left(G\,V\,F + F\,V\,G\right)$. It turns out that the stress energy momentum vector has the same analytical structure in matter as in vacuum~\eqref{T_V}. Moreover, for the gradient $V=\nabla$, the stress energy momentum vector~\eqref{Poynting's theorem STA matter ter} is written as $\dot{T}\left(\dot{\nabla}\right) = \frac{1}{c}\,J \cdot F = f$ where $f$ is the electromagnetic force density vector in space-time. Finally, in the space-time algebra (STA), the Maxwell equation in matter~\eqref{Maxwell eq A matter} can be written as a wave equation for the potential vector $A$, namely $\nabla^2 A = \mu_0\,\tilde{J}$ since $F = \nabla \wedge A$.


\section*{Acknowledgements}

The author would like to thank Jean-Philippe Ansermet and Fran\c{c}ois A. Reuse for insightful discussions as well as his PhD thesis adviser Anthony Lasenby for introducing him to geometric algebra.


\pagebreak

\appendix

\renewcommand*{\thesection}{\Alph{section}}

\section{Spatial algebra (SA)}
\label{Spatial algebra}

\noindent We consider an orthonormal vector spatial frame $\{\boldsymbol{e}_1,\boldsymbol{e}_2,\boldsymbol{e}_3\}$. The geometric product of two basis vectors reads,~\cite{Lasenby:2003}
\begin{equation}\label{geometric product space 0}
\boldsymbol{e}_{i}\,\boldsymbol{e}_{j} = \boldsymbol{e}_{i}\cdot\boldsymbol{e}_{j} + \boldsymbol{e}_{i}\wedge\boldsymbol{e}_{j} \qquad\text{with}\qquad i,j = 1,2,3
\end{equation}
where,
\begin{equation}\label{square space}
\boldsymbol{e}_{1}^2 = \boldsymbol{e}_{2}^2 = \boldsymbol{e}_{3}^2 = 1 \qquad\text{and}\qquad \boldsymbol{e}_{i}\wedge\boldsymbol{e}_{i} = \boldsymbol{0}
\end{equation}
The inner product of two basis vectors is symmetric and defined as,
\begin{equation}\label{inner product space}
\boldsymbol{e}_{i}\cdot\boldsymbol{e}_{j} = \boldsymbol{e}_{j}\cdot\boldsymbol{e}_{i} = \frac{1}{2}\left(\boldsymbol{e}_{i}\,\boldsymbol{e}_{j} + \boldsymbol{e}_{j}\,\boldsymbol{e}_{i}\right)
\end{equation}
and the outer product is antisymmetric,
\begin{equation}\label{outer product space}
\boldsymbol{e}_{i}\wedge\boldsymbol{e}_{j} = -\,\boldsymbol{e}_{j}\wedge\boldsymbol{e}_{i} = \frac{1}{2}\left(\boldsymbol{e}_{i}\,\boldsymbol{e}_{j} -\,\boldsymbol{e}_{j}\,\boldsymbol{e}_{i}\right)
\end{equation}
The $8$ basis elements of the geometric algebra (GA) $\mathbb{G}^{3}$ called the spatial algebra are :
\begin{itemize}
\item one scalar : $\ 1$
\item three vectors : $\ \boldsymbol{e}_1$, $\ \boldsymbol{e}_2$, $\ \boldsymbol{e}_3$
\item three bivectors : $\ \boldsymbol{e}_1\wedge\boldsymbol{e}_2$, $\ \boldsymbol{e}_2\wedge\boldsymbol{e}_3$, $\ \boldsymbol{e}_3\wedge\boldsymbol{e}_1$
\item one pseudoscalar : $\ \boldsymbol{e}_1\wedge\boldsymbol{e}_2\wedge\boldsymbol{e}_3$
\end{itemize}
where the number of elements of each type $1$, $3$, $3$, $1$ are the binomial coefficients in three dimensions. In $\mathbb{G}^{3}$, there are $2^3$ basis elements. The geometric interpretation of the basis elements is clear. A scalar $s$ is oriented point, a vector is an oriented line element, a bivector is an oriented plane element and a pseudoscalar is an oriented volume element. The notation used is the following : scalars $s$ are written in lower case letters, vectors $\boldsymbol{v}$ are written in lower case bold letters, bivectors $\boldsymbol{B}$ are written in upper case bold letters and pseudoscalars are written as $s^{\prime}I$. A multivector $M$, which is linear combination of elements of $\mathbb{G}^{3}$, is written in upper case letters,
\begin{equation}\label{multivector}
M = s + \boldsymbol{v} + \boldsymbol{B} + s^{\prime}I
\end{equation}
Taking into account the definition~\eqref{geometric product space 0}, the bivectors and the pseudoscalar are also written as geometric products of the basis unit vectors $\boldsymbol{e}_1$, $\boldsymbol{e}_2$, $\boldsymbol{e}_3$ :
\begin{itemize}
\item bivectors : $\ \boldsymbol{e}_1\,\boldsymbol{e}_2$, $\ \boldsymbol{e}_2\,\boldsymbol{e}_3$, $\ \boldsymbol{e}_3\,\boldsymbol{e}_1$
\item pseudoscalar : $\ \boldsymbol{e}_1\,\boldsymbol{e}_2\,\boldsymbol{e}_3$
\end{itemize}
The pseudoscalar $I = \boldsymbol{e}_1\,\boldsymbol{e}_2\,\boldsymbol{e}_3$ behaves as an imaginary number since,
\begin{equation}\label{pseudoscalar squared}
I^2 = \left(\boldsymbol{e}_1\,\boldsymbol{e}_2\,\boldsymbol{e}_3\right)\left(\boldsymbol{e}_1\,\boldsymbol{e}_2\,\boldsymbol{e}_3\right) = \left(\boldsymbol{e}_1\,\boldsymbol{e}_2\,\boldsymbol{e}_1\,\boldsymbol{e}_2\right)\left(\boldsymbol{e}_3\,\boldsymbol{e}_3\right) = -\,\left(\boldsymbol{e}_1\,\boldsymbol{e}_1\right)\left(\boldsymbol{e}_2\,\boldsymbol{e}_2\right) = -\,1
\end{equation}

The geometric product of two basis vectors $\boldsymbol{u}$ and $\boldsymbol{v}$ reads,
\begin{equation}\label{geometric product space u v}
\boldsymbol{u}\,\boldsymbol{v} = \boldsymbol{u}\cdot\boldsymbol{v} + \boldsymbol{u}\wedge\boldsymbol{v}
\end{equation}
The inner product between the two basis vectors is symmetric,
\begin{equation}\label{inner product space u v}
\boldsymbol{u}\cdot\boldsymbol{v} = \boldsymbol{v}\cdot\boldsymbol{u} = \frac{1}{2}\left(\boldsymbol{u}\,\boldsymbol{v} + \boldsymbol{v}\,\boldsymbol{u}\right)
\end{equation}
and the outer product is antisymmetric,
\begin{equation}\label{outer product space u v}
\boldsymbol{u}\wedge\boldsymbol{v} = -\,\boldsymbol{v}\wedge\boldsymbol{u} = \frac{1}{2}\left(\boldsymbol{u}\,\boldsymbol{v} -\,\boldsymbol{v}\,\boldsymbol{u}\right)
\end{equation}
The geometric product of a vector $\boldsymbol{v}$ and a bivector $\boldsymbol{B}$ is the sum of the inner product of the outer product,
\begin{equation}\label{geometric product space v B}
\boldsymbol{v}\,\boldsymbol{B} = \boldsymbol{v}\cdot\boldsymbol{B} + \boldsymbol{v}\wedge\boldsymbol{B}
\end{equation}
and the geometric product of a bivector $\boldsymbol{B}$ and a vector $\boldsymbol{v}$ is the sum of the inner product of the outer product,
\begin{equation}\label{geometric product space B v}
\boldsymbol{B}\,\boldsymbol{v} = \boldsymbol{B}\cdot\boldsymbol{v} + \boldsymbol{B}\wedge\boldsymbol{v}
\end{equation}
To determine the symmetry of the these products, the spatial frame is oriented such that the bivector reads $\boldsymbol{B} = B_{12}\,\boldsymbol{e}_1\wedge\boldsymbol{e}_2 = B_{12}\,\boldsymbol{e}_1\,\boldsymbol{e}_2$ and the orientation of the vector 
$\boldsymbol{v} = v_1\,\boldsymbol{e}_1 + v_2\,\boldsymbol{e}_2 + v_3\,\boldsymbol{e}_3$ is arbitrary. The geometric product of the vector $\boldsymbol{v}$ and the bivector $\boldsymbol{B}$ reads,
\begin{equation}\label{geometric product v B}
\begin{split}
&\boldsymbol{v}\,\boldsymbol{B} = \left(v_1\,\boldsymbol{e}_1 + v_2\,\boldsymbol{e}_2 + v_3\,\boldsymbol{e}_3\right)\,\left(B_{12}\,\boldsymbol{e}_1\,\boldsymbol{e}_2\right)\\
&\phantom{\boldsymbol{v}\,\boldsymbol{B}} = v_1\,B_{12}\,\boldsymbol{e}_1\,\boldsymbol{e}_1\,\boldsymbol{e}_2 + v_2\,B_{12}\,\boldsymbol{e}_2\,\boldsymbol{e}_1\,\boldsymbol{e}_2 + v_3\,B_{12}\,\boldsymbol{e}_3\,\boldsymbol{e}_1\,\boldsymbol{e}_2\\
&\phantom{\boldsymbol{v}\,\boldsymbol{B}} = B_{12}\,v_{1}\,\boldsymbol{e}_2 -\,B_{12}\,v_{2}\,\boldsymbol{e}_1 + B_{12}\,v_{3}\,\boldsymbol{e}_1\,\boldsymbol{e}_2\,\boldsymbol{e}_3
\end{split}
\end{equation}
The geometric product of the bivector $\boldsymbol{B}$ and the vector $\boldsymbol{v}$ reads,
\begin{equation}\label{geometric product B v}
\begin{split}
&\boldsymbol{B}\,\boldsymbol{v} = \left(B_{12}\,\boldsymbol{e}_1\,\boldsymbol{e}_2\right)\left(v_1\,\boldsymbol{e}_1 + v_2\,\boldsymbol{e}_2 + v_3\,\boldsymbol{e}_3\right)\\
&\phantom{\boldsymbol{B}\,\boldsymbol{v}} = B_{12}\,v_{1}\,\boldsymbol{e}_1\,\boldsymbol{e}_2\,\boldsymbol{e}_1 + B_{12}\,v_{2}\,\boldsymbol{e}_1\,\boldsymbol{e}_2\,\boldsymbol{e}_2 + B_{12}\,v_{3}\,\boldsymbol{e}_1\,\boldsymbol{e}_2\,\boldsymbol{e}_3\\
&\phantom{\boldsymbol{B}\,\boldsymbol{v}} = -\,B_{12}\,v_{1}\,\boldsymbol{e}_2 + B_{12}\,v_{2}\,\boldsymbol{e}_1 + B_{12}\,v_{3}\,\boldsymbol{e}_1\,\boldsymbol{e}_2\,\boldsymbol{e}_3
\end{split}
\end{equation}
The outer product of the vector $\boldsymbol{v}$ and the bivector $\boldsymbol{B}$ reads,
\begin{align}\label{outer product v B}
&\boldsymbol{v}\wedge\boldsymbol{B} = \left(v_1\,\boldsymbol{e}_1 + v_2\,\boldsymbol{e}_2 + v_3\,\boldsymbol{e}_3\right)\wedge\left(B_{12}\,\boldsymbol{e}_1\wedge\boldsymbol{e}_2\right)\nonumber\\
&\phantom{\boldsymbol{v}\wedge\boldsymbol{B}} = v_1\,B_{12}\,\boldsymbol{e}_1\wedge\boldsymbol{e}_1\wedge\boldsymbol{e}_2 + v_2\,B_{12}\,\boldsymbol{e}_2\wedge\boldsymbol{e}_1\wedge\boldsymbol{e}_2 + v_3\,B_{12}\,\boldsymbol{e}_3\wedge\boldsymbol{e}_1\wedge\boldsymbol{e}_2\nonumber\\
&\phantom{\boldsymbol{v}\wedge\boldsymbol{B}} = v_3\,B_{12}\,\boldsymbol{e}_1\wedge\boldsymbol{e}_2\wedge\boldsymbol{e}_3
\end{align}
The outer product of the bivector $\boldsymbol{B}$ and the vector $\boldsymbol{v}$ reads,
\begin{align}\label{outer product B v}
&\boldsymbol{B}\wedge\boldsymbol{v} = \left(B_{12}\,\boldsymbol{e}_1\wedge\boldsymbol{e}_2\right)\wedge\left(v_1\,\boldsymbol{e}_1 + v_2\,\boldsymbol{e}_2 + v_3\,\boldsymbol{e}_3\right)\nonumber\\
&\phantom{\boldsymbol{B}\wedge\boldsymbol{v}} = B_{12}\,v_{1}\,\boldsymbol{e}_1\wedge\boldsymbol{e}_2\wedge\boldsymbol{e}_1 + B_{12}\,v_{2}\,\boldsymbol{e}_1\wedge\boldsymbol{e}_2\wedge\boldsymbol{e}_2 + B_{12}\,v_{3}\,\boldsymbol{e}_1\wedge\boldsymbol{e}_2\wedge\boldsymbol{e}_3\nonumber\\
&\phantom{\boldsymbol{B}\wedge\boldsymbol{v}} = B_{12}\,v_{3}\,\boldsymbol{e}_1\wedge\boldsymbol{e}_2\wedge\boldsymbol{e}_3
\end{align}
From relations~\eqref{outer product v B} and~\eqref{outer product B v} we conclude that the outer product of a vector $\boldsymbol{v}$ and a bivector $\boldsymbol{B}$ is symmetric,
\begin{equation}\label{outer product v B symmetric}
\boldsymbol{v}\wedge\boldsymbol{B} = \boldsymbol{B}\wedge\boldsymbol{v}
\end{equation}
In view of relations\eqref{geometric product space v B},~\eqref{geometric product v B} and~\eqref{outer product v B},
\begin{equation}\label{inner product v B}
\boldsymbol{v}\cdot\boldsymbol{B} = \boldsymbol{v}\,\boldsymbol{B} -\,\boldsymbol{v}\wedge\boldsymbol{B} = -\,B_{12}\,v_{2}\,\boldsymbol{e}_1 + B_{12}\,v_{1}\,\boldsymbol{e}_2 
\end{equation}
In view of relations\eqref{geometric product space B v},~\eqref{geometric product B v} and~\eqref{outer product B v},
\begin{equation}\label{inner product B v}
\boldsymbol{B}\cdot\boldsymbol{v} = \boldsymbol{B}\,\boldsymbol{v} -\,\boldsymbol{B}\wedge\boldsymbol{v} = B_{12}\,v_{2}\,\boldsymbol{e}_1 -\,B_{12}\,v_{1}\,\boldsymbol{e}_2 
\end{equation}
From relations~\eqref{inner product v B} and~\eqref{inner product B v} we conclude that the inner product of a vector $\boldsymbol{v}$ and a bivector $\boldsymbol{B}$ is antisymmetric,
\begin{equation}\label{inner product v B antisymmetric}
\boldsymbol{v}\cdot\boldsymbol{B} = -\,\boldsymbol{B}\cdot\boldsymbol{v}
\end{equation}
In view of relations~\eqref{geometric product v B} and~\eqref{geometric product B v},~\eqref{outer product v B symmetric} and~\eqref{inner product v B antisymmetric},
\begin{align}
\label{inner product antisymmetry v B}
&\boldsymbol{v}\cdot\boldsymbol{B} = \frac{1}{2}\left(\boldsymbol{v}\,\boldsymbol{B} -\,\boldsymbol{B}\,\boldsymbol{v}\right)\\
\label{outer product symmetry v B}
&\boldsymbol{v}\wedge\boldsymbol{B} = \frac{1}{2}\left(\boldsymbol{v}\,\boldsymbol{B} + \boldsymbol{B}\,\boldsymbol{v}\right)
\end{align}
To determine the geometric product of two bivectors $\boldsymbol{A}$ and $\boldsymbol{B}$, we write them as $\boldsymbol{A} = A_{12}\,\boldsymbol{e}_1\,\boldsymbol{e}_2 + A_{23}\,\boldsymbol{e}_2\,\boldsymbol{e}_3 + A_{31}\,\boldsymbol{e}_3\,\boldsymbol{e}_1$ and $\boldsymbol{B} = B_{12}\,\boldsymbol{e}_1\,\boldsymbol{e}_2 + B_{23}\,\boldsymbol{e}_2\,\boldsymbol{e}_3 + B_{31}\,\boldsymbol{e}_3\,\boldsymbol{e}_1$. The geometric product of the bivectors $\boldsymbol{A}$ and $\boldsymbol{B}$ reads,
\begin{align}\label{geometric product A V 0}
&\boldsymbol{A}\,\boldsymbol{B} = \left(A_{12}\,\boldsymbol{e}_1\,\boldsymbol{e}_2 + A_{23}\,\boldsymbol{e}_2\,\boldsymbol{e}_3 + A_{31}\,\boldsymbol{e}_3\,\boldsymbol{e}_1\right)\left(B_{12}\,\boldsymbol{e}_1\,\boldsymbol{e}_2 + B_{23}\,\boldsymbol{e}_2\,\boldsymbol{e}_3 + B_{31}\,\boldsymbol{e}_3\,\boldsymbol{e}_1\right)\nonumber\\
&\phantom{\boldsymbol{A}\,\boldsymbol{B}} = A_{12}\,B_{12}\,\boldsymbol{e}_1\,\boldsymbol{e}_2\,\boldsymbol{e}_1\,\boldsymbol{e}_2 + A_{12}\,B_{23}\,\boldsymbol{e}_1\,\boldsymbol{e}_2\,\boldsymbol{e}_2\,\boldsymbol{e}_3 + A_{12}\,B_{31}\,\boldsymbol{e}_1\,\boldsymbol{e}_2\,\boldsymbol{e}_3\,\boldsymbol{e}_1\\
&\phantom{\boldsymbol{A}\,\boldsymbol{B} =} + A_{23}\,B_{12}\,\boldsymbol{e}_2\,\boldsymbol{e}_3\,\boldsymbol{e}_1\,\boldsymbol{e}_2 + A_{23}\,B_{23}\,\boldsymbol{e}_2\,\boldsymbol{e}_3\,\boldsymbol{e}_2\,\boldsymbol{e}_3 + A_{23}\,B_{31}\,\boldsymbol{e}_2\,\boldsymbol{e}_3\,\boldsymbol{e}_3\,\boldsymbol{e}_1\nonumber\\
&\phantom{\boldsymbol{A}\,\boldsymbol{B} =} + A_{31}\,B_{12}\,\boldsymbol{e}_3\,\boldsymbol{e}_1\,\boldsymbol{e}_1\,\boldsymbol{e}_2 + A_{31}\,B_{23}\,\boldsymbol{e}_3\,\boldsymbol{e}_1\,\boldsymbol{e}_2\,\boldsymbol{e}_3 + A_{31}\,B_{31}\,\boldsymbol{e}_3\,\boldsymbol{e}_1\,\boldsymbol{e}_3\,\boldsymbol{e}_1\nonumber
\end{align}
which reduces to,
\begin{equation}\label{geometric product A V}
\begin{split}
&\boldsymbol{A}\,\boldsymbol{B} = -\,\left(A_{12}\,B_{12} + A_{23}\,B_{23} + A_{31}\,B_{31}\right) + \left(A_{31}\,B_{23} -\,A_{23}\,B_{31}\right)\boldsymbol{e}_1\,\boldsymbol{e}_2\\
&\phantom{\boldsymbol{A}\,\boldsymbol{B} =} + \left(A_{12}\,B_{31} -\,A_{31}\,B_{12}\right)\boldsymbol{e}_2\,\boldsymbol{e}_3 + \left(A_{23}\,B_{12} -\,A_{12}\,B_{23}\right)\boldsymbol{e}_3\,\boldsymbol{e}_1 
\end{split}
\end{equation}
The geometric product of the bivectors $\boldsymbol{B}$ and $\boldsymbol{A}$ reads,
\begin{align}\label{geometric product B V 0}
&\boldsymbol{B}\,\boldsymbol{A} = \left(B_{12}\,\boldsymbol{e}_1\,\boldsymbol{e}_2 + B_{23}\,\boldsymbol{e}_2\,\boldsymbol{e}_3 + B_{31}\,\boldsymbol{e}_3\,\boldsymbol{e}_1\right)\left(A_{12}\,\boldsymbol{e}_1\,\boldsymbol{e}_2 + A_{23}\,\boldsymbol{e}_2\,\boldsymbol{e}_3 + A_{31}\,\boldsymbol{e}_3\,\boldsymbol{e}_1\right)\nonumber\\
&\phantom{\boldsymbol{B}\,\boldsymbol{A}} = A_{12}\,B_{12}\,\boldsymbol{e}_1\,\boldsymbol{e}_2\,\boldsymbol{e}_1\,\boldsymbol{e}_2 + A_{23}\,B_{12}\,\boldsymbol{e}_1\,\boldsymbol{e}_2\,\boldsymbol{e}_2\,\boldsymbol{e}_3 + A_{12}\,B_{31}\,\boldsymbol{e}_1\,\boldsymbol{e}_2\,\boldsymbol{e}_3\,\boldsymbol{e}_1\\
&\phantom{\boldsymbol{B}\,\boldsymbol{A} =} + A_{12}\,B_{23}\,\boldsymbol{e}_2\,\boldsymbol{e}_3\,\boldsymbol{e}_1\,\boldsymbol{e}_2 + A_{23}\,B_{23}\,\boldsymbol{e}_2\,\boldsymbol{e}_3\,\boldsymbol{e}_2\,\boldsymbol{e}_3 + A_{31}\,B_{23}\,\boldsymbol{e}_2\,\boldsymbol{e}_3\,\boldsymbol{e}_3\,\boldsymbol{e}_1\nonumber\\
&\phantom{\boldsymbol{B}\,\boldsymbol{A} =} + A_{12}\,B_{31}\,\boldsymbol{e}_3\,\boldsymbol{e}_1\,\boldsymbol{e}_1\,\boldsymbol{e}_2 + A_{23}\,B_{31}\,\boldsymbol{e}_3\,\boldsymbol{e}_1\,\boldsymbol{e}_2\,\boldsymbol{e}_3 + A_{31}\,B_{31}\,\boldsymbol{e}_3\,\boldsymbol{e}_1\,\boldsymbol{e}_3\,\boldsymbol{e}_1\nonumber
\end{align}
which reduces to,
\begin{equation}\label{geometric product B V}
\begin{split}
&\boldsymbol{B}\,\boldsymbol{A} = -\,\left(A_{12}\,B_{12} + A_{23}\,B_{23} + A_{31}\,B_{31}\right) + \left(A_{23}\,B_{31} -\,A_{31}\,B_{23}\right)\boldsymbol{e}_1\,\boldsymbol{e}_2\\
&\phantom{\boldsymbol{B}\,\boldsymbol{A} =} + \left(A_{31}\,B_{12} -\,A_{12}\,B_{31}\right)\boldsymbol{e}_2\,\boldsymbol{e}_3 + \left(A_{12}\,B_{23} -\,A_{23}\,B_{12}\right)\boldsymbol{e}_3\,\boldsymbol{e}_1 
\end{split}
\end{equation}
The geometric product~\eqref{geometric product A V} of two bivectors $\boldsymbol{A}$ and $\boldsymbol{B}$ is the sum an inner product and a cross product,
\begin{equation}\label{geometric product bivectors}
\boldsymbol{A}\,\boldsymbol{B} = \boldsymbol{A}\cdot\boldsymbol{B} + \boldsymbol{A}\times\boldsymbol{B}
\end{equation}
According to relations~\eqref{geometric product A V} and~\eqref{geometric product B V}, the symmetric inner product of two bivectors $\boldsymbol{A}$ and $\boldsymbol{B}$ yields a scalar,
\begin{equation}\label{geometric product bivectors inner}
\boldsymbol{A}\cdot\boldsymbol{B} = \boldsymbol{B}\cdot\boldsymbol{A} = \frac{1}{2}\left(\boldsymbol{A}\,\boldsymbol{B} + \boldsymbol{B}\,\boldsymbol{A}\right) =  -\,\left(A_{12}\,B_{12} + A_{23}\,B_{23} + A_{31}\,B_{31}\right)
\end{equation}
According to relations~\eqref{geometric product A V} and~\eqref{geometric product B V}, the antisymmetric cross product of two bivectors $\boldsymbol{A}$ and $\boldsymbol{B}$ yields a bivector,
\begin{align}\label{geometric product bivectors outer}
&\boldsymbol{A}\times\boldsymbol{B} = -\,\boldsymbol{B}\times\boldsymbol{A} = \frac{1}{2}\left(\boldsymbol{A}\,\boldsymbol{B} -\,\boldsymbol{B}\,\boldsymbol{A}\right)\nonumber\\
&= \left(A_{31}\,B_{23} -\,A_{23}\,B_{31}\right)\boldsymbol{e}_1\,\boldsymbol{e}_2 + \left(A_{12}\,B_{31} -\,A_{31}\,B_{12}\right)\boldsymbol{e}_2\,\boldsymbol{e}_3\\
&\phantom{=} + \left(A_{23}\,B_{12} -\,A_{12}\,B_{23}\right)\boldsymbol{e}_3\,\boldsymbol{e}_1\nonumber
\end{align}
The fact that an inner product of two bivectors yields a scalar and the cross product of two bivectors yields a bivector is specific to the spatial algebra $\mathbb{G}^{3}$. To determine the geometric product of a vector $\boldsymbol{v}$ and the pseudoscalar $I$, we write them as $\boldsymbol{B} = B_{12}\,\boldsymbol{e}_1\,\boldsymbol{e}_2 + B_{23}\,\boldsymbol{e}_2\,\boldsymbol{e}_3 + B_{31}\,\boldsymbol{e}_3\,\boldsymbol{e}_1$ and $I = \boldsymbol{e}_1\,\boldsymbol{e}_2\,\boldsymbol{e}_3$. The geometric product of the vector $\boldsymbol{v}$ and the pseudoscalar $I$ reads,
\begin{align}
\label{v I}
&\boldsymbol{v}\,I = \left(v_1\,\boldsymbol{e}_1 + v_2\,\boldsymbol{e}_2 + v_3\,\boldsymbol{e}_3\right)\left(\boldsymbol{e}_1\,\boldsymbol{e}_2\,\boldsymbol{e}_3\right) = v_1\,\boldsymbol{e}_2\,\boldsymbol{e}_3 + v_2\,\boldsymbol{e}_3\,\boldsymbol{e}_1 + v_3\,\boldsymbol{e}_1\,\boldsymbol{e}_2\\
\label{I v}
&I\,\boldsymbol{v} = \left(\boldsymbol{e}_1\,\boldsymbol{e}_2\,\boldsymbol{e}_3\right)\left(v_1\,\boldsymbol{e}_1 + v_2\,\boldsymbol{e}_2 + v_3\,\boldsymbol{e}_3\right) = v_1\,\boldsymbol{e}_2\,\boldsymbol{e}_3 + v_2\,\boldsymbol{e}_3\,\boldsymbol{e}_1 + v_3\,\boldsymbol{e}_1\,\boldsymbol{e}_2
\end{align}
According to relation~\eqref{v I} and~\eqref{I v}, the vector $\boldsymbol{v}$ commutes with the pseudoscalar $I$,
\begin{equation}\label{commutation v I}
\boldsymbol{v}\,I = I\,\boldsymbol{v}
\end{equation}
To determine the geometric product of a vector $\boldsymbol{v}$ and the pseudoscalar $I$, we write them as $\boldsymbol{v} = v_1\,\boldsymbol{e}_1 + v_2\,\boldsymbol{e}_2 + v_3\,\boldsymbol{e}_3$ and $I = \boldsymbol{e}_1\,\boldsymbol{e}_2\,\boldsymbol{e}_3$.The geometric product of the bivector $\boldsymbol{B}$ and the pseudoscalar $I$ reads,
\begin{align}
\label{B I}
&\boldsymbol{B}\,I = \left(B_{12}\,\boldsymbol{e}_1\,\boldsymbol{e}_2 + B_{23}\,\boldsymbol{e}_2\,\boldsymbol{e}_3 + B_{31}\,\boldsymbol{e}_3\,\boldsymbol{e}_1\right)\left(\boldsymbol{e}_1\,\boldsymbol{e}_2\,\boldsymbol{e}_3\right)\nonumber\\
&\phantom{\boldsymbol{B}\,I} = -\left(B_{23}\,\boldsymbol{e}_1 + B_{31}\,\boldsymbol{e}_2 + B_{12}\,\boldsymbol{e}_3\right)\\
\label{I B}
&I\,\boldsymbol{B} = \left(\boldsymbol{e}_1\,\boldsymbol{e}_2\,\boldsymbol{e}_3\right)\left(B_{12}\,\boldsymbol{e}_1\,\boldsymbol{e}_2 + B_{23}\,\boldsymbol{e}_2\,\boldsymbol{e}_3 + B_{31}\,\boldsymbol{e}_3\,\boldsymbol{e}_1\right) = -\,B_{12}\,\boldsymbol{e}_3\nonumber\\
&\phantom{I\,\boldsymbol{B}} = -\left(B_{23}\,\boldsymbol{e}_1 + B_{31}\,\boldsymbol{e}_2 + B_{12}\,\boldsymbol{e}_3\right)
\end{align}
According to relation~\eqref{B I} and~\eqref{I B}, the bivector $\boldsymbol{B}$ commutes with the pseudoscalar $I$,
\begin{equation}\label{commutation B I}
\boldsymbol{B}\,I = I\,\boldsymbol{B}
\end{equation}

\section{Duality in spatial algebra (SA)}
\label{Duality in spatial algebra}

\noindent The reverse of the scalar $s$, the vector $\boldsymbol{v} = v_3\,\boldsymbol{e}_3$, of the bivector $\boldsymbol{B} = B_{12}\,\boldsymbol{e}_1\,\boldsymbol{e}_2$ and of the pseudoscalar $I$ is obtained by reversing the order of the basis vectors,~\cite{Macdonald:2011}
\begin{align}
\label{reverse s}
&\boldsymbol{s}^{\dag} = s\\
\label{reverse v}
&\boldsymbol{v}^{\dag} = v_3\,\boldsymbol{e}_3 = \boldsymbol{v}\\
\label{reverse B}
&\boldsymbol{B}^{\dag} = B_{12}\,\boldsymbol{e}_2\,\boldsymbol{e}_1 = -\,B_{12}\,\boldsymbol{e}_1\,\boldsymbol{e}_2 = -\,\boldsymbol{B}\\
\label{reverse I}
&I^{\dag} = \boldsymbol{e}_3\,\boldsymbol{e}_2\,\boldsymbol{e}_1 = -\,\boldsymbol{e}_1\,\boldsymbol{e}_2\,\boldsymbol{e}_3 = -\,I
\end{align}
Thus, the reverse of the multivector~\eqref{multivector} is given by,
\begin{equation}\label{multivector reverse}
M^{\dag} = s^{\dag} + \boldsymbol{v}^{\dag} + \boldsymbol{B}^{\dag} + s^{\prime}I^{\dag} = s + \boldsymbol{v} -\,\boldsymbol{B} -\,s^{\prime}I
\end{equation}
The square of the modulus of a vector $\boldsymbol{v}$, a bivector $\boldsymbol{B}$ and a pseudovector $I$ are defined as,
\begin{align}
\label{modulus v}
&\left|\boldsymbol{v}\right|^2 = \boldsymbol{v}^{\dag}\cdot\boldsymbol{v} = \boldsymbol{v}\cdot\boldsymbol{v} = \boldsymbol{v}^2\\
\label{modulus B}
&\left|\boldsymbol{B}\right|^2 = \boldsymbol{B}^{\dag}\cdot\boldsymbol{B} = -\,\boldsymbol{B}\cdot\boldsymbol{B} = -\,\boldsymbol{B}^2\\
\label{modulus I}
&\left|I\right|^2 = I^{\dag}\cdot I = -\,I\cdot I = -\,I^2 = 1
\end{align}
The geometric interpretation of the modulus is clear : the modulus of a vector $\left|\boldsymbol{v}\right|$ is the length of a line element, the modulus of a bivector $\left|\boldsymbol{B}\right|$ is the surface of a plane element and the modulus of a pseudoscalar $\left|s^{\prime}I\right|$ is the volume of a space element. Thus, the modulus of the bivector obtained by taking the outer product of a vector $\boldsymbol{u}$ and a vector $\boldsymbol{v}$ is the surface of the parallelogram spanned by these vectors,
\begin{equation}\label{area bivector}
\left|\boldsymbol{u}\wedge\boldsymbol{v}\right| = \left|\boldsymbol{u}\right|\,\left|\boldsymbol{v}\right|\,\sin\theta
\end{equation}
where $\theta$ is the acute angle between these vectors. This modulus is the same as the modulus of the cross product of these vectors,
\begin{equation}\label{cross product length}
\left|\boldsymbol{u}\times\boldsymbol{v}\right| = \left|\boldsymbol{u}\right|\,\left|\boldsymbol{v}\right|\,\sin\theta
\end{equation}
Thus,
\begin{equation}\label{modulus duality}
\left|\boldsymbol{u}\wedge\boldsymbol{v}\right| = \left|\boldsymbol{u}\times\boldsymbol{v}\right|
\end{equation}
The inverse of a vector $\boldsymbol{v}$, a bivector $\boldsymbol{B}$ and a pseudovector $I$ are defined as,
\begin{align}
\label{inverse v}
&\boldsymbol{v}^{-1} = \frac{\boldsymbol{v}}{\boldsymbol{v}^2} = \frac{\boldsymbol{v}^{\dag}}{\left|\boldsymbol{v}\right|^2}\\
\label{inverse B}
&\boldsymbol{B}^{-1} = \frac{\boldsymbol{B}}{\boldsymbol{B}^2} = \frac{\boldsymbol{B}^{\dag}}{\left|\boldsymbol{B}\right|^2}\\
\label{inverse I}
&I^{-1} = \frac{I}{I^2} = \frac{I^{\dag}}{\left|I\right|^2} = -\,I
\end{align}
The dual of a vector $\boldsymbol{v}$, a bivector $\boldsymbol{B}$ and a pseudovector $I$ are defined as,~\cite{Macdonald:2011}
\begin{align}
\label{dual v}
&\boldsymbol{v}^{\ast} = \frac{\boldsymbol{v}}{I} = \boldsymbol{v}\,I^{-1} = -\,\boldsymbol{v}\,I\\
\label{dual B}
&\boldsymbol{B}^{\ast} = \frac{\boldsymbol{B}}{I} = \boldsymbol{B}\,I^{-1} = -\,\boldsymbol{B}\,I\\
\label{dual I}
&I^{\ast} = \frac{I}{I} = I\,I^{-1} = 1
\end{align}
This duality is the transformation as the Hodge duality in differential forms. The dual of the dual of a vector $\boldsymbol{v}$, a bivector $\boldsymbol{B}$ and a pseudovector $I$ are their opposite,
\begin{align}
\label{dual dual v}
&\left(\boldsymbol{v}^{\ast}\right)^{\ast} = -\,\boldsymbol{v}^{\ast}\,I = \boldsymbol{v}\,I^2 = -\,\boldsymbol{v}\\
\label{dual dual B}
&\left(\boldsymbol{B}^{\ast}\right)^{\ast} = -\,\boldsymbol{B}^{\ast}\,I = \boldsymbol{B}\,I^2 = -\,\boldsymbol{B}\\
\label{dual dual I}
&\left(I^{\ast}\right)^{\ast} = -\,I^{\ast}\,I = I\,I^2 = -\,I
\end{align}
The dual of the vector $\boldsymbol{v} = v_3\,\boldsymbol{e}_3$ is,
\begin{equation}\label{dual vector v 0}
\boldsymbol{v}^{\ast} = -\,\boldsymbol{v}\,I = -\,\left(v_3\,\boldsymbol{e}_3\right)\left(\boldsymbol{e}_1\,\boldsymbol{e}_2\,\boldsymbol{e}_3\right) = -\,v_3\,\boldsymbol{e}_1\,\boldsymbol{e}_2
\end{equation}
Defining the bivector $\boldsymbol{V}$ as,
\begin{equation}\label{dual vector v 1}
\boldsymbol{V} = v_{12}\,\boldsymbol{e}_1\,\boldsymbol{e}_2 \qquad\text{where}\qquad \left|\boldsymbol{V}\right| = \left|\boldsymbol{v}\right| \qquad\text{and thus}\qquad v_{12} = v_3
\end{equation}
it satisfies the duality,
\begin{equation}\label{dual vector v}
\boldsymbol{v}^{\ast} = -\,\boldsymbol{V} = \left(\boldsymbol{V}^{\ast}\right)^{\ast} \qquad\text{and thus}\qquad \boldsymbol{V}^{\ast} = \boldsymbol{v}
\end{equation}
The dual of the bivector $\boldsymbol{B} = B_{12}\,\boldsymbol{e}_1\,\boldsymbol{e}_2$ is,
\begin{equation}\label{dual bivector B 0}
\boldsymbol{B}^{\ast} = -\,\boldsymbol{B}\,I = -\,\left(B_{12}\,\boldsymbol{e}_1\,\boldsymbol{e}_2\right)\left(\boldsymbol{e}_1\,\boldsymbol{e}_2\,\boldsymbol{e}_3\right) = B_{12}\,\boldsymbol{e}_3
\end{equation}
Defining the vector $\boldsymbol{b}$ as,
\begin{equation}\label{dual bivector B 1}
\boldsymbol{b} = b_{3}\,\boldsymbol{e}_3 \qquad\text{where}\qquad \left|\boldsymbol{b}\right| = \left|\boldsymbol{B}\right| \qquad\text{and thus}\qquad b_3 = B_{12}
\end{equation}
it satisfies the duality,
\begin{equation}\label{dual bivector B}
\boldsymbol{B}^{\ast} = \boldsymbol{b} = -\,\left(\boldsymbol{b}^{\ast}\right)^{\ast} \qquad\text{and thus}\qquad \boldsymbol{b}^{\ast} = -\,\boldsymbol{B}
\end{equation}
There is a duality between a vector and a bivector of same modulus and there is a duality between a scalar and a pseudoscalar of same modulus. In view of relations~\eqref{dual v} and~\eqref{dual bivector B},
\begin{equation}\label{dual bivector B bis}
\boldsymbol{B} = -\,\boldsymbol{b}^{\ast} = \boldsymbol{b}\,I
\end{equation}
which implies that by duality the multivector~\eqref{multivector} is recast as,~\cite{Hestenes:2015}
\begin{equation}\label{multivector bis}
M = \left(s + s^{\prime}I\right) + \left(\boldsymbol{v} + \boldsymbol{b}\,I\right)
\end{equation}
and the reverse of the multivector~\eqref{multivector reverse} is recast as,
\begin{equation}\label{multivector reverse bis}
M^{\dag} = \left(s -\,s^{\prime}I\right) + \left(\boldsymbol{v} -\,\boldsymbol{b}\,I\right)
\end{equation}
which means that the reverse of a multivector is like a complex conjugate where the pseudoscalar $I$ is like the imaginary number $i$. The bivector $\boldsymbol{W}$ and the vector $\boldsymbol{w}$ are defined as the wedge product and the cross product of two vectors $\boldsymbol{u}$ and $\boldsymbol{v}$ respectively,
\begin{equation}\label{wedge and cross products}
\boldsymbol{W} = \boldsymbol{u}\wedge\boldsymbol{v} \qquad\text{and}\qquad \boldsymbol{w} = \boldsymbol{u}\times\boldsymbol{v}
\end{equation}
According to relation~\eqref{modulus duality} the modulus of the bivector $\boldsymbol{W}$ and vector $\boldsymbol{w}$ are equal, which means that the vector $\boldsymbol{w}$ is the dual of the bivector $\boldsymbol{W}$,
\begin{equation}\label{wedge and cross products duality 0}
\left|\boldsymbol{W}\right| = \left|\boldsymbol{w}\right| \qquad\text{and thus}\qquad \boldsymbol{W}^{\ast} = \boldsymbol{w} \qquad\text{and}\qquad \boldsymbol{w}^{\ast} = -\,\boldsymbol{W}
\end{equation}
Thus, the cross product of the vectors $\boldsymbol{u}$ and $\boldsymbol{v}$ is the dual of the wedge product of these vectors,
\begin{equation}\label{wedge and cross products duality}
\left(\boldsymbol{u}\wedge\boldsymbol{v}\right)^{\ast} = \boldsymbol{u}\times\boldsymbol{v} \qquad\text{and}\qquad \left(\boldsymbol{u}\times\boldsymbol{v}\right)^{\ast} = -\,\boldsymbol{u}\wedge\boldsymbol{v} 
\end{equation}
To establish the duality between the inner and outer product of two vectors, we choose a vector $\boldsymbol{u} = u_1\,\boldsymbol{e}_1 + u_2\,\boldsymbol{e}_2$ and a vector $\boldsymbol{v} = v_1\,\boldsymbol{e}_1 + v_2\,\boldsymbol{e}_2$ in the same plane. The dual of the outer product of the vectors $\boldsymbol{u}$ and $\boldsymbol{v}$ yields,
\begin{equation}\label{duality vectors outer}
\begin{split}
&\left(\boldsymbol{u}\wedge\boldsymbol{v}\right)^{\ast} = -\,\Big(\left(u_1\,\boldsymbol{e}_1 + u_2\,\boldsymbol{e}_2\right)\wedge\left(v_1\,\boldsymbol{e}_1 + v_2\,\boldsymbol{e}_2\right)\Big)\left(\boldsymbol{e}_1\,\boldsymbol{e}_2\,\boldsymbol{e}_3\right)\\
&\phantom{\left(\boldsymbol{u}\wedge\boldsymbol{v}\right)^{\ast}} = -\,\left(u_1\,v_2\,\boldsymbol{e}_1\,\boldsymbol{e}_2 + u_2\,v_1\,\boldsymbol{e}_2\,\boldsymbol{e}_1\right)\left(\boldsymbol{e}_1\,\boldsymbol{e}_2\,\boldsymbol{e}_3\right)\\
&\phantom{\left(\boldsymbol{u}\wedge\boldsymbol{v}\right)^{\ast}} = \left(u_1\,v_2 -\,u_2\,v_1\right)\boldsymbol{e}_3
\end{split}
\end{equation}
The inner product of the vector $\boldsymbol{u}$ and $\boldsymbol{v}^{\ast}$ yields,
\begin{equation}\label{duality vectors outer bis}
\begin{split}
&\boldsymbol{u}\cdot\boldsymbol{v}^{\ast} = -\,\left(u_1\,\boldsymbol{e}_1 + u_2\,\boldsymbol{e}_2\right)\cdot\Big(\left(v_1\,\boldsymbol{e}_1 + v_2\,\boldsymbol{e}_2\right)\left(\boldsymbol{e}_1\,\boldsymbol{e}_2\,\boldsymbol{e}_3\right)\Big)\\
&\phantom{\boldsymbol{u}\cdot\boldsymbol{v}^{\ast}} = -\,\left(u_1\,\boldsymbol{e}_1 + u_2\,\boldsymbol{e}_2\right)\cdot\left(v_1\,\boldsymbol{e}_2\,\boldsymbol{e}_3 -\,v_2\,\boldsymbol{e}_1\,\boldsymbol{e}_3\right)\\
&\phantom{\boldsymbol{u}\cdot\boldsymbol{v}^{\ast}} = \left(u_1\,v_2 -\,u_2\,v_1\right)\boldsymbol{e}_3
\end{split}
\end{equation}
The identification of relations~\eqref{duality vectors outer} and~\eqref{duality vectors outer bis} yields the vectorial duality,
\begin{equation}\label{vector duality}
\left(\boldsymbol{u}\wedge\boldsymbol{v}\right)^{\ast} = \boldsymbol{u}\cdot\boldsymbol{v}^{\ast}
\end{equation}
which is recast in terms of the dual bivector $\boldsymbol{V} = -\,\boldsymbol{v}^{\ast}$ as,
\begin{equation}\label{vector duality bis 0}
\left(\boldsymbol{u}\wedge\boldsymbol{v}\right)^{\ast} = -\,\boldsymbol{u}\cdot\boldsymbol{V} = \boldsymbol{V}\cdot\boldsymbol{u}
\end{equation}
In view of relations~\eqref{wedge and cross products duality} and~\eqref{vector duality}, we obtain,
\begin{equation}\label{wedge and cross products duality bis}
\boldsymbol{u}\times\boldsymbol{v} = \boldsymbol{u}\cdot\boldsymbol{v}^{\ast}
\end{equation}
which is recast in terms of the dual bivector $\boldsymbol{V} = -\,\boldsymbol{v}^{\ast}$ as,
\begin{equation}\label{wedge and cross products duality ter}
\boldsymbol{u}\times\boldsymbol{v} = -\,\boldsymbol{u}\cdot\boldsymbol{V} = \boldsymbol{V}\cdot\boldsymbol{u}
\end{equation}
The dual of the inner product of the vectors $\boldsymbol{u}$ and $\boldsymbol{v}$ yields,
\begin{equation}\label{duality vectors inner}
\begin{split}
&\left(\boldsymbol{u}\cdot\boldsymbol{v}\right)^{\ast} = -\,\Big(\left(u_1\,\boldsymbol{e}_1 + u_2\,\boldsymbol{e}_2\right)\cdot\left(v_1\,\boldsymbol{e}_1 + v_2\,\boldsymbol{e}_2\right)\Big)\left(\boldsymbol{e}_1\,\boldsymbol{e}_2\,\boldsymbol{e}_3\right)\\
&\phantom{\left(\boldsymbol{u}\cdot\boldsymbol{v}\right)^{\ast}} = -\,\left(u_1\,v_1 + u_2\,v_2\right)\boldsymbol{e}_1\,\boldsymbol{e}_2\,\boldsymbol{e}_3
\end{split}
\end{equation}
The outer product of the vectors $\boldsymbol{u}$ and $\boldsymbol{v}^{\ast}$ yields,
\begin{equation}\label{duality vectors inner bis}
\begin{split}
&\boldsymbol{u}\wedge\boldsymbol{v}^{\ast} = -\,\left(u_1\,\boldsymbol{e}_1 + u_2\,\boldsymbol{e}_2\right)\wedge\Big(\left(v_1\,\boldsymbol{e}_1 + v_2\,\boldsymbol{e}_2\right)\left(\boldsymbol{e}_1\,\boldsymbol{e}_2\,\boldsymbol{e}_3\right)\Big)\\
&\phantom{\boldsymbol{u}\wedge\boldsymbol{v}^{\ast}} = -\,\left(u_1\,\boldsymbol{e}_1 + u_2\,\boldsymbol{e}_2\right)\wedge\left(v_1\,\boldsymbol{e}_2\,\boldsymbol{e}_3 -\,v_2\,\boldsymbol{e}_1\,\boldsymbol{e}_3\right)\\
&\phantom{\boldsymbol{u}\wedge\boldsymbol{v}^{\ast}} = \left(u_1\,v_1 + u_2\,v_2\right)\boldsymbol{e}_1\,\boldsymbol{e}_2\,\boldsymbol{e}_3
\end{split}
\end{equation}
The identification of relations~\eqref{duality vectors inner} and~\eqref{duality vectors inner bis} yields the pseudoscalar duality,
\begin{equation}\label{vector duality bis}
\left(\boldsymbol{u}\cdot\boldsymbol{v}\right)^{\ast} = \boldsymbol{u}\wedge\boldsymbol{v}^{\ast}
\end{equation}
which is recast in terms of the dual bivector $\boldsymbol{V} = -\,\boldsymbol{v}^{\ast}$ as,
\begin{equation}\label{vector duality ter}
\left(\boldsymbol{u}\cdot\boldsymbol{v}\right)^{\ast} = -\,\boldsymbol{u}\wedge\boldsymbol{V}
\end{equation}
To establish the duality between the inner and outer product of a bivector and a vector, the spatial frame is oriented such that the bivector is given by $\boldsymbol{B} = B_{12}\,\boldsymbol{e}_1\,\boldsymbol{e}_2$ and the vector is written as $\boldsymbol{v} = v_1\,\boldsymbol{e}_1 + v_2\,\boldsymbol{e}_2 + v_3\,\boldsymbol{e}_3$. The dual of the outer product of the vector $\boldsymbol{u}$ and the bivector $\boldsymbol{B}$ yields,
\begin{equation}\label{duality vector bivector outer}
\begin{split}
&\left(\boldsymbol{u}\wedge\boldsymbol{B}\right)^{\ast} = -\,\Big(\left(u_1\,\boldsymbol{e}_1 + u_2\,\boldsymbol{e}_2 + u_3\,\boldsymbol{e}_3\right)\wedge\left(B_{12}\,\boldsymbol{e}_1\,\boldsymbol{e}_2\right)\Big)\left(\boldsymbol{e}_1\,\boldsymbol{e}_2\,\boldsymbol{e}_3\right)\\
&\phantom{\left(\boldsymbol{u}\wedge\boldsymbol{B}\right)^{\ast}} = -\,\left(u_3\,B_{12}\,\boldsymbol{e}_1\,\boldsymbol{e}_2\,\boldsymbol{e}_3\right)\left(\boldsymbol{e}_1\,\boldsymbol{e}_2\,\boldsymbol{e}_3\right)\\
&\phantom{\left(\boldsymbol{u}\wedge\boldsymbol{B}\right)^{\ast}} = u_3\,B_{12}
\end{split}
\end{equation}
The inner product of the vector $\boldsymbol{u}$ and the bivector $\boldsymbol{B}^{\ast}$ yields,
\begin{equation}\label{duality vector bivector outer bis}
\begin{split}
&\boldsymbol{u}\cdot\boldsymbol{B}^{\ast} = -\,\left(u_1\,\boldsymbol{e}_1 + u_2\,\boldsymbol{e}_2 + u_3\,\boldsymbol{e}_3\right)\cdot\Big(\left(B_{12}\,\boldsymbol{e}_1\,\boldsymbol{e}_2\right)\Big)\left(\boldsymbol{e}_1\,\boldsymbol{e}_2\,\boldsymbol{e}_3\right)\Big)\\
&\phantom{\boldsymbol{u}\cdot\boldsymbol{B}^{\ast}} = \left(u_1\,\boldsymbol{e}_1 + u_2\,\boldsymbol{e}_2 + u_3\,\boldsymbol{e}_3\right)\cdot\left(B_{12}\,\boldsymbol{e}_3\right)\\
&\phantom{\boldsymbol{u}\cdot\boldsymbol{B}^{\ast}} = u_3\,B_{12}
\end{split}
\end{equation}
The identification of relations~\eqref{duality vector bivector outer} and~\eqref{duality vector bivector outer bis} yields the scalar duality,
\begin{equation}\label{scalar duality}
\left(\boldsymbol{u}\wedge\boldsymbol{B}\right)^{\ast} = \boldsymbol{u}\cdot\boldsymbol{B}^{\ast}
\end{equation}
which is recast in terms of the dual vector $\boldsymbol{b} = \boldsymbol{B}^{\ast}$ as,
\begin{equation}\label{scalar duality bis}
\left(\boldsymbol{u}\wedge\boldsymbol{B}\right)^{\ast} = \boldsymbol{u}\cdot\boldsymbol{b}
\end{equation}
and is the dual of the pseudoscalar duality~\eqref{vector duality ter} for $\boldsymbol{b} = \boldsymbol{v}$ and $\boldsymbol{B} = \boldsymbol{V}$. The dual of the inner product of the vector $\boldsymbol{u}$ and the bivector $\boldsymbol{B}$ yields,
\begin{equation}\label{duality vector bivector inner}
\begin{split}
&\left(\boldsymbol{u}\cdot\boldsymbol{B}\right)^{\ast} = -\,\Big(\left(u_1\,\boldsymbol{e}_1 + u_2\,\boldsymbol{e}_2 + u_3\,\boldsymbol{e}_3\right)\cdot\left(B_{12}\,\boldsymbol{e}_1\,\boldsymbol{e}_2\right)\Big)\left(\boldsymbol{e}_1\,\boldsymbol{e}_2\,\boldsymbol{e}_3\right)\\
&\phantom{\left(\boldsymbol{u}\cdot\boldsymbol{B}\right)^{\ast}} = -\,\left(u_1\,B_{12}\,\boldsymbol{e}_2 -\,u_2\,B_{12}\,\boldsymbol{e}_1\right)\left(\boldsymbol{e}_1\,\boldsymbol{e}_2\,\boldsymbol{e}_3\right)\\
&\phantom{\left(\boldsymbol{u}\cdot\boldsymbol{B}\right)^{\ast}} = u_1\,B_{12}\,\boldsymbol{e}_1\,\boldsymbol{e}_3 + u_2\,B_{12}\,\boldsymbol{e}_2\,\boldsymbol{e}_3
\end{split}
\end{equation}
The outer product of the vector $\boldsymbol{u}$ and the dual of the bivector $\boldsymbol{B}^{\ast}$ yields,
\begin{equation}\label{duality vector bivector inner bis}
\begin{split}
&\boldsymbol{u}\wedge\boldsymbol{B}^{\ast} = -\,\left(u_1\,\boldsymbol{e}_1 + u_2\,\boldsymbol{e}_2 + u_3\,\boldsymbol{e}_3\right)\wedge\Big(\left(B_{12}\,\boldsymbol{e}_1\,\boldsymbol{e}_2\right)\left(\boldsymbol{e}_1\,\boldsymbol{e}_2\,\boldsymbol{e}_3\right)\Big)\\
&\phantom{\boldsymbol{u}\wedge\boldsymbol{B}^{\ast}} = \left(u_1\,\boldsymbol{e}_1 + u_2\,\boldsymbol{e}_2 + u_3\,\boldsymbol{e}_3\right)\wedge\left(B_{12}\,\boldsymbol{e}_3\right)\\
&\phantom{\boldsymbol{u}\wedge\boldsymbol{B}^{\ast}} = u_1\,B_{12}\,\boldsymbol{e}_1\,\boldsymbol{e}_3 + u_2\,B_{12}\,\boldsymbol{e}_2\,\boldsymbol{e}_3
\end{split}
\end{equation}
The identification of relations~\eqref{duality vector bivector inner} and~\eqref{duality vector bivector inner bis} yields the bivectorial duality,
\begin{equation}\label{bivectorial duality}
\left(\boldsymbol{u}\cdot\boldsymbol{B}\right)^{\ast} = \boldsymbol{u}\wedge\boldsymbol{B}^{\ast}
\end{equation}
which is recast in terms of the dual vector $\boldsymbol{b} = \boldsymbol{B}^{\ast}$ as,
\begin{equation}\label{bivectorial duality bis}
\left(\boldsymbol{u}\cdot\boldsymbol{B}\right)^{\ast} = \boldsymbol{u}\wedge\boldsymbol{b}
\end{equation}
and is the dual of the vectorial duality~\eqref{vector duality bis 0} for $\boldsymbol{b} = \boldsymbol{v}$ and $\boldsymbol{B} = \boldsymbol{V}$. To establish the duality between the inner and outer product of bivectors in the same plane, the spatial frame is oriented such that the bivectors $\boldsymbol{A}$ and $\boldsymbol{B}$ are written as $\boldsymbol{A} = A_{12}\,\boldsymbol{e}_1\,\boldsymbol{e}_2$ and $\boldsymbol{B} = B_{12}\,\boldsymbol{e}_2\,\boldsymbol{e}_3$. The dual of the inner product of the bivectors $\boldsymbol{A}$ and $\boldsymbol{B}$ yields,
\begin{equation}\label{duality bivectors inner}
\begin{split}
&\left(\boldsymbol{A}\cdot\boldsymbol{B}\right)^{\ast} = -\,\Big(\left(A_{12}\,\boldsymbol{e}_1\,\boldsymbol{e}_2\right)\cdot\left(B_{12}\,\boldsymbol{e}_1\,\boldsymbol{e}_2\right)\Big)\left(\boldsymbol{e}_1\,\boldsymbol{e}_2\,\boldsymbol{e}_3\right)\\
&\phantom{\left(\boldsymbol{A}\cdot\boldsymbol{B}\right)^{\ast}} = A_{12}\,B_{12}\,\boldsymbol{e}_1\,\boldsymbol{e}_2\,\boldsymbol{e}_3
\end{split}
\end{equation}
The outer product of the bivectors $\boldsymbol{A}$ and $\boldsymbol{B}^{\ast}$ yields,
\begin{equation}\label{duality bivectors inner bis}
\begin{split}
&\boldsymbol{A}\wedge\boldsymbol{B}^{\ast} = -\,\left(A_{12}\,\boldsymbol{e}_1\,\boldsymbol{e}_2\right)\wedge\Big(\left(B_{12}\,\boldsymbol{e}_1\,\boldsymbol{e}_2\right)\left(\boldsymbol{e}_1\,\boldsymbol{e}_2\,\boldsymbol{e}_3\right)\Big)\\
&\phantom{\boldsymbol{A}\wedge\boldsymbol{B}^{\ast}} = \left(A_{12}\,\boldsymbol{e}_1\,\boldsymbol{e}_2\right)\wedge\left(B_{12}\,\boldsymbol{e}_3\right)\\
&\phantom{\boldsymbol{A}\wedge\boldsymbol{B}^{\ast}} = A_{12}\,B_{12}\,\boldsymbol{e}_1\,\boldsymbol{e}_2\,\boldsymbol{e}_3
\end{split}
\end{equation}
The identification of relations~\eqref{duality bivectors inner} and~\eqref{duality bivectors inner bis} yields the pseudoscalar duality,
\begin{equation}\label{bivectorial duality bi}
\left(\boldsymbol{A}\cdot\boldsymbol{B}\right)^{\ast} = \boldsymbol{A}\wedge\boldsymbol{B}^{\ast}
\end{equation}
The pseudoscalar duality~\eqref{bivectorial duality bi} is expressed in terms of the dual vector $\boldsymbol{b} = \boldsymbol{B}^{\ast}$,
\begin{equation}\label{bivectorial duality bi bis}
\left(\boldsymbol{A}\cdot\boldsymbol{B}\right)^{\ast} = \boldsymbol{A}\wedge\boldsymbol{b}
\end{equation}
In view of the identities~\eqref{outer product v B symmetric},~\eqref{bivectorial duality}, the inner product of two bivectors $\boldsymbol{A}$ and $\boldsymbol{B}$ is expressed in terms of the dual vectors $\boldsymbol{a} = \boldsymbol{A}^{\ast}$ and $\boldsymbol{b} = \boldsymbol{B}^{\ast}$ as,
\begin{equation}\label{grad bivec dual}
\begin{split}
&\left(\boldsymbol{A}\cdot\boldsymbol{B}\right)^{\ast} = \boldsymbol{A}\wedge\boldsymbol{b} = \boldsymbol{b}\wedge\boldsymbol{A} = -\,\boldsymbol{b}\wedge\boldsymbol{a}^{\ast}\\
&\phantom{\left(\boldsymbol{A}\cdot\boldsymbol{B}\right)^{\ast}} = -\,\left(\boldsymbol{b}\cdot\boldsymbol{a}\right)^{\ast} = -\,\left(\boldsymbol{a}\cdot\boldsymbol{b}\right)^{\ast}
\end{split}
\end{equation}
The dual of identity~\eqref{grad bivec dual} yields,
\begin{equation}\label{grad bivec}
\boldsymbol{A}\cdot\boldsymbol{B} = -\,\boldsymbol{a}\cdot\boldsymbol{b}
\end{equation}

\section{Differential duality in spatial algebra (SA)}
\label{Differential duality in spatial algebra}

\noindent For the gradient operator $\boldsymbol{u} = \boldsymbol{\nabla}$, the duality~\eqref{wedge and cross products duality} for the curl of a vector $\boldsymbol{v}$ is expressed as,
\begin{equation}\label{curl duality}
\left(\boldsymbol{\nabla}\wedge\boldsymbol{v}\right)^{\ast} = \boldsymbol{\nabla}\times\boldsymbol{v} \qquad\text{and}\qquad \left(\boldsymbol{\nabla}\times\boldsymbol{v}\right)^{\ast} = -\,\boldsymbol{\nabla}\wedge\boldsymbol{v} 
\end{equation}
and the vectorial duality~\eqref{vector duality} yields,
\begin{equation}\label{curl divergence duality}
\left(\boldsymbol{\nabla}\wedge\boldsymbol{v}\right)^{\ast} = \boldsymbol{\nabla}\cdot\boldsymbol{v}^{\ast}
\end{equation}
In view of relations~\eqref{curl duality} and~\eqref{curl divergence duality}, we obtain,
\begin{equation}\label{curl divergence duality bis}
\boldsymbol{\nabla}\times\boldsymbol{v} = \boldsymbol{\nabla}\cdot\boldsymbol{v}^{\ast}
\end{equation}
which is recast in terms of the dual bivector $\boldsymbol{V} = -\,\boldsymbol{v}^{\ast}$ as,
\begin{equation}\label{curl divergence duality ter}
\boldsymbol{\nabla}\times\boldsymbol{v} = -\,\boldsymbol{\nabla}\cdot\boldsymbol{V}
\end{equation}
For the gradient operator $\boldsymbol{u} = \boldsymbol{\nabla}$, the pseudoscalar duality~\eqref{vector duality bis 0} is expressed as,
\begin{equation}\label{curl divergence duality more}
\left(\boldsymbol{\nabla}\cdot\boldsymbol{v}\right)^{\ast} = \boldsymbol{\nabla}\wedge\boldsymbol{v}^{\ast}
\end{equation}
which is recast in terms of the dual bivector $\boldsymbol{V} = -\,\boldsymbol{v}^{\ast}$ as,
\begin{equation}\label{curl divergence duality quad}
\left(\boldsymbol{\nabla}\cdot\boldsymbol{v}\right)^{\ast} = -\,\boldsymbol{\nabla}\wedge\boldsymbol{V}
\end{equation}
For the gradient operator $\boldsymbol{u} = \boldsymbol{\nabla}$, the scalar duality~\eqref{scalar duality} is expressed as,
\begin{equation}\label{curl divergence scalar duality}
\left(\boldsymbol{\nabla}\wedge\boldsymbol{B}\right)^{\ast} = \boldsymbol{\nabla}\cdot\boldsymbol{B}^{\ast}
\end{equation}
which is recast in terms of the dual vector $\boldsymbol{b} = \boldsymbol{B}^{\ast}$ as,
\begin{equation}\label{curl divergence scalar duality bis}
\left(\boldsymbol{\nabla}\wedge\boldsymbol{B}\right)^{\ast} = \boldsymbol{\nabla}\cdot\boldsymbol{b}
\end{equation}
and is the dual of the pseudoscalar duality~\eqref{curl divergence duality quad} for $\boldsymbol{b} = \boldsymbol{v}$ and $\boldsymbol{B} = \boldsymbol{V}$. For the gradient operator $\boldsymbol{u} = \boldsymbol{\nabla}$, the bivectorial duality~\eqref{bivectorial duality} is expressed as,
\begin{equation}\label{curl divergence bivectorial duality}
\left(\boldsymbol{\nabla}\cdot\boldsymbol{B}\right)^{\ast} = \boldsymbol{\nabla}\wedge\boldsymbol{B}^{\ast}
\end{equation}
which is recast in terms of the dual vector $\boldsymbol{b} = \boldsymbol{B}^{\ast}$ as,
\begin{equation}\label{curl divergence scalar bivectorial bis}
\left(\boldsymbol{\nabla}\cdot\boldsymbol{B}\right)^{\ast} = \boldsymbol{\nabla}\wedge\boldsymbol{b}
\end{equation}
and is the dual of the vectorial duality~\eqref{curl divergence duality ter} for $\boldsymbol{b} = \boldsymbol{v}$ and $\boldsymbol{B} = \boldsymbol{V}$.

\section{Algebraic identities in spatial algebra (SA)}
\label{Algebraic identities in spatial algebra}

\noindent The double cross product of the three vectors $\boldsymbol{u}$, $\boldsymbol{v}$ and $\boldsymbol{w}$ is written as,
\begin{equation}\label{double cross product vectors}
\boldsymbol{u}\times\left(\boldsymbol{v}\times\boldsymbol{w}\right) = \left(\boldsymbol{u}\cdot\boldsymbol{w}\right)\boldsymbol{v} -\,\left(\boldsymbol{u}\cdot\boldsymbol{v}\right)\boldsymbol{w}
\end{equation}
In view of the duality~\eqref{wedge and cross products duality} between the cross and wedge products, the vectorial duality~\eqref{wedge and cross products duality bis}, the double duality~\eqref{dual dual B} and the antisymmetry of the outer product~\eqref{outer product space u v}, the left-hand side of relation~\eqref{double cross product vectors} is recast as,
\begin{equation}\label{id cross}
\boldsymbol{u}\times\left(\boldsymbol{v}\times\boldsymbol{w}\right) = \boldsymbol{u}\times\left(\boldsymbol{v}\wedge\boldsymbol{w}\right)^{\ast} = \boldsymbol{u}\cdot\Big(\left(\boldsymbol{v}\wedge\boldsymbol{w}\right)^{\ast}\Big)^{\ast} = -\,\boldsymbol{u}\cdot\left(\boldsymbol{v}\wedge\boldsymbol{w}\right) = \boldsymbol{u}\cdot\left(\boldsymbol{w}\wedge\boldsymbol{v}\right)
\end{equation}
According to the identity~\eqref{id cross}, the double cross product~\eqref{double cross product vectors} yields the triple product,
\begin{equation}\label{triple product vectors}
\boldsymbol{u}\cdot\left(\boldsymbol{v}\wedge\boldsymbol{w}\right) = \left(\boldsymbol{u}\cdot\boldsymbol{v}\right)\boldsymbol{w} -\,\left(\boldsymbol{u}\cdot\boldsymbol{w}\right)\boldsymbol{v}
\end{equation}
This triple product~\eqref{triple product vectors} represents the projection of the vector $\boldsymbol{u}$ on the bivector $\boldsymbol{v}\wedge\boldsymbol{w}$. The vectors $\boldsymbol{u}$, $\boldsymbol{v}$ and $\boldsymbol{w}$ satisfy the identity,
\begin{equation}\label{id dot cross}
\boldsymbol{u}\cdot\left(\boldsymbol{v}\times\boldsymbol{w}\right) = \boldsymbol{v}\cdot\left(\boldsymbol{w}\times\boldsymbol{u}\right) = \boldsymbol{w}\cdot\left(\boldsymbol{u}\times\boldsymbol{v}\right)
\end{equation}
In view of the duality~\eqref{wedge and cross products duality} between cross product and wedge product and the scalar duality~\eqref{scalar duality}, the three terms of identity~\eqref{id dot cross} are recast as,
\begin{equation}\label{id dot cross wedge}
\begin{split}
&\boldsymbol{u}\cdot\left(\boldsymbol{v}\times\boldsymbol{w}\right) = \boldsymbol{u}\cdot\left(\boldsymbol{v}\wedge\boldsymbol{w}\right)^{\ast} = \left(\boldsymbol{u}\wedge\boldsymbol{v}\wedge\boldsymbol{w}\right)^{\ast}\\
&\boldsymbol{v}\cdot\left(\boldsymbol{w}\times\boldsymbol{u}\right) = \boldsymbol{v}\cdot\left(\boldsymbol{w}\wedge\boldsymbol{u}\right)^{\ast} = \left(\boldsymbol{v}\wedge\boldsymbol{w}\wedge\boldsymbol{u}\right)^{\ast}\\
&\boldsymbol{w}\cdot\left(\boldsymbol{u}\times\boldsymbol{v}\right) = \boldsymbol{w}\cdot\left(\boldsymbol{u}\wedge\boldsymbol{v}\right)^{\ast} = \left(\boldsymbol{w}\wedge\boldsymbol{u}\wedge\boldsymbol{v}\right)^{\ast}
\end{split}
\end{equation}
According to the relations~\eqref{id dot cross wedge}, the scalar identity~\eqref{id dot cross} becomes,
\begin{equation}\label{id dot cross bis}
\left(\boldsymbol{u}\wedge\boldsymbol{v}\wedge\boldsymbol{w}\right)^{\ast} = \left(\boldsymbol{v}\wedge\boldsymbol{w}\wedge\boldsymbol{u}\right)^{\ast} = \left(\boldsymbol{w}\wedge\boldsymbol{u}\wedge\boldsymbol{v}\right)^{\ast}
\end{equation}
The dual of the scalar identity~\eqref{id dot cross bis} yields the pseudoscalar identity,
\begin{equation}\label{id dot cross ter}
\boldsymbol{u}\wedge\boldsymbol{v}\wedge\boldsymbol{w} = \boldsymbol{v}\wedge\boldsymbol{w}\wedge\boldsymbol{u} = \boldsymbol{w}\wedge\boldsymbol{u}\wedge\boldsymbol{v}
\end{equation}
This identity represents the fact the oriented volume of the parallelepiped spanned by the vectors $\boldsymbol{u}$, $\boldsymbol{v}$, $\boldsymbol{w}$ is the same as that of the parallelepiped spanned by the vectors $\boldsymbol{v}$, $\boldsymbol{w}$, $\boldsymbol{u}$ and also the same as that of the parallelepiped spanned by the vectors $\boldsymbol{w}$, $\boldsymbol{u}$, $\boldsymbol{v}$ because these three parallelepipeds have the same orientation. According to relations~\eqref{inner product antisymmetry v B} and~\eqref{outer product symmetry v B} for $\boldsymbol{B} = \boldsymbol{v}\wedge\boldsymbol{w}$, the inner and outer products of the vector $\boldsymbol{u}$ and the bivector $\boldsymbol{v}\wedge\boldsymbol{w}$ are given by,
\begin{align}
\label{inner product vec bivec}
&\boldsymbol{u}\cdot\left(\boldsymbol{v}\wedge\boldsymbol{w}\right) = \frac{1}{2}\Big(\boldsymbol{u}\left(\boldsymbol{v}\wedge\boldsymbol{w}\right) -\,\left(\boldsymbol{v}\wedge\boldsymbol{w}\right)\boldsymbol{u}\Big)\\
\label{outer product vec bivec}
&\boldsymbol{u}\wedge\left(\boldsymbol{v}\wedge\boldsymbol{w}\right) = \frac{1}{2}\Big(\boldsymbol{u}\left(\boldsymbol{v}\wedge\boldsymbol{w}\right) + \left(\boldsymbol{v}\wedge\boldsymbol{w}\right)\boldsymbol{u}\Big)
\end{align}
To determine the mixed product of two vectors $\boldsymbol{u}$, $\boldsymbol{v}$ and a bivector $\boldsymbol{B}$, we use the dual vector defined as $\boldsymbol{b} = -\,\boldsymbol{B}\,I$. Using the identity~\eqref{inner product v B antisymmetric} and the triple product of vectors~\eqref{id dot cross}, we obtain,
\begin{equation}\label{triple product u v B}
\left(\boldsymbol{u}\wedge\boldsymbol{v}\right)\cdot\boldsymbol{B} = \left(\boldsymbol{u}\wedge\boldsymbol{v}\right)\cdot\boldsymbol{b}\,I = -\,\boldsymbol{b}\cdot\left(\boldsymbol{u}\wedge\boldsymbol{v}\right)\,I = -\,I\left(\boldsymbol{b}\cdot\boldsymbol{u}\right)\boldsymbol{v} + I\left(\boldsymbol{b}\cdot\boldsymbol{v}\right)\boldsymbol{u}
\end{equation}
which reduces to,
\begin{equation}\label{triple product u v B bis}
\left(\boldsymbol{u}\wedge\boldsymbol{v}\right)\cdot\boldsymbol{B} = \left(\boldsymbol{B}\cdot\boldsymbol{u}\right)\cdot\boldsymbol{v} -\,\left(\boldsymbol{B}\cdot\boldsymbol{v}\right)\cdot\boldsymbol{u} = \boldsymbol{u}\cdot\left(\boldsymbol{v}\cdot\boldsymbol{B}\right) -\,\boldsymbol{v}\cdot\left(\boldsymbol{u}\cdot\boldsymbol{B}\right)
\end{equation}
To determine the first kind of mixed product of two bivectors $\boldsymbol{A}$, $\boldsymbol{B}$ and a vector $\boldsymbol{v}$, we use the dual vectors defined as $\boldsymbol{a} = -\,\boldsymbol{A}\,I$ and $\boldsymbol{b} = -\,\boldsymbol{B}\,I$. Using the identities~\eqref{inner product v B antisymmetric} and $I^2 = -1$, we obtain,
\begin{equation}\label{triple product A v B 1st}
\left(\boldsymbol{A}\wedge\boldsymbol{v}\right)\cdot\boldsymbol{B} = \left(\boldsymbol{a}\,I\wedge\boldsymbol{v}\right)\cdot\boldsymbol{b}\,I = -\,\left(\boldsymbol{a}\wedge\boldsymbol{v}\right)\cdot\boldsymbol{b} = \boldsymbol{b}\cdot\left(\boldsymbol{a}\wedge\boldsymbol{v}\right)
\end{equation}
Using the triple product of vectors~\eqref{id dot cross}, the mixed product is recast as,
\begin{equation}\label{triple product A v B 1st bis}
\begin{split}
&\left(\boldsymbol{A}\wedge\boldsymbol{v}\right)\cdot\boldsymbol{B} = \boldsymbol{b}\cdot\left(\boldsymbol{a}\wedge\boldsymbol{v}\right) = \left(\boldsymbol{a}\cdot\boldsymbol{b}\right)\boldsymbol{v} -\,\left(\boldsymbol{b}\cdot\boldsymbol{v}\right)\boldsymbol{a}\\
&\phantom{\left(\boldsymbol{A}\wedge\boldsymbol{v}\right)\cdot\boldsymbol{B}} = -\,\left(\boldsymbol{A}\cdot\boldsymbol{B}\right)\boldsymbol{v} + \left(\boldsymbol{B}\cdot\boldsymbol{v}\right)\cdot\boldsymbol{A}
\end{split}
\end{equation}
To determine the first kind of mixed product of two bivectors $\boldsymbol{A}$, $\boldsymbol{B}$ and a vector $\boldsymbol{v}$, we use the dual vector defined as $\boldsymbol{a} = \boldsymbol{A}^{\ast} = -\,\boldsymbol{A}\,I$ and the duality $\left(\boldsymbol{v}\wedge\boldsymbol{a}\right)^{\ast} = -\left(\boldsymbol{v}\wedge\boldsymbol{a}\right)I$. Using the identities~\eqref{inner product v B antisymmetric},~\eqref{vector duality} and $I^2 = -1$, we obtain,
\begin{equation}\label{triple product A v B 2nd}
\begin{split}
&\left(\boldsymbol{A}\cdot\boldsymbol{v}\right)\wedge\boldsymbol{B} = -\,\left(\boldsymbol{v}\cdot\boldsymbol{A}\right)\wedge\boldsymbol{B} = \left(\boldsymbol{v}\cdot\boldsymbol{a}^{\ast}\right)\wedge\boldsymbol{B} = \left(\boldsymbol{v}\wedge\boldsymbol{a}\right)^{\ast}\wedge\boldsymbol{B}\\
&\phantom{\left(\boldsymbol{A}\cdot\boldsymbol{v}\right)\wedge\boldsymbol{B}} = -\,\left(\boldsymbol{v}\wedge\boldsymbol{a}\right)I\wedge\boldsymbol{B} = \boldsymbol{v}\wedge\left(\boldsymbol{A}\wedge\boldsymbol{B}\right)
\end{split}
\end{equation}

\section{Differential algebraic identities in spatial algebra (SA)}
\label{Differential algebraic identities in spatial algebra}

\noindent For the gradient operator $\boldsymbol{u} = \boldsymbol{\nabla}$, the geometric product~\eqref{geometric product space u v} implies that the gradient of the vector $\boldsymbol{v}$ is the sum of its divergence and its curl,
\begin{equation}\label{gradient vec}
\boldsymbol{\nabla}\,\boldsymbol{v} = \boldsymbol{\nabla}\cdot\boldsymbol{v} + \boldsymbol{\nabla}\wedge\boldsymbol{v}
\end{equation}
For the gradient operator $\boldsymbol{v} = \boldsymbol{\nabla}$, the geometric product~\eqref{geometric product space v B} implies that the gradient of the bivector $\boldsymbol{B}$ is the sum of its divergence and its curl,
\begin{equation}\label{gradient bivec}
\boldsymbol{\nabla}\,\boldsymbol{B} = \boldsymbol{\nabla}\cdot\boldsymbol{B} + \boldsymbol{\nabla}\wedge\boldsymbol{B}
\end{equation}
The gradient of the product of two scalars $r$ and $s$ is written as,
\begin{equation}\label{gradient scalars}
\boldsymbol{\nabla}\left(r\,s\right) = r\,\boldsymbol{\nabla}\,s + s\,\boldsymbol{\nabla}\,r
\end{equation}
The divergence of the product of a scalar $s$ and a vector $\boldsymbol{v}$ reads,
\begin{equation}\label{divergence scalar vec}
\boldsymbol{\nabla}\left(s\,\boldsymbol{v}\right) = s\,\boldsymbol{\nabla}\cdot\boldsymbol{v} + \boldsymbol{v}\cdot\boldsymbol{\nabla}\,s
\end{equation}
The curl of the product of a scalar $s$ and a vector $\boldsymbol{v}$ reads,
\begin{equation}\label{curl scalar vec 0}
\boldsymbol{\nabla}\times\left(s\,\boldsymbol{v}\right) = s\,\boldsymbol{\nabla}\times\boldsymbol{v} -\,\boldsymbol{v}\times\boldsymbol{\nabla}\,s
\end{equation}
In view of the duality~\eqref{wedge and cross products duality} between the cross and wedge products, relation~\eqref{curl scalar vec 0} is recast as,
\begin{equation}\label{curl scalar vec bis}
\Big(\boldsymbol{\nabla}\wedge\left(s\,\boldsymbol{v}\right)\Big)^{\ast} = \left(s\,\boldsymbol{\nabla}\wedge\boldsymbol{v}\right)^{\ast} -\,\left(\boldsymbol{v}\wedge\boldsymbol{\nabla}\,s\right)^{\ast}
\end{equation}
The dual of relation~\eqref{curl scalar vec bis} is given by,
\begin{equation}\label{curl scalar vec}
\boldsymbol{\nabla}\wedge\left(s\,\boldsymbol{v}\right) = s\,\boldsymbol{\nabla}\wedge\boldsymbol{v} -\,\boldsymbol{v}\wedge\boldsymbol{\nabla}\,s
\end{equation}
The gradient of the inner product of two vectors $\boldsymbol{u}$ and $\boldsymbol{v}$ is written as,
\begin{equation}\label{gradient inner 0}
\boldsymbol{\nabla}\left(\boldsymbol{u}\cdot\boldsymbol{v}\right) = \boldsymbol{u}\times\left(\boldsymbol{\nabla}\times\boldsymbol{v}\right) + \boldsymbol{v}\times\left(\boldsymbol{\nabla}\times\boldsymbol{u}\right) + \left(\boldsymbol{u}\cdot\boldsymbol{\nabla}\right)\boldsymbol{v} + \left(\boldsymbol{v}\cdot\boldsymbol{\nabla}\right)\boldsymbol{u}
\end{equation}
In view of the duality~\eqref{wedge and cross products duality} between the cross and wedge products, the vectorial duality~\eqref{wedge and cross products duality bis}, the double duality~\eqref{dual dual B} and the antisymmetry of the inner product~\eqref{inner product v B antisymmetric}, the first two terms on the right-hand side of relation~\eqref{gradient inner 0} are recast as,
\begin{equation}\label{id cross nabla cross}
\begin{split}
&\boldsymbol{u}\times\left(\boldsymbol{\nabla}\times\boldsymbol{v}\right) = \boldsymbol{u}\times\left(\boldsymbol{\nabla}\wedge\boldsymbol{v}\right)^{\ast} = \boldsymbol{u}\cdot\Big(\left(\boldsymbol{\nabla}\wedge\boldsymbol{v}\right)^{\ast}\Big)^{\ast} = -\,\boldsymbol{u}\cdot\left(\boldsymbol{\nabla}\wedge\boldsymbol{v}\right) = \left(\boldsymbol{\nabla}\wedge\boldsymbol{v}\right)\cdot\boldsymbol{u}\\
&\boldsymbol{v}\times\left(\boldsymbol{\nabla}\times\boldsymbol{u}\right) = \boldsymbol{v}\times\left(\boldsymbol{\nabla}\wedge\boldsymbol{u}\right)^{\ast} = \boldsymbol{v}\cdot\Big(\left(\boldsymbol{\nabla}\wedge\boldsymbol{u}\right)^{\ast}\Big)^{\ast} = -\,\boldsymbol{v}\cdot\left(\boldsymbol{\nabla}\wedge\boldsymbol{u}\right) = \left(\boldsymbol{\nabla}\wedge\boldsymbol{u}\right)\cdot\boldsymbol{v}
\end{split}
\end{equation}
According to relations~\eqref{id cross nabla cross}, the gradient~\eqref{gradient inner 0} of the inner product of vectors $\boldsymbol{u}$ and $\boldsymbol{v}$ is recast as,
\begin{equation}\label{gradient inner}
\boldsymbol{\nabla}\left(\boldsymbol{u}\cdot\boldsymbol{v}\right) = \left(\boldsymbol{\nabla}\wedge\boldsymbol{u}\right)\cdot\boldsymbol{v} + \left(\boldsymbol{\nabla}\wedge\boldsymbol{v}\right)\cdot\boldsymbol{u} + \left(\boldsymbol{u}\cdot\boldsymbol{\nabla}\right)\boldsymbol{v} + \left(\boldsymbol{v}\cdot\boldsymbol{\nabla}\right)\boldsymbol{u}
\end{equation}
The curl of the cross product of two vectors $\boldsymbol{u}$ and $\boldsymbol{v}$ is written as,
\begin{equation}\label{curl cross}
\boldsymbol{\nabla}\times\left(\boldsymbol{u}\times\boldsymbol{v}\right) = -\,\left(\boldsymbol{\nabla}\cdot\boldsymbol{u}\right)\boldsymbol{v} + \left(\boldsymbol{\nabla}\cdot\boldsymbol{v}\right)\boldsymbol{u} -\,\left(\boldsymbol{u}\cdot\boldsymbol{\nabla}\right)\boldsymbol{v} + \left(\boldsymbol{v}\cdot\boldsymbol{\nabla}\right)\boldsymbol{u}
\end{equation}
In view of the duality~\eqref{wedge and cross products duality} between the cross and wedge products, the vectorial duality~\eqref{wedge and cross products duality bis} and the double duality~\eqref{dual dual B}, the left-hand side of relation~\eqref{curl cross} is recast as,
\begin{equation}\label{id nabla cross cross}
\boldsymbol{\nabla}\times\left(\boldsymbol{u}\times\boldsymbol{v}\right) = \boldsymbol{\nabla}\times\left(\boldsymbol{u}\wedge\boldsymbol{v}\right)^{\ast} = \boldsymbol{\nabla}\cdot\Big(\left(\boldsymbol{u}\wedge\boldsymbol{v}\right)^{\ast}\Big)^{\ast} = -\,\boldsymbol{\nabla}\cdot\left(\boldsymbol{u}\wedge\boldsymbol{v}\right)
\end{equation}
According to relation~\eqref{id nabla cross cross}, the curl~\eqref{curl cross} of the cross product of two vectors $\boldsymbol{u}$ and $\boldsymbol{v}$ is recast as the divergence of the outer product of these vectors,
\begin{equation}\label{divergence outer}
\boldsymbol{\nabla}\cdot\left(\boldsymbol{u}\wedge\boldsymbol{v}\right) = \left(\boldsymbol{\nabla}\cdot\boldsymbol{u}\right)\boldsymbol{v}  -\,\left(\boldsymbol{\nabla}\cdot\boldsymbol{v}\right)\boldsymbol{u} + \left(\boldsymbol{u}\cdot\boldsymbol{\nabla}\right)\boldsymbol{v} -\,\left(\boldsymbol{v}\cdot\boldsymbol{\nabla}\right)\boldsymbol{u}
\end{equation}
The divergence of the cross product of two vectors $\boldsymbol{u}$ and $\boldsymbol{v}$ is written as,
\begin{equation}\label{div cross}
\boldsymbol{\nabla}\cdot\left(\boldsymbol{u}\times\boldsymbol{v}\right) = -\,\boldsymbol{u}\cdot\left(\boldsymbol{\nabla}\times\boldsymbol{v}\right) + \boldsymbol{v}\cdot\left(\boldsymbol{\nabla}\times\boldsymbol{u}\right)
\end{equation}
In view of the duality~\eqref{wedge and cross products duality} between the cross and wedge products, the vectorial duality~\eqref{wedge and cross products duality bis} and the double duality~\eqref{dual dual B}, the three terms of relation~\eqref{div cross} are recast as,
\begin{equation}\label{id cross duality}
\begin{split}
&\boldsymbol{\nabla}\cdot\left(\boldsymbol{u}\times\boldsymbol{v}\right) = \boldsymbol{\nabla}\cdot\left(\boldsymbol{u}\wedge\boldsymbol{v}\right)^{\ast} = \Big(\boldsymbol{\nabla}\wedge\left(\boldsymbol{u}\wedge\boldsymbol{v}\right)\Big)^{\ast}\\
&\boldsymbol{u}\cdot\left(\boldsymbol{\nabla}\times\boldsymbol{v}\right) = \boldsymbol{u}\cdot\left(\boldsymbol{\nabla}\wedge\boldsymbol{v}\right)^{\ast} = \Big(\boldsymbol{u}\wedge\left(\boldsymbol{\nabla}\wedge\boldsymbol{v}\right)\Big)^{\ast}\\
&\boldsymbol{v}\cdot\left(\boldsymbol{\nabla}\times\boldsymbol{u}\right) = \boldsymbol{v}\cdot\left(\boldsymbol{\nabla}\wedge\boldsymbol{u}\right)^{\ast} = \Big(\boldsymbol{v}\wedge\left(\boldsymbol{\nabla}\wedge\boldsymbol{u}\right)\Big)^{\ast}
\end{split}
\end{equation}
According to relations~\eqref{id cross duality}, the divergence~\eqref{div cross} of the cross product of two vectors $\boldsymbol{u}$ and $\boldsymbol{v}$ is recast as the dual of the curl of the wedge product of these vectors,
\begin{equation}\label{curl wedge dual}
\Big(\boldsymbol{\nabla}\wedge\left(\boldsymbol{u}\wedge\boldsymbol{v}\right)\Big)^{\ast} = -\,\Big(\boldsymbol{u}\wedge\left(\boldsymbol{\nabla}\wedge\boldsymbol{v}\right)\Big)^{\ast} + \Big(\boldsymbol{v}\wedge\left(\boldsymbol{\nabla}\wedge\boldsymbol{u}\right)\Big)^{\ast}
\end{equation}
The dual of relation~\eqref{curl wedge dual} yields,
\begin{equation}\label{curl wedge}
\boldsymbol{\nabla}\wedge\left(\boldsymbol{u}\wedge\boldsymbol{v}\right) = -\,\boldsymbol{u}\left(\wedge\boldsymbol{\nabla}\wedge\boldsymbol{v}\right) + \boldsymbol{v}\wedge\left(\boldsymbol{\nabla}\wedge\boldsymbol{u}\right)
\end{equation}
In view of the identities~\eqref{vector duality},~\eqref{curl divergence duality} and~\eqref{curl divergence scalar duality}, the divergence of the inner product of a vector $\boldsymbol{v}$ and a bivector $\boldsymbol{B}$ is recast in terms of the dual vector $\boldsymbol{b} = \boldsymbol{B}^{\ast}$ as,
\begin{equation}\label{div vec bivec}
\boldsymbol{\nabla}\cdot\left(\boldsymbol{v}\cdot\boldsymbol{B}\right) = -\,\boldsymbol{\nabla}\cdot\left(\boldsymbol{v}\cdot\boldsymbol{b}^{\ast}\right) = -\,\boldsymbol{\nabla}\cdot\left(\boldsymbol{v}\wedge\boldsymbol{b}\right)^{\ast} = -\,\Big(\boldsymbol{\nabla}\wedge\left(\boldsymbol{v}\wedge\boldsymbol{b}\right)\Big)^{\ast}
\end{equation}
Taking into account relation~\eqref{curl wedge dual}, relation~\eqref{div vec bivec} becomes,
\begin{equation}\label{div vec bivec bis}
\boldsymbol{\nabla}\cdot\left(\boldsymbol{v}\cdot\boldsymbol{B}\right) = \Big(\boldsymbol{v}\wedge\left(\boldsymbol{\nabla}\wedge\boldsymbol{b}\right)\Big)^{\ast} -\,\Big(\boldsymbol{b}\wedge\left(\boldsymbol{\nabla}\wedge\boldsymbol{v}\right)\Big)^{\ast}
\end{equation}
According to relations~\eqref{vector duality} and~\eqref{curl divergence duality}, the first term on the right-hand side of relation~\eqref{div vec bivec bis} is recast as,
\begin{equation}\label{div vec bivec ter}
\Big(\boldsymbol{v}\wedge\left(\boldsymbol{\nabla}\wedge\boldsymbol{b}\right)\Big)^{\ast} = \boldsymbol{v}\cdot\left(\boldsymbol{\nabla}\wedge\boldsymbol{b}\right)^{\ast} = \boldsymbol{v}\cdot\left(\boldsymbol{\nabla}\cdot\boldsymbol{b}^{\ast}\right) = -\,\boldsymbol{v}\cdot\left(\boldsymbol{\nabla}\cdot\boldsymbol{B}\right)
\end{equation}
According to relations~\eqref{outer product v B symmetric} and~\eqref{vector duality} the second term on the right-hand side of relation~\eqref{div vec bivec bis} is recast as,
\begin{equation}\label{div vec bivec quad}
\Big(\boldsymbol{b}\wedge\left(\boldsymbol{\nabla}\wedge\boldsymbol{v}\right)\Big)^{\ast} = \Big(\left(\boldsymbol{\nabla}\wedge\boldsymbol{v}\right)\wedge\boldsymbol{b}\Big)^{\ast} = \left(\boldsymbol{\nabla}\wedge\boldsymbol{v}\right)\cdot\boldsymbol{b}^{\ast} = -\,\left(\boldsymbol{\nabla}\wedge\boldsymbol{v}\right)\cdot\boldsymbol{B}
\end{equation}
In view of relations~\eqref{div vec bivec ter} and~\eqref{div vec bivec quad}, the divergence~\eqref{div vec bivec bis} of the inner product of a vector $\boldsymbol{v}$ and a bivector $\boldsymbol{B}$ becomes,
\begin{equation}\label{div vec bivec pent}
\boldsymbol{\nabla}\cdot\left(\boldsymbol{v}\cdot\boldsymbol{B}\right) = \left(\boldsymbol{\nabla}\wedge\boldsymbol{v}\right)\cdot\boldsymbol{B} -\,\boldsymbol{v}\cdot\left(\boldsymbol{\nabla}\cdot\boldsymbol{B}\right)
\end{equation}
According to the antisymmetry~\eqref{inner product v B antisymmetric} of the inner product of a vector and a bivector and the symmetry~\eqref{geometric product bivectors inner} of the inner product of two bivectors, the identity~\eqref{div vec bivec pent} is recast as,
\begin{equation}\label{div vec bivec hex}
\boldsymbol{\nabla}\cdot\left(\boldsymbol{B}\cdot\boldsymbol{v}\right) = \left(\boldsymbol{\nabla}\cdot\boldsymbol{B}\right)\cdot\boldsymbol{v} -\,\boldsymbol{B}\cdot\left(\boldsymbol{\nabla}\wedge\boldsymbol{v}\right)
\end{equation}
In view of identities~\eqref{vector duality},~\eqref{curl divergence duality} and~\eqref{curl divergence bivectorial duality}, the curl of the inner product of a vector $\boldsymbol{v}$ and a bivector $\boldsymbol{B}$ is recast in terms of the dual vector $\boldsymbol{b} = \boldsymbol{B}^{\ast}$ as,
\begin{equation}\label{curl vec bivec}
\boldsymbol{\nabla}\wedge\left(\boldsymbol{v}\cdot\boldsymbol{B}\right) = -\,\boldsymbol{\nabla}\wedge\left(\boldsymbol{v}\cdot\boldsymbol{b}^{\ast}\right) = -\,\boldsymbol{\nabla}\wedge\left(\boldsymbol{v}\wedge\boldsymbol{b}\right)^{\ast} = -\,\Big(\boldsymbol{\nabla}\cdot\left(\boldsymbol{v}\wedge\boldsymbol{b}\right)\Big)^{\ast}
\end{equation}
Taking into account relation~\eqref{divergence outer}, relation~\eqref{curl vec bivec} becomes,
\begin{equation}\label{curl vec bivec bis}
\boldsymbol{\nabla}\wedge\left(\boldsymbol{v}\cdot\boldsymbol{B}\right) =
-\left(\boldsymbol{\nabla}\cdot\boldsymbol{v}\right)\boldsymbol{b}^{\ast} + \left(\boldsymbol{\nabla}\cdot\boldsymbol{b}\right)\boldsymbol{v}^{\ast} -\left(\boldsymbol{v}\cdot\boldsymbol{\nabla}\right)\boldsymbol{b}^{\ast} + \left(\boldsymbol{b}\cdot\boldsymbol{\nabla}\right)\boldsymbol{v}^{\ast}
\end{equation}
which is recast as,
\begin{equation}\label{curl vec bivec ter}
\boldsymbol{\nabla}\wedge\left(\boldsymbol{v}\cdot\boldsymbol{B}\right) =
\left(\boldsymbol{\nabla}\cdot\boldsymbol{v}\right)\boldsymbol{B} + \left(\boldsymbol{\nabla}\cdot\boldsymbol{B}^{\ast}\right)\boldsymbol{v}^{\ast} + \left(\boldsymbol{v}\cdot\boldsymbol{\nabla}\right)\boldsymbol{B} + \left(\boldsymbol{B}^{\ast}\cdot\boldsymbol{\nabla}\right)\boldsymbol{v}^{\ast}
\end{equation}
In view of the dualities~\eqref{curl divergence scalar duality},~\eqref{dual v},~\eqref{dual B} and~\eqref{pseudoscalar squared},
\begin{align}
\label{curl vec bivec 1}
&\left(\boldsymbol{\nabla}\cdot\boldsymbol{B}^{\ast}\right)\boldsymbol{v}^{\ast} = \left(\boldsymbol{\nabla}\wedge\boldsymbol{B}\right)^{\ast}\boldsymbol{v}^{\ast} = \left(\boldsymbol{\nabla}\wedge\boldsymbol{B}\right)I\,I\,\boldsymbol{v} = -\,\left(\boldsymbol{\nabla}\wedge\boldsymbol{B}\right)\cdot\boldsymbol{v}\\
\label{curl vec bivec 2}
&\left(\boldsymbol{B}^{\ast}\cdot\boldsymbol{\nabla}\right)\boldsymbol{v}^{\ast} = \left(\boldsymbol{B}\wedge\boldsymbol{\nabla}\right)^{\ast}\boldsymbol{v}^{\ast} = \left(\boldsymbol{B}\wedge\boldsymbol{\nabla}\right)I\,I\,\boldsymbol{v} = -\,\left(\boldsymbol{B}\wedge\boldsymbol{\nabla}\right)\cdot\boldsymbol{v}
\end{align}
Using the identities~\eqref{curl vec bivec 1} and~\eqref{curl vec bivec 2}, the curl of the inner product of a vector $\boldsymbol{v}$ and a bivector $\boldsymbol{B}$~\eqref{curl vec bivec ter} is recast as,
\begin{equation}\label{curl vec bivec quad}
\boldsymbol{\nabla}\wedge\left(\boldsymbol{v}\cdot\boldsymbol{B}\right) =
\left(\boldsymbol{\nabla}\cdot\boldsymbol{v}\right)\boldsymbol{B} + \left(\boldsymbol{v}\cdot\boldsymbol{\nabla}\right)\boldsymbol{B} -\,\left(\boldsymbol{\nabla}\wedge\boldsymbol{B}\right)\cdot\boldsymbol{v} -\,\left(\boldsymbol{B}\wedge\boldsymbol{\nabla}\right)\cdot\boldsymbol{v}
\end{equation}
According to the antisymmetry~\eqref{inner product v B antisymmetric} of the inner product of a vector and a bivector, the identity~\eqref{curl vec bivec quad} is recast as,
\begin{equation}\label{curl vec bivec pent}
\boldsymbol{\nabla}\wedge\left(\boldsymbol{B}\cdot\boldsymbol{v}\right) = \left(\boldsymbol{\nabla}\wedge\boldsymbol{B}\right)\cdot\boldsymbol{v} + \left(\boldsymbol{B}\wedge\boldsymbol{\nabla}\right)\cdot\boldsymbol{v}
-\,\left(\boldsymbol{\nabla}\cdot\boldsymbol{v}\right)\boldsymbol{B} -\,\left(\boldsymbol{v}\cdot\boldsymbol{\nabla}\right)\boldsymbol{B} 
\end{equation}
According to the duality~\eqref{grad bivec} and the differential identity~\eqref{gradient inner}, the divergence of a bivector $\boldsymbol{A}$ with a bivector $\boldsymbol{B}$ is expressed in terms of the dual vectors $\boldsymbol{a} = \boldsymbol{A}^{\ast}$ and $\boldsymbol{b} = \boldsymbol{A}^{\ast}$ as,
\begin{equation}\label{gradient inner bivectors 0}
\boldsymbol{\nabla}\left(\boldsymbol{A}\cdot\boldsymbol{B}\right) = -\,\left(\boldsymbol{\nabla}\wedge\boldsymbol{a}\right)\cdot\boldsymbol{b} -\,\left(\boldsymbol{\nabla}\wedge\boldsymbol{b}\right)\cdot\boldsymbol{a} -\,\left(\boldsymbol{a}\cdot\boldsymbol{\nabla}\right)\boldsymbol{b} -\,\left(\boldsymbol{b}\cdot\boldsymbol{\nabla}\right)\boldsymbol{a}
\end{equation}
In view of the identities~\eqref{curl divergence bivectorial duality},~\eqref{inner product v B antisymmetric},~\eqref{vector duality},~\eqref{outer product space u v},~\eqref{bivectorial duality}, we obtain the relations,
\begin{equation*}
\begin{split}
&\left(\boldsymbol{\nabla}\wedge\boldsymbol{a}\right)\cdot\boldsymbol{b} = \left(\boldsymbol{\nabla}\wedge\boldsymbol{A}^{\ast}\right)\cdot\boldsymbol{b} = \left(\boldsymbol{\nabla}\cdot\boldsymbol{A}\right)^{\ast}\cdot\boldsymbol{b} = -\,\boldsymbol{b}\cdot\left(\boldsymbol{\nabla}\cdot\boldsymbol{A}\right)^{\ast} = -\,\Big(\boldsymbol{b}\wedge\left(\boldsymbol{\nabla}\cdot\boldsymbol{A}\right)\Big)^{\ast}\\
&= \Big(\left(\boldsymbol{\nabla}\cdot\boldsymbol{A}\right)\wedge\boldsymbol{b}\Big)^{\ast} = \Big(\left(\boldsymbol{\nabla}\cdot\boldsymbol{A}\right)\wedge\boldsymbol{B}^{\ast}\Big)^{\ast} = -\,\left(\boldsymbol{\nabla}\cdot\boldsymbol{A}\right)\cdot\boldsymbol{B} = \boldsymbol{B}\cdot\left(\boldsymbol{\nabla}\cdot\boldsymbol{A}\right)
\end{split}
\end{equation*}
and
\begin{equation*}
\begin{split}
&\left(\boldsymbol{\nabla}\wedge\boldsymbol{b}\right)\cdot\boldsymbol{a} = \left(\boldsymbol{\nabla}\wedge\boldsymbol{B}^{\ast}\right)\cdot\boldsymbol{a} = \left(\boldsymbol{\nabla}\cdot\boldsymbol{B}\right)^{\ast}\cdot\boldsymbol{a} = -\,\boldsymbol{a}\cdot\left(\boldsymbol{\nabla}\cdot\boldsymbol{B}\right)^{\ast} = -\,\Big(\boldsymbol{a}\wedge\left(\boldsymbol{\nabla}\cdot\boldsymbol{B}\right)\Big)^{\ast}\\
&= \Big(\left(\boldsymbol{\nabla}\cdot\boldsymbol{B}\right)\wedge\boldsymbol{a}\Big)^{\ast} = \Big(\left(\boldsymbol{\nabla}\cdot\boldsymbol{B}\right)\wedge\boldsymbol{A}^{\ast}\Big)^{\ast} = -\,\left(\boldsymbol{\nabla}\cdot\boldsymbol{B}\right)\cdot\boldsymbol{A} = \boldsymbol{A}\cdot\left(\boldsymbol{\nabla}\cdot\boldsymbol{B}\right)
\end{split}
\end{equation*}
which implies that the divergence~\eqref{gradient inner bivectors 0} of a bivector $\boldsymbol{A}$ with a bivector $\boldsymbol{B}$ is recast as,
\begin{equation}\label{gradient inner bivectors 1}
\boldsymbol{\nabla}\left(\boldsymbol{A}\cdot\boldsymbol{B}\right) = -\,\boldsymbol{A}\cdot\left(\boldsymbol{\nabla}\cdot\boldsymbol{B}\right) -\,\boldsymbol{B}\cdot\left(\boldsymbol{\nabla}\cdot\boldsymbol{A}\right) -\,\left(\boldsymbol{A}^{\ast}\cdot\boldsymbol{\nabla}\right)\boldsymbol{B}^{\ast} -\,\left(\boldsymbol{B}^{\ast}\cdot\boldsymbol{\nabla}\right)\boldsymbol{A}^{\ast}
\end{equation}
According to the dualities~\eqref{curl divergence scalar duality},~\eqref{dual B} and~\eqref{pseudoscalar squared},
\begin{align}
\label{id bivectors A B dual 1}
&\left(\boldsymbol{A}^{\ast}\cdot\boldsymbol{\nabla}\right)\boldsymbol{B}^{\ast} = \left(\boldsymbol{A}\wedge\boldsymbol{\nabla}\right)^{\ast}\boldsymbol{B}^{\ast} = \left(\boldsymbol{A}\wedge\boldsymbol{\nabla}\right)I\,I\,\boldsymbol{B} = -\,\left(\boldsymbol{A}\wedge\boldsymbol{\nabla}\right)\cdot\boldsymbol{B}\\
\label{id bivectors A B dual 2}
&\left(\boldsymbol{B}^{\ast}\cdot\boldsymbol{\nabla}\right)\boldsymbol{A}^{\ast} = \left(\boldsymbol{B}\wedge\boldsymbol{\nabla}\right)^{\ast}\boldsymbol{A}^{\ast} = \left(\boldsymbol{B}\wedge\boldsymbol{\nabla}\right)I\,I\,\boldsymbol{A} = -\,\left(\boldsymbol{B}\wedge\boldsymbol{\nabla}\right)\cdot\boldsymbol{A}
\end{align}
In view of the identities~\eqref{id bivectors A B dual 1} and~\eqref{id bivectors A B dual 2}, the divergence~\eqref{gradient inner bivectors 1} of a bivector $\boldsymbol{A}$ with a bivector $\boldsymbol{B}$ is recast as,
\begin{equation}\label{gradient inner bivectors}
\boldsymbol{\nabla}\left(\boldsymbol{A}\cdot\boldsymbol{B}\right) = -\,\boldsymbol{A}\cdot\left(\boldsymbol{\nabla}\cdot\boldsymbol{B}\right) -\,\boldsymbol{B}\cdot\left(\boldsymbol{\nabla}\cdot\boldsymbol{A}\right) + \left(\boldsymbol{A}\wedge\boldsymbol{\nabla}\right)\cdot\boldsymbol{B} + \left(\boldsymbol{B}\wedge\boldsymbol{\nabla}\right)\cdot\boldsymbol{A}
\end{equation}
The curl of the curl of a vector $\boldsymbol{v}$ vanishes,
\begin{equation}\label{curl curl vec}
\boldsymbol{\nabla}\wedge\left(\boldsymbol{\nabla}\wedge\boldsymbol{v}\right) = \boldsymbol{e}^i\,\partial_{i}\wedge\left(\boldsymbol{e}^j\,\partial_{j}\wedge\boldsymbol{v}\right) = \left(\boldsymbol{e}^i\wedge\boldsymbol{e}^j\right)\partial_{i}\,\partial_{j}\,\boldsymbol{v} = 0
\end{equation}
In view of relations,~\eqref{curl divergence duality},~\eqref{curl divergence scalar duality} and~\eqref{curl curl vec}, the divergence of the divergence of a bivector $\boldsymbol{B}$ with a dual vector $\boldsymbol{b} = \boldsymbol{B}^{\ast}$ vanishes,
\begin{equation}\label{div div bivec}
\boldsymbol{\nabla}\cdot\left(\boldsymbol{\nabla}\cdot\boldsymbol{B}\right) = -\,\boldsymbol{\nabla}\cdot\left(\boldsymbol{\nabla}\cdot\boldsymbol{b}^{\ast}\right) = -\,\boldsymbol{\nabla}\cdot\left(\boldsymbol{\nabla}\wedge\boldsymbol{b}\right)^{\ast} = -\,\Big(\boldsymbol{\nabla}\wedge\left(\boldsymbol{\nabla}\wedge\boldsymbol{b}\right)\Big)^{\ast} = 0
\end{equation}

\section{Space-time algebra (STA)}
\label{Space-time algebra}

\noindent We consider an orthonormal vector frame $\{e_0,e_1,e_2,e_3\}$ in space-time. The geometric product of two basis vectors reads,~\cite{Hestenes:2015}
\begin{equation}\label{geometric product space-time}
e_{\mu}\,e_{\nu} = e_{\mu}\cdot e_{\nu} + e_{\mu}\wedge e_{\nu} \qquad\text{where}\qquad \mu,\nu = 0,1,2,3
\end{equation}
The inner product between two basis vectors is symmetric and defined by the Minkowski metric with the signature convention $\left(+,-,-,-\right)$,
\begin{equation}\label{inner product space-time}
e_{\mu}\cdot e_{\nu} = e_{\nu}\cdot e_{\mu} = \eta_{\mu\nu} = 2\,\delta_{\mu 0}\,\delta_{\nu 0} -\,\delta_{\mu\nu}
\end{equation}
which implies that,
\begin{equation}\label{square space-time}
e_{0}^2 = 1\,,\qquad e_{i}^2 = -\,1\,,\qquad e_{0}\cdot e_{i} = 0\,,\qquad e_{i}\cdot e_{j} = -\,\delta_{ij}
\end{equation}
and the outer product is antisymmetric,
\begin{equation}\label{outer product space-time}
e_{\mu} \wedge e_{\nu} = -\,e_{\nu} \wedge e_{\mu}
\end{equation}
The definitions of the geometric product~\eqref{geometric product space-time}, the inner product~\eqref{inner product space-time} and the outer product~\eqref{outer product space-time} yield the anti-commutation relation of the space-time algebra $\mathbb{G}^{1,3}$,
\begin{equation}\label{space-time algebra}
\frac{1}{2}\left(e_{\mu}\,e_{\nu} + e_{\nu}\,e_{\mu}\right) = \eta_{\mu\nu}
\end{equation}
which is the Dirac algebra. A relative spatial orthonormal vector frame $\{\boldsymbol{e}_1,\boldsymbol{e}_2,\boldsymbol{e}_3\}$ attached to an observer consists of space-time bivectors defined as,~\cite{Hestenes:2015}
\begin{equation}\label{space-time bivectors}
\boldsymbol{e}_i = e_{i}\,e_{0} = e_{i} \wedge e_{0} \qquad\text{where}\qquad i = 1,2,3
\end{equation}
The geometric product of two basis spatial vectors reads,
\begin{equation}\label{geometric product space}
\boldsymbol{e}_i\,\boldsymbol{e}_j = \boldsymbol{e}_i\cdot \boldsymbol{e}_j + \boldsymbol{e}_i\wedge \boldsymbol{e}_j
\end{equation}
In view of relations~\eqref{square space-time},~\eqref{outer product space-time} and~\eqref{space-time bivectors}, the symmetric inner product of the spatial orthonormal basis vectors yields,
\begin{align}\label{space-time bivectors inner product}
&\boldsymbol{e}_i\cdot\boldsymbol{e}_j = \frac{1}{2}\left(\boldsymbol{e}_i\cdot\boldsymbol{e}_j + \boldsymbol{e}_j\cdot\boldsymbol{e}_i\right) = \frac{1}{2}\Big(\left(e_{i}\,e_{0}\right)\left(e_{j}\,e_{0}\right) + \left(e_{j}\,e_{0}\right)\left(e_{i}\,e_{0}\right)\Big)\\
&\phantom{\boldsymbol{e}_i\cdot\boldsymbol{e}_j} = -\,\frac{1}{2}\Big(\left(e_{i}\,e_{j}\right)\left(e_{0}\,e_{0}\right) + \left(e_{j}\,e_{i}\right)\left(e_{0}\,e_{0}\right)\Big) = -\,\frac{1}{2}\left(e_{i}\cdot e_{j} + e_{j}\cdot e_{i}\right) = \delta_{ij}\nonumber
\end{align}
and the pseudoscalar is given by,
\begin{equation}\label{space-time pseudoscalar}
I = \boldsymbol{e}_1\,\boldsymbol{e}_2\,\boldsymbol{e}_3 = \left(e_{1}\,e_{0}\right)\left(e_{2}\,e_{0}\right)\left(e_{3}\,e_{0}\right) = \left(e_{0}\,e_{1}\,e_{2}\,e_{3}\right)\left(e_{0}\,e_{0}\right) = e_{0}\,e_{1}\,e_{2}\,e_{3}
\end{equation}
In view of relations~\eqref{square space-time},~\eqref{outer product space-time} and~\eqref{space-time bivectors}, the geometric product of two different spatial basis vectors yields,
\begin{equation}\label{geometric product spatial vectors}
\begin{split}
&\boldsymbol{e}_1\,\boldsymbol{e}_2 = \left(\boldsymbol{e}_1\,\boldsymbol{e}_2\,\boldsymbol{e}_3\right)\boldsymbol{e}_3 = I\,\boldsymbol{e}_3 = -\,\boldsymbol{e}_2\,\boldsymbol{e}_1\\
&\boldsymbol{e}_2\,\boldsymbol{e}_3 = \boldsymbol{e}_1\left(\boldsymbol{e}_1\,\boldsymbol{e}_2\,\boldsymbol{e}_3\right) = \boldsymbol{e}_1\,I = I\,\boldsymbol{e}_1 = -\,\boldsymbol{e}_3\,\boldsymbol{e}_2\\
&\boldsymbol{e}_3\,\boldsymbol{e}_1 = -\left(\boldsymbol{e}_1\,\boldsymbol{e}_3\right)\left(\boldsymbol{e}_2\,\boldsymbol{e}_2\right) = \left(\boldsymbol{e}_1\,\boldsymbol{e}_2\,\boldsymbol{e}_3\right)\boldsymbol{e}_2 = I\,\boldsymbol{e}_2 = -\,\boldsymbol{e}_1\,\boldsymbol{e}_3
\end{split}
\end{equation}
which can be written as,
\begin{equation}\label{geometric product spatial vectors bis}
\boldsymbol{e}_i\,\boldsymbol{e}_j = \varepsilon_{ijk}\,I\,\boldsymbol{e}_k \qquad\text{where}\qquad i \neq j \neq k \neq i  
\end{equation}
The inner product~\eqref{geometric product space-time} and the geometric product~\eqref{geometric product spatial vectors bis} yield the anti-commutation and commutation relations of the spatial algebra $\mathbb{G}^{3}$,
\begin{equation}\label{spatial algebra}
\begin{split}
&\frac{1}{2}\left(\boldsymbol{e}_i\,\boldsymbol{e}_j + \boldsymbol{e}_j\,\boldsymbol{e}_i\right) = \delta_{ij}\\
&\frac{1}{2}\left(\boldsymbol{e}_i\,\boldsymbol{e}_j -\,\boldsymbol{e}_j\,\boldsymbol{e}_i\right) = \varepsilon_{ijk}\,I\,\boldsymbol{e}_k
\end{split}
\end{equation}
which is the Pauli algebra. The spatial algebra $\mathbb{G}^{3}$ (Pauli algebra) is the even subalgebra of the space-time algebra $\mathbb{G}^{1,3}$ (Dirac algebra). In view of relations~\eqref{square space-time} and~\eqref{outer product space-time}, the pseudoscalar anticommutes with the basis vectors in space-time,
\begin{align}\label{pseudoscalar anticommutation vector}
&I\,e_0 = \left(e_0\,e_1\,e_2\,e_3\right)e_0 = -\,e_0\left(e_0\,e_1\,e_2\,e_3\right) = -\,e_0\,I \nonumber\\
&I\,e_1 = \left(e_0\,e_1\,e_2\,e_3\right)e_1 = -\,e_1\left(e_0\,e_1\,e_2\,e_3\right) = -\,e_1\,I \nonumber\\
&I\,e_2 = \left(e_0\,e_1\,e_2\,e_3\right)e_0 = -\,e_2\left(e_0\,e_1\,e_2\,e_3\right) = -\,e_2\,I \\
&I\,e_3 = \left(e_0\,e_1\,e_2\,e_3\right)e_0 = -\,e_3\left(e_0\,e_1\,e_2\,e_3\right) = -\,e_3\,I \nonumber
\end{align}
In view of relations~\eqref{square space-time} and~\eqref{outer product space-time}, the pseudoscalar commutes with the basis bivector in space-time,
\begin{align}\label{pseudoscalar commutation bivector}
&I\,e_0\,e_1 = \left(e_0\,e_1\,e_2\,e_3\right)\left(e_0\,e_1\right) = \left(e_0\,e_1\right)\left(e_0\,e_1\,e_2\,e_3\right) = e_0\,e_1\,I \nonumber\\
&I\,e_0\,e_2 = \left(e_0\,e_1\,e_2\,e_3\right)\left(e_0\,e_2\right) = \left(e_0\,e_2\right)\left(e_0\,e_1\,e_2\,e_3\right) = e_0\,e_2\,I \nonumber\\
&I\,e_0\,e_3 = \left(e_0\,e_1\,e_2\,e_3\right)\left(e_0\,e_3\right) = \left(e_0\,e_3\right)\left(e_0\,e_1\,e_2\,e_3\right) = e_0\,e_3\,I \nonumber\\
&I\,e_1\,e_2 = \left(e_0\,e_1\,e_2\,e_3\right)\left(e_1\,e_2\right) = \left(e_1\,e_2\right)\left(e_0\,e_1\,e_2\,e_3\right) = e_1\,e_2\,I \\
&I\,e_2\,e_3 = \left(e_0\,e_1\,e_2\,e_3\right)\left(e_2\,e_3\right) = \left(e_2\,e_3\right)\left(e_0\,e_1\,e_2\,e_3\right) = e_2\,e_3\,I \nonumber\\
&I\,e_3\,e_1 = \left(e_0\,e_1\,e_2\,e_3\right)\left(e_3\,e_1\right) = \left(e_3\,e_1\right)\left(e_0\,e_1\,e_2\,e_3\right) = e_3\,e_1\,I \nonumber
\end{align}
In view of relations~\eqref{square space-time} and~\eqref{outer product space-time} and~\eqref{pseudoscalar anticommutation vector}, the pseudoscalar anticommutes with the basis trivector in space-time,
\begin{align}\label{pseudoscalar anticommutation trivector}
&I\,e_0\,e_1\,e_2 = I^2\,e_3 = -\,I\,e_3\,I = -\,e_0\,e_1\,e_2\,I \nonumber\\
&I\,e_1\,e_2\,e_3 = I\,e_0\,I = -\,e_0\,I^2 = -\,e_1\,e_2\,e_3\,I \\
&I\,e_2\,e_3\,e_0 = \left(e_0\,e_1\,e_2\,e_3\right)\left(e_2\,e_3\,e_0\right) = -\,\left(e_2\,e_3\,e_0\right)\left(e_0\,e_1\,e_2\,e_3\right) = -\,e_2\,e_3\,e_0\,I\nonumber\\
&I\,e_3\,e_0\,e_1 = \left(e_0\,e_1\,e_2\,e_3\right)\left(e_3\,e_0\,e_1\right) = -\,\left(e_3\,e_0\,e_1\right)\left(e_0\,e_1\,e_2\,e_3\right) = -\,e_3\,e_0\,e_1\,I\nonumber
\end{align}
A contravariant vector $V$ in space-time is written in coordinates with respect to the orthonormal vector frame $\{e_0,e_1,e_2,e_3\}$ as,
\begin{equation}\label{contravariant vector}
V = V^{\mu}\,e_{\mu} = V^{0}\,e_{0} + V^{i}\,e_{i} \qquad\text{where}\qquad i = 1,2,3
\end{equation}
where we used the Einstein summation convention. In view of relation~\eqref{square space-time}, the inner product of the vector $V$ with the time vector $e_{0}$ yields,
\begin{equation}\label{inner product contravariant vector}
V \cdot e_{0} = V^{0}\left(e_{0} \cdot e_{0}\right) + V^{i}\left(e_{i} \cdot e_{0}\right) = V^{0}
\end{equation}
In view of relations~\eqref{square space-time} and~\eqref{space-time bivectors}, the outer product of the vector $V$ with the time vector $e_{0}$ is given by,
\begin{equation}\label{outer product contravariant vector}
V \wedge e_{0} = V^{0}\left(e_{0} \wedge e_{0}\right) + V^{i}\left(e_{i} \wedge e_{0}\right) = V^{i}\,\boldsymbol{e}_{i} = \boldsymbol{v}
\end{equation}
In view of the inner product~\eqref{inner product contravariant vector} and the outer product~\eqref{outer product contravariant vector}, the geometric product of the contravariant vector $V$ and the time vector $e_{0}$ yields,
\begin{equation}\label{geometric product contravariant vector}
V\,e_{0} = V \cdot e_{0} + V \wedge e_{0} = V^{0} + \boldsymbol{v}
\end{equation}
Taking into account the properties~\eqref{inner product space-time} and~\eqref{outer product space-time}, the geometric product in reverse order is given by,
\begin{equation}\label{geometric product contravariant vector bis}
e_{0}\,V = e_{0} \cdot V + e_{0} \wedge V = V \cdot e_{0} -\,V \wedge e_{0} = V^{0} -\,\boldsymbol{v}
\end{equation}
According to relations~\eqref{pseudoscalar anticommutation vector}, the contravariant vector $V = V^{\mu}\,e_{\mu}$ anticommutes with the pseudoscalar $I$,
\begin{equation}\label{contravariant vector anticommutes}
V\,I = -\,I\,V
\end{equation}
The orthonormal comoving vector frame $\{e^0,e^1,e^2,e^3\}$ in space-time is related to the orthonormal vector frame $\{e_0,e_1,e_2,e_3\}$ through the Minkowski metric,
\begin{equation}\label{vectors of space time frames}
e_{\mu} = \eta_{\mu\nu}\,e^{\nu} \qquad\text{and}\qquad e^{\mu} = \eta^{\mu\nu}\,e_{\nu}
\end{equation}
Thus, in view of the definition~\eqref{inner product space-time}, the inner product of two different basis vectors is written as,
\begin{equation}\label{inner product different vectors}
e^{\mu} \cdot e_{\nu} = \eta^{\mu\rho}\,e_{\rho} \cdot e_{\nu} = \eta^{\mu\rho}\,\eta_{\rho\nu} = \delta^{\mu}_{\nu}
\end{equation}
and the outer product of two different basis vectors is written as,
\begin{equation}\label{outer product different vectors}
e^{\mu} \wedge e_{\nu} = \eta^{\mu\rho}\,e_{\rho} \wedge e_{\nu}
\end{equation}
A relative spatial comoving orthonormal vector frame $\{\boldsymbol{e}^1,\boldsymbol{e}^2,\boldsymbol{e}^3\}$ attached to an observer consists of space-time bivectors defined as,
\begin{equation}\label{space-time bivectors comoving}
\boldsymbol{e}^i = e^{i}\,e^{0} = e^{i} \wedge e^{0}
\end{equation}
In view of the identities~\eqref{vectors of space time frames} and~\eqref{outer product different vectors} and the definitions~\eqref{space-time bivectors} and~\eqref{space-time bivectors comoving} of the basis vectors, a comoving basis vector is the opposite of the corresponding basis vector,
\begin{equation}\label{basis vector comoving}
\boldsymbol{e}^i = e^{i} \wedge e^{0} = \eta^{ij}\,\eta^{00}\,e_{j} \wedge e_{0} = -\,\delta^{ij}\,\boldsymbol{e}_j = -\,\boldsymbol{e}_i
\end{equation}
A covariant vector $U$ in space-time is written in coordinates with respect to the comoving orthonormal vector frame $\{e^0,e^1,e^2,e^3\}$ as,
\begin{equation}\label{covariant vector}
U = U_{\mu}\,e^{\mu} = U_{0}\,e^{0} + U_{i}\,e^{i}
\end{equation}
where we used the Einstein summation convention. In view of relation~\eqref{inner product different vectors}, the inner product of the vector $U$ with the time vector $e_{0}$ yields,
\begin{equation}\label{inner product covariant vector}
U \cdot e_{0} = U_{0}\left(e^{0} \cdot e_{0}\right) + U_{i}\left(e^{i} \cdot e_{0}\right) = U_{0}
\end{equation}
In view of relations~\eqref{square space-time} and~\eqref{outer product different vectors}, the outer product of the vector $U$ with the time vector $e_{0}$ is given by,
\begin{equation}\label{outer product covariant vector}
\begin{split}
&U \wedge e_{0} = U_{0}\left(e^{0} \wedge e_{0}\right) + U_{i}\left(e^{i} \wedge e_{0}\right)\\
&\phantom{U \wedge e_{0}} = U_{i}\,\eta_{00}\left(e^{i} \wedge e^{0}\right) = U_{i}\,\boldsymbol{e}^{i} = -\,U^{i}\,\boldsymbol{e}_{i} = -\,\boldsymbol{u}
\end{split}
\end{equation}
In view of the inner product~\eqref{inner product covariant vector} and the outer product~\eqref{outer product covariant vector}, the geometric product of the covariant vector $U$ and the time vector $e_{0}$ yields,
\begin{equation}\label{geometric product covariant vector}
U\,e_{0} = U \cdot e_{0} + U \wedge e_{0} = U_{0} -\,\boldsymbol{u}
\end{equation}
Taking into account the properties~\eqref{inner product space-time} and~\eqref{outer product space-time}, the geometric product in reverse order is given by,
\begin{equation}\label{geometric product coavariant vector bis}
e_{0}\,U = e_{0} \cdot U + e_{0} \wedge U = U \cdot e_{0} -\,U \wedge e_{0} = U_{0} + \boldsymbol{u}
\end{equation}
In view of relations~\eqref{contravariant vector},~\eqref{covariant vector} and~\eqref{inner product different vectors}, the inner product of the covariant vector $U$ with the contravariant vector $V$ yields,
\begin{equation}\label{inner product coavariant contravariant}
\begin{split}
&U \cdot V = \left(U_{0}\,e^{0} + U_{i}\,e^{i}\right)\cdot\left(V^{0}\,e_{0} + V^{j}\,e_{j}\right)\\
&\phantom{U \cdot V} = U_{0}\,V^{0}\left(e^{0} \cdot e_{0}\right) + U_{i}\,V^{j}\left(e^{i}\cdot e_{j}\right) = U_{0}\,V^{0} + U_{i}\,V^{i} 
\end{split}
\end{equation}
According to relations~\eqref{space-time bivectors inner product},~\eqref{outer product contravariant vector} and~\eqref{outer product covariant vector}, the inner product~\eqref{inner product coavariant contravariant} is recast as,
\begin{equation}\label{inner product coavariant contravariant bis}
U \cdot V = U_{0}\,V^{0} + U^{i}\,V^{j}\left(\boldsymbol{e}_{i}\cdot\boldsymbol{e}_{j}\right) = U_{0}\,V^{0} + \boldsymbol{u}\cdot\boldsymbol{v}
\end{equation}
In view of relations~\eqref{outer product space-time},~\eqref{outer product different vectors},~\eqref{contravariant vector} and~\eqref{covariant vector}, the outer product of the covariant vector $U$ with the contravariant vector $V$ yields,
\begin{equation}\label{outer product coavariant contravariant}
\begin{split}
&U \wedge V = \left(U_{0}\,e^{0} + U_{i}\,e^{i}\right)\wedge\left(V^{0}\,e_{0} + V^{j}\,e_{j}\right)\\
&\phantom{U \wedge V} = U_{0}\,V^{j}\left(e^{0} \wedge e_{j}\right) + U_{i}\,V^{0}\left(e^{i}\wedge e_{0}\right) + U_{i}\,V^{j}\left(e^{i} \wedge e_{j}\right)\\
&\phantom{U \wedge V} = -\,U^{0}\,V^{j}\left(e_{j} \wedge e_{0}\right) -\,V^{0}\,U^{i}\left(e_{i}\wedge e_{0}\right) -\,U^{i}\,V^{j}\left(e_{i} \wedge e_{j}\right)
\end{split}
\end{equation}
According to the identities~\eqref{square space-time},~\eqref{outer product space-time} and~\eqref{space-time bivectors}, the outer product of two relative spatial vectors,
\begin{equation}\label{outer product spatial vectors}
\begin{split}
&\boldsymbol{e}_{i} \wedge \boldsymbol{e}_{j} = \left(e_{i} \wedge e_{0}\right)\wedge\left(e_{j} \wedge e_{0}\right) = \left(e_{i}\,e_{0}\right)\left(e_{i}\,e_{0}\right)\\
&\phantom{\boldsymbol{e}_{i} \wedge \boldsymbol{e}_{j}} = -\,\left(e_{i}\,e_{j}\right)\left(e_{0}\,e_{0}\right) = -\,e_{i}\,e_{j} = -\,e_{i} \wedge e_{j}
\end{split}
\end{equation}
According to relations~\eqref{space-time bivectors},~\eqref{outer product contravariant vector},~\eqref{outer product covariant vector} and~\eqref{outer product spatial vectors} the outer product~\eqref{outer product coavariant contravariant} is recast as,
\begin{equation}\label{outer product coavariant contravariant bis}
\begin{split}
&U \wedge V = -\,U^{0}\left(V^{j}\,\boldsymbol{e}_{j}\right) -\,\left(U^{i}\,\boldsymbol{e}_{i}\right)V^{0} + \left(U^{i}\,\boldsymbol{e}_{i}\right)\wedge \left(V^{j}\,\boldsymbol{e}_{j}\right)\\
&\phantom{U \wedge V} = -\,U^{0}\,\boldsymbol{v} -\,\boldsymbol{u}\,V^{0} + \boldsymbol{u} \wedge \boldsymbol{v}
\end{split}
\end{equation}
In view of relations~\eqref{contravariant vector},~\eqref{covariant vector} and~\eqref{inner product different vectors}, the scalar part of the geometric product of two vectors $U$ and $V$, denoted by angle brackets, is written as,
\begin{equation}\label{scalar part vector product}
\begin{split}
&\langle\,U\,V\,\rangle = \langle\,U_{\mu}\,e^{\mu}\,V^{\nu}\,e_{\nu}\,\rangle = U_{\mu}\,V^{\nu}\,\langle\,e^{\mu}\,e_{\nu}\,\rangle = U_{\mu}\,V^{\nu}\,e^{\mu} \cdot e_{\nu} = U_{\mu}\,V^{\mu}\\
&\langle\,V\,U\,\rangle = \langle\,V^{\mu}\,e_{\mu}\,U_{\nu}\,e^{\nu}\,\rangle = V^{\mu}\,U_{\nu}\,\langle\,e_{\mu}\,e^{\nu}\,\rangle = V^{\mu}\,U_{\nu}\,e_{\mu} \cdot e^{\nu} = V^{\mu}\,U_{\mu}
\end{split}
\end{equation}
Thus, the scalar part of this geometric product is symmetric,
\begin{equation}\label{scalar part vector product sym}
\langle\,U\,V\,\rangle = \langle\,V\,U\,\rangle
\end{equation}
We now consider a contravariant multivector in the space-time algebra $\mathbb{G}^{1,3}$ written as,
\begin{equation}\label{spatial multivector contravariant}
M = m + M^{\mu}\,e_{\mu} + \frac{1}{2}\,M^{\mu\nu}\,e_{\mu} \wedge e_{\nu} + \frac{1}{6}\,M^{\mu\nu\rho}\,e_{\mu} \wedge e_{\nu} \wedge e_{\rho} + m^{\prime}\,I
\end{equation}
and a covariant multivector given by,
\begin{equation}\label{spatial multivector covariant}
N = n + N_{\mu}\,e^{\mu} + \frac{1}{2}\,N_{\mu\nu}\,e^{\mu} \wedge e^{\nu} + \frac{1}{6}\,N_{\mu\nu\rho}\,e^{\mu} \wedge e^{\nu} \wedge e^{\rho} + n^{\prime}\,I
\end{equation}
Using the algebraic relation~\eqref{geometric product spatial vectors bis}, the pseudoscalar~\eqref{space-time pseudoscalar} and the identities~\eqref{vectors of space time frames} and~\eqref{inner product different vectors} for different indices $i$, $j$ and $k$, we obtain,
\begin{align}\label{pseudoscalar relations bis}
&e_{i} \wedge e_{0} = -\,e_{0} \wedge e_{i} = \boldsymbol{e}_i\nonumber\\
&e^{i} \wedge e^{0} = -\,e^{0} \wedge e^{i} = \eta^{00}\,\eta^{ij}\,e_j \wedge e_{0} = -\,\boldsymbol{e}_i\nonumber\\
&e_{i} \wedge e_{j} = -\,\left(e_{i} \wedge e_{0}\right)\wedge\left(e_{j} \wedge e_{0}\right) = -\,\boldsymbol{e}_i\wedge\boldsymbol{e}_j = -\,\left(\boldsymbol{e}_i\wedge\boldsymbol{e}_j\wedge\boldsymbol{e}_k\right)\boldsymbol{e}_k = -\,\varepsilon_{ijk}\,I\,\boldsymbol{e}_k\nonumber\\
&e^{i} \wedge e^{j} = \eta^{ik}\,\eta^{j\ell}\,e_k \wedge e_{\ell} = e_{i} \wedge e_{j} = -\,\varepsilon_{ijk}\,I\,\boldsymbol{e}_k
\end{align}
which implies that the bivector part of the multivectors~\eqref{spatial multivector contravariant} and~\eqref{spatial multivector covariant} can be recast as,
\begin{equation}\label{bivector part}
\begin{split}
&\frac{1}{2}\,M^{\mu\nu}\,e_{\mu} \wedge e_{\nu} = \left(M^{i} + M^{\prime\,i}\,I\right)\boldsymbol{e}_{i}\\
&\frac{1}{2}\,N_{\mu\nu}\,e^{\mu} \wedge e^{\nu} = \left(N_{i} + N^{\prime}_{i}\,I\right)\boldsymbol{e}_{i}
\end{split}
\end{equation}
Using the algebraic relation~\eqref{inner product space-time}, the pseudoscalar~\eqref{space-time pseudoscalar} and the identities~\eqref{vectors of space time frames} and~\eqref{inner product different vectors} for different indices $\mu$, $\nu$, $\rho$ and $\alpha$, we obtain, 
\begin{equation}\label{pseudoscalar relations}
\begin{split}
&e_{\mu} \wedge e_{\nu} \wedge e_{\rho} = \left(e_{\mu} \wedge e_{\nu} \wedge e_{\rho} \wedge e_{\alpha}\right)e_{\beta}\,\eta_{\alpha\beta} = I\,e_{\beta}\,\eta_{\alpha\beta}\\
&e^{\mu} \wedge e^{\nu} \wedge e^{\rho} = \left(e^{\mu} \wedge e^{\nu} \wedge e^{\rho} \wedge e^{\alpha}\right)e_{\alpha} = I\,e^{\beta}\,\eta^{\alpha\beta}
\end{split}
\end{equation}
which implies that the trivector part of the multivectors~\eqref{spatial multivector contravariant} and~\eqref{spatial multivector covariant} can be recast as,
\begin{equation}\label{trivector part}
\begin{split}
&\frac{1}{6}\,M^{\mu\nu\rho}\,e_{\mu} \wedge e_{\nu} \wedge e_{\rho} = M^{\mu}\,I\,e_{\mu}\\
&\frac{1}{6}\,N_{\mu\nu\rho}\,e^{\mu} \wedge e^{\nu} \wedge e^{\rho} = N_{\mu}\,I\,e^{\mu}
\end{split}
\end{equation}
In view of relations~\eqref{trivector part} and~\eqref{bivector part}, the contravariant multivector~\eqref{spatial multivector contravariant} is recast as,
\begin{equation}\label{spatial multivector contravariant bis}
M = \left(m + m^{\prime}\,I\right) + \left(M^{i} + M^{\prime\,i}\,I\right)\boldsymbol{e}_{i} + \left(M^{\mu} + M^{\prime\mu}\,I\right)e_{\mu} 
\end{equation}
and the covariant multivector~\eqref{spatial multivector covariant} is recast as,
\begin{equation}\label{spatial multivector covariant bis}
N = \left(n + n^{\prime}\,I\right) + \left(N_{i} + N^{\prime}_{i}\,I\right)\boldsymbol{e}_{i} + \left(N_{\mu} + N^{\prime}_{\mu}\,I\right)e^{\mu} 
\end{equation}
In view of identities~\eqref{space-time bivectors inner product} and~\eqref{inner product different vectors}, the scalar part of the geometric products of two multivectors $M$ and $N$ is written as,
\begin{align}\label{scalar part multivector product}
&\langle\,N\,M\,\rangle = \left(n + n^{\prime}\,I\right)\left(m + m^{\prime}\,I\right) + \left(N_{i} + N^{\prime}_{i}\,I\right)\left(M^{i} + M^{\prime\,i}\,I\right)\nonumber\\
&\phantom{\langle\,N\,M\,\rangle =} + \left(N_{\mu} + N^{\prime}_{\mu}\,I\right)\left(M^{\mu} + M^{\prime\mu}\,I\right)\\
&\langle\,M\,N\,\rangle = \left(m + m^{\prime}\,I\right)\left(n + n^{\prime}\,I\right) + \left(M^{i} + M^{\prime\,i}\,I\right)\left(N_{i} + N^{\prime}_{i}\,I\right)\nonumber\\
&\phantom{\langle\,M\,N\,\rangle =} + \left(M^{\mu} + M^{\prime\mu}\,I\right)\left(N_{\mu} + N^{\prime}_{\mu}\,I\right)
\end{align}
Thus, the scalar part of the geometric product of two multivectors $M$ and $N$ is symmetric,
\begin{equation}\label{multivectors symmetric}
\langle\,M\,N\,\rangle = \langle\,N\,M\,\rangle
\end{equation}
Writing the multivector $N$ as the geometric product of two multivectors $R$ and $S$ in the space-time algebra $\mathbb{G}^{1,3}$, relation~\eqref{multivectors symmetric} yields the cyclic permutation identity,
\begin{equation}\label{multivectors cyclic permutation}
\langle\,M\,R\,S\,\rangle = \langle\,R\,S\,M\,\rangle = \langle\,S\,M\,R\,\rangle
\end{equation}
Writing the multivector $R$ as the geometric product of two multivectors $N$ and $Q$ in the space-time algebra $\mathbb{G}^{1,3}$, relation~\eqref{multivectors cyclic permutation} yields the cyclic permutation identity,
\begin{equation}\label{multivectors cyclic permutation bis}
\langle\,M\,N\,Q\,S\,\rangle = \langle\,N\,Q\,S\,M\,\rangle = \langle\,Q\,S\,M\,N\,\rangle = \langle\,S\,M\,N\,Q\,\rangle
\end{equation}
In the spatial algebra $\mathbb{G}^3$, the triple product~\eqref{triple product vectors} for the orthonormal basis vectors $\boldsymbol{u} = \boldsymbol{e}^{i}$, $\boldsymbol{v} = \boldsymbol{e}_{j}$ and $\boldsymbol{w} = \boldsymbol{e}_{k}$ is written as,
\begin{equation}\label{triple product basis vectors}
\boldsymbol{e}^{i}\cdot\left(\boldsymbol{e}_{j}\wedge\boldsymbol{e}_{k}\right) = \left(\boldsymbol{e}^{i}\cdot\boldsymbol{e}_{j}\right)\boldsymbol{e}_{k} -\,\left(\boldsymbol{e}^{i}\cdot\boldsymbol{e}_{k}\right)\boldsymbol{e}_{j}
\end{equation}
In the space-time algebra $\mathbb{G}^{1,3}$, the triple product for the orthonormal basis vectors is given by,
\begin{equation}\label{triple product basis vectors STA}
e^{\mu}\cdot\left(e_{\nu} \wedge e_{\rho}\right) = \left(e^{\mu} \cdot e_{\nu}\right)e_{\rho} -\,\left(e^{\mu} \cdot e_{\rho}\right)e_{\nu}
\end{equation}
or equivalently by,
\begin{equation}\label{triple product basis vectors STA bis}
e_{\mu}\cdot\left(e^{\nu} \wedge e^{\rho}\right) = \left(e_{\mu} \cdot e^{\nu}\right)e^{\rho} -\,\left(e_{\mu} \cdot e^{\rho}\right)e^{\nu}
\end{equation}
In the spatial algebra $\mathbb{G}^3$, using the antisymmetry~\eqref{inner product v B antisymmetric}, the triple product~\eqref{triple product basis vectors} is recast as,
\begin{equation}\label{triple product basis vectors bis}
\left(\boldsymbol{e}_{j}\wedge\boldsymbol{e}_{k}\right)\cdot\boldsymbol{e}^{i} = \left(\boldsymbol{e}^{i}\cdot\boldsymbol{e}_{k}\right)\boldsymbol{e}_{j} -\,\left(\boldsymbol{e}^{i}\cdot\boldsymbol{e}_{j}\right)\boldsymbol{e}_{k}
\end{equation}
In the space-time algebra $\mathbb{G}^{1,3}$, the triple product for the orthonormal basis vectors is given by,
\begin{equation}\label{triple product basis vectors STA ter}
\left(e_{\nu} \wedge e_{\rho}\right)\cdot e^{\mu} = \left(e^{\mu}\cdot e_{\rho}\right)e_{\nu} -\,\left(e^{\mu} \cdot e_{\nu}\right)e_{\rho}
\end{equation}
or equivalently by,
\begin{equation}\label{triple product basis vectors STA quad}
\left(e^{\nu} \wedge e^{\rho}\right)\cdot e_{\mu} = \left(e_{\mu}\cdot e^{\rho}\right)e^{\nu} -\,\left(e_{\mu} \cdot e^{\nu}\right)e^{\rho}
\end{equation}
In view of identities~\eqref{triple product basis vectors STA} and~\eqref{triple product basis vectors STA ter},
\begin{equation}\label{triple product basis vectors STA anti 1}
e^{\mu}\cdot\left(e_{\nu} \wedge e_{\rho}\right) = -\,\left(e_{\nu} \wedge e_{\rho}\right)\cdot e^{\mu}
\end{equation}
and in view of identities~\eqref{triple product basis vectors STA bis} and~\eqref{triple product basis vectors STA quad},
\begin{equation}\label{triple product basis vectors STA anti 2}
e_{\mu}\cdot\left(e^{\nu} \wedge e^{\rho}\right) = -\,\left(e^{\nu} \wedge e^{\rho}\right)\cdot e_{\mu}
\end{equation}
In view of relation~\eqref{triple product basis vectors STA anti 1}, the inner product of a covariant vector $U = U_{\mu}\,e^{\mu}$ and a contravariant bivector $A = \frac{1}{2}\,A^{\nu\rho}\,e_{\mu} \wedge e_{\rho}$ is antisymmetric,
\begin{equation}\label{inner product antisym STA}
U \cdot A = \frac{1}{2}\,U_{\mu}\,A^{\nu\rho}\,\Big(e^{\mu}\cdot\left(e_{\nu} \wedge e_{\rho}\right)\Big) = -\,\frac{1}{2}\,A^{\nu\rho}\,U_{\mu}\,\Big(\left(e_{\nu} \wedge e_{\rho}\right)\cdot e^{\mu}\Big) = -\,A \cdot U
\end{equation}
which implies that,
\begin{equation}\label{inner product antisym STA bis}
U \cdot A = \frac{1}{2}\left(U\,A -\,A\,U\right)
\end{equation}
In view of relation~\eqref{triple product basis vectors STA anti 2}, the inner product of a contravariant vector $V = V^{\mu}\,e_{\mu}$ and a covariant bivector $B = \frac{1}{2}\,B_{\nu\rho}\,e^{\mu} \wedge e^{\rho}$ is antisymmetric,
\begin{equation}\label{inner product antisym STA ter}
V \cdot B = \frac{1}{2}\,V^{\mu}\,B_{\nu\rho}\,\Big(e_{\mu}\cdot\left(e^{\nu} \wedge e^{\rho}\right)\Big) = -\,\frac{1}{2}\,B_{\nu\rho}\,V^{\mu}\,\Big(\left(e^{\nu} \wedge e^{\rho}\right)\cdot e_{\mu}\Big) = -\,B \cdot V
\end{equation}
which implies that,
\begin{equation}\label{inner product antisym STA quad}
V \cdot B = \frac{1}{2}\left(V\,B -\,B\,V\right)
\end{equation}
In view of relation~\eqref{outer product space-time}, the outer product of a contravariant vector $V = V^{\mu}\,e_{\mu}$ and a contravariant bivector $A = \frac{1}{2}\,A^{\nu\rho}\,e_{\mu} \wedge e_{\rho}$ is symmetric,
\begin{equation}\label{outer product sym STA}
V \wedge A = \frac{1}{2}\,V^{\mu}\,A^{\nu\rho}\,\Big(e_{\mu}\wedge\left(e_{\nu} \wedge e_{\rho}\right)\Big) = \frac{1}{2}\,A^{\nu\rho}\,V^{\mu}\,\Big(\left(e_{\nu} \wedge e_{\rho}\right)\wedge e_{\mu}\Big) = A \wedge V
\end{equation}
which implies that,
\begin{equation}\label{outer product sym STA bis}
V \wedge A = \frac{1}{2}\left(V\,A + A\,V\right)
\end{equation}
In view of relations~\eqref{outer product space-time} and~\eqref{outer product different vectors}, the outer product of a covariant vector $U = U_{\mu}\,e^{\mu}$ and a covariant bivector $B = \frac{1}{2}\,B_{\nu\rho}\,e^{\mu} \wedge e^{\rho}$ is symmetric,
\begin{equation}\label{outer product sym STA ter}
U \wedge B = \frac{1}{2}\,U_{\mu}\,B_{\nu\rho}\,\Big(e^{\mu}\wedge\left(e^{\nu} \wedge e^{\rho}\right)\Big) = \frac{1}{2}\,B^{\nu\rho}\,U^{\mu}\,\Big(\left(e^{\nu} \wedge e^{\rho}\right)\wedge e^{\mu}\Big) = B \wedge U
\end{equation}
which implies that,
\begin{equation}\label{outer product sym STA quad}
U \wedge B = \frac{1}{2}\left(U\,B + B\,U\right)
\end{equation}
In the space-time algebra $\mathbb{G}^{1,3}$, the inner product of the bivectors $A$ and $B$ is defined as the symmetric scalar part of the geometric product of the bivectors,
\begin{equation}\label{inner product STA bivectors}
A \cdot B = \langle\,A\,B\,\rangle = \frac{1}{2}\,\langle\,A\,B + B\,A\,\rangle
\end{equation}
The inner product of the orthonormal basis bivectors is given by,
\begin{equation}\label{inner product basis bivectors}
\left(e^{\mu} \wedge e^{\nu}\right)\cdot\left(e_{\rho} \wedge e_{\sigma}\right) = \left(e^{\nu} \cdot e_{\rho}\right)\left(e^{\mu} \cdot e_{\sigma}\right) -\,\left(e^{\nu} \cdot e_{\sigma}\right)\left(e^{\mu} \cdot e_{\rho}\right)
\end{equation}
or equivalently by,
\begin{equation}\label{inner product basis bivectors bis}
\left(e_{\rho} \wedge e_{\sigma}\right)\cdot\left(e^{\mu} \wedge e^{\nu}\right) = \left(e_{\sigma}\cdot e^{\mu}\right)\left(e_{\rho} \cdot e^{\nu}\right) -\,\left(e_{\sigma} \cdot e^{\nu}\right)\left(e_{\rho} \cdot e^{\mu}\right)
\end{equation}
which yields the symmetry,
\begin{equation}\label{inner product basis bivectors symmetry}
\left(e^{\mu} \wedge e^{\nu}\right)\cdot\left(e_{\rho} \wedge e_{\sigma}\right) = \left(e_{\rho} \wedge e_{\sigma}\right)\cdot\left(e^{\mu} \wedge e^{\nu}\right)
\end{equation}
In view of relations~\eqref{inner product basis bivectors bis},~\eqref{inner product basis bivectors symmetry}, the inner product of a covariant bivector $A = \frac{1}{2}\,A_{\mu\nu}\,e^{\mu} \wedge e^{\nu}$ and a contravariant bivector $B = \frac{1}{2}\,B^{\rho\sigma}\,e_{\rho} \wedge e_{\sigma}$ is a scalar,
\begin{equation}\label{inner product sym STA bivectors}
\begin{split}
&A \cdot B = \frac{1}{4}\,A_{\mu\nu}\,B^{\rho\sigma}\,\Big(\left(e^{\mu} \wedge e^{\nu}\right)\cdot\left(e_{\rho} \wedge e_{\sigma}\right)\Big)\\
&\phantom{A \cdot B} = \frac{1}{4}\,B^{\rho\sigma}\,A_{\mu\nu}\,\Big(\left(e_{\rho} \wedge e_{\sigma}\right)\cdot\left(e^{\mu} \wedge e^{\nu}\right)\Big) = B \cdot A
\end{split}
\end{equation}
In the space-time algebra $\mathbb{G}^{1,3}$, the commutator of the bivectors $A$ and $B$ is defined as the antisymmetric part of the geometric product of the bivectors,
\begin{equation}\label{commutator STA bivectors}
A \times B = \frac{1}{2}\left(AB-\,BA\right)
\end{equation}
The commutator of the orthonormal basis bivectors is given by,
\begin{equation}\label{commutator basis bivectors}
\begin{split}
&\left(e^{\mu} \wedge e^{\nu}\right)\times\left(e_{\rho} \wedge e_{\sigma}\right) = \left(e^{\nu} \cdot e_{\rho}\right)\left(e^{\mu} \wedge e_{\sigma}\right) -\,\left(e^{\nu} \cdot e_{\sigma}\right)\left(e^{\mu} \wedge e_{\rho}\right)\\
&\phantom{\left(e^{\mu} \wedge e^{\nu}\right)\times\left(e_{\rho} \wedge e_{\sigma}\right) =} -\,\left(e^{\mu} \cdot e_{\rho}\right)\left(e^{\nu} \wedge e_{\sigma}\right) + \left(e^{\mu} \cdot e_{\sigma}\right)\left(e^{\nu} \wedge e_{\rho}\right)
\end{split}
\end{equation}
or equivalently by,
\begin{equation}\label{commutator basis bivectors bis}
\begin{split}
&\left(e_{\rho} \wedge e_{\sigma}\right)\times\left(e^{\mu} \wedge e^{\nu}\right) = \left(e_{\sigma}\cdot e^{\mu}\right)\left(e_{\rho} \wedge e^{\nu}\right) -\,\left(e_{\sigma} \cdot e^{\nu}\right)\left(e_{\rho} \wedge e^{\mu}\right)\\
&\phantom{\left(e_{\rho} \wedge e_{\sigma}\right)\times\left(e^{\mu} \wedge e^{\nu}\right) =} -\,\left(e_{\rho} \cdot e^{\mu}\right)\left(e_{\sigma} \wedge e^{\nu}\right) + \left(e_{\rho} \cdot e^{\nu}\right)\left(e_{\sigma} \wedge e^{\mu}\right)
\end{split}
\end{equation}
which yields the antisymmetry,
\begin{equation}\label{commutator basis bivectors antisymmetry}
\left(e^{\mu} \wedge e^{\nu}\right)\times\left(e_{\rho} \wedge e_{\sigma}\right) = -\,\left(e_{\rho} \wedge e_{\sigma}\right)\times\left(e^{\mu} \wedge e^{\nu}\right)
\end{equation}
In the space-time algebra $\mathbb{G}^{1,3}$, the outer product of the bivectors $A$ and $B$ is the symmetric pseudoscalar part of the geometric product of the bivectors,
\begin{equation}\label{outer product STA bivectors}
A \wedge B = \frac{1}{2}\left(A\,B + B\,A\right) -\,\frac{1}{2}\,\langle\,A\,B + B\,A\,\rangle
\end{equation}
In view of relations~\eqref{commutator STA bivectors},~\eqref{commutator basis bivectors antisymmetry}, the commutator of a covariant bivector $A = \frac{1}{2}\,A_{\mu\nu}\,e^{\mu} \wedge e^{\nu}$ and a contravariant bivector $B = \frac{1}{2}\,B^{\rho\sigma}\,e_{\rho} \wedge e_{\sigma}$ is a antisymmetric bivector,
\begin{equation}\label{commutator basis STA bivectors}
\begin{split}
&A \times B = \frac{1}{4}\,A_{\mu\nu}\,B^{\rho\sigma}\,\Big(\left(e^{\mu} \wedge e^{\nu}\right)\times\left(e_{\rho} \wedge e_{\sigma}\right)\Big)\\
&\phantom{A \times B} = -\,\frac{1}{4}\,B^{\rho\sigma}\,A_{\mu\nu}\,\Big(\left(e_{\rho} \wedge e_{\sigma}\right)\times\left(e^{\mu} \wedge e^{\nu}\right)\Big) = -\,B \times A
\end{split}
\end{equation}
In view of relation~\eqref{outer product space-time}, the outer product of a covariant bivector $A = \frac{1}{2}\,A_{\mu\nu}\,e^{\mu} \wedge e^{\nu}$ and a contravariant bivector $B = \frac{1}{2}\,B^{\rho\sigma}\,e_{\rho} \wedge e_{\sigma}$ is a symmetric pseudoscalar,
\begin{equation}\label{outer inner product sym STA bivectors}
\begin{split}
&A \wedge B = \frac{1}{4}\,A_{\mu\nu}\,B^{\rho\sigma}\,\Big(\left(e^{\mu} \wedge e^{\nu}\right)\wedge\left(e_{\rho} \wedge e_{\sigma}\right)\Big)\\
&\phantom{A \wedge B} = \frac{1}{4}\,B^{\rho\sigma}\,A_{\mu\nu}\,\Big(\left(e_{\rho} \wedge e_{\sigma}\right)\wedge\left(e^{\mu} \wedge e^{\nu}\right)\Big) = B \wedge A
\end{split}
\end{equation}
In view of relations~\eqref{inner product STA bivectors} and~\eqref{outer inner product sym STA bivectors}, the sum of the inner product and the outer product of two bivectors is written as,
\begin{equation}\label{other products bivectors symmetry STA}
A \cdot B + A \wedge B = \frac{1}{2}\left(A\,B + B\,A\right)
\end{equation}
Thus, according to relations~\eqref{commutator STA bivectors} and~\eqref{other products bivectors symmetry STA}, the geometric product of two bivectors in the space-time algebra $\mathbb{G}^{1,3}$ is written as,
\begin{equation}\label{geometric product bivectors STA}
A\,B = A \cdot B + A \times B + A \wedge B
\end{equation}
According to relations~\eqref{pseudoscalar commutation bivector}, the contravariant bivector $A = A^{\mu\nu}\,e_{\mu} \wedge e_{\nu}$ commutes with the pseudoscalar $I$,
\begin{equation}\label{contravariant bivector commutes}
A\,I = I\,A
\end{equation}
According to relations~\eqref{pseudoscalar anticommutation trivector}, the contravariant trivector $T = T^{\mu\nu\rho}\,e_{\mu} \wedge e_{\nu} \wedge e_{\rho}$ anticommutes with the pseudoscalar $I$,
\begin{equation}\label{contravariant triivector anticommutes}
T\,I = -\,I\,T
\end{equation}
%


\bibliography{references}
\bibliographystyle{plainnat} 

\end{document}